\documentclass[11pt,a4paper]{article}
\usepackage{longtable}
\usepackage{amsmath}
\usepackage{amssymb}
\usepackage{graphicx}
\usepackage{bm}
\usepackage{footmisc}
\usepackage{afterpage}

\newcommand*{\no}{\noindent}
\newcommand*{\bea}{\begin{eqnarray}}
\newcommand*{\eea}{\end{eqnarray}}
\newcommand*{\be}{\begin{equation}}
\newcommand*{\ee}{\end{equation}}
\newcommand*{\pd}{\partial}
\newcommand*{\pdm}{\pd_{\mu}}

\newcommand*{\pref}[1]{(\ref{#1})}

\newcommand*{\prefr}[2]{(\ref{#1}-\ref{#2})} 
\newcommand*{\nn}{\nonumber}

\newcommand*{\tr}{\mathrm{tr}}

 \usepackage[square,comma,sort&compress,numbers]{natbib}

\title{Some more details of minimal-Landau-gauge SU(2) Yang-Mills propagators}

\author{Axel Maas\\
Institute of Physics, University of Graz,\\
Universit\"atsplatz 5, A-8010 Graz, Austria}

\begin{document}

 \renewcommand{\topfraction}{1}	
     \renewcommand{\bottomfraction}{1}	
     \setcounter{topnumber}{2}
     \setcounter{bottomnumber}{2}
     \setcounter{totalnumber}{2}     
     \setcounter{dbltopnumber}{2}    
     \renewcommand{\dbltopfraction}{0.99}	
     \renewcommand{\textfraction}{0.00}	
     \renewcommand{\floatpagefraction}{0.99}	
   \renewcommand{\dblfloatpagefraction}{0.99}	

\maketitle

\begin{abstract}
The propagators of the elementary degrees of freedom of (minimal-)Landau-gauge Yang-Mills theory have been a useful tool in various investigations. However, in lattice calculations they show severe dependencies on lattice artifacts. This problem has been addressed for various subsets of lattice artifacts and various subsets of propagators over the time. Here, an extended study of all propagators in momentum space, and for the gluon also in position space, as well as derived quantities like the running coupling, is provided simultaneously for two, three, and four dimensions over one or more orders of magnitude in both physical volume and lattice spacing, in lower dimensions also over more than two orders of magnitude for the gauge group SU(2). Most of the known qualitative results are confirmed, but two quantities also indicate a slight, but possibly interesting deviation.
\end{abstract}

\section{Introduction}

The propagators of the elementary degrees of freedom of (minimal Landau-gauge) Yang-Mills theory, the gluon propagator and the ghost propagator, have been a subject of intense study over the last 15 years, see  \cite{Maas:2011se,Alkofer:2000wg,Fischer:2006ub,Binosi:2009qm,Boucaud:2011ug,Vandersickel:2012tg} for reviews. The reason for this interest is that the propagators hold themselves interesting information on the structure of the theory and are also building blocks, e.\ g.\ in functional methods, to determine experimentally accessible quantities. Lattice calculations have been one of the main methods to study them, ever since the pioneering works of \cite{Mandula:1987rh,Suman:1995zg}.

Unfortunately, it turns out that these propagators are quite sensitive to lattice artifacts, especially in terms of volume, but also in terms of the discretization. This has been established in a long series of investigations \cite{Maas:2011se,Boucaud:2011ug,Cucchieri:1997dx,Cucchieri:1999sz,Cucchieri:2003di,Bloch:2003sk,Silva:2005hb,Maas:2007uv,Sternbeck:2007ug,Bogolubsky:2007ud,Cucchieri:2007rg,Fischer:2007pf,Cucchieri:2008fc,Sternbeck:2008mv,Desoto:2009aw,Cucchieri:2009zt,Maas:2009ph,Bogolubsky:2009dc,Bornyakov:2009ug,Oliveira:2012eh,Bornyakov:2013ysa,Simeth:2013ima,Bornyakov:2013pha}. The result of these investigations was that the gluon propagator is infrared finite in three and four dimensions, but vanishing in two dimensions. At the same time, the ghost propagator is in three and four dimensions close to the one of a free particle, while it is stronger enhanced in two dimensions. These results are reviewed in \cite{Maas:2011se,Alkofer:2000wg,Fischer:2006ub,Binosi:2009qm,Boucaud:2011ug,Vandersickel:2012tg}, including also the wealth of results from continuum methods, which will not be detailed here. This result applies to the case of minimal Landau gauge \cite{Maas:2011se}. A different way of choosing Gribov copies may lead to different results, but since any alternative is much more expensive in terms of computational time, results have not yet been available for as large a range of lattice parameters as for the minimal Landau gauge. This will also not be the subject of the present investigation, but will be treated in an upcoming separate one \cite{Maas:unpublished}. 

Most of these investigations have concentrated on the gluon propagator in momentum space. Compared to the gluon propagator, lattice artifacts for the ghost propagator and the gluon Schwinger function have much less been studied. As a consequence, also the running coupling, which can be determined from the propagators, has not yet been explored in as great a detail, except at rather large momenta on (comparatively) small lattices.

While it is not expected that there is still any qualitative change to be found, this situation is not satisfactory from the point of view of applications. Especially the running coupling is of prime importance when coupling to quarks \cite{Fischer:2006ub,Alkofer:2000wg,Binosi:2009qm}. Precision information are therefore highly valuable, e.\ g.\ to constrain quantitatively truncations in functional methods.

To provide such a systematic study is therefore the aim of the present work. The lattice setups and methods will be presented in section \ref{ssetup}, with renormalization treated in section \ref{srenorm}. Two particular kinds of derived lattice artifacts are then highlighted in sections \ref{scdep} and \ref{srot}, before the gluon and ghost propagator results are discussed in detail in section \ref{sgp} and \ref{sghp}. The running coupling, and in four dimensions also the $\beta$-function, are finally analyzed in section \ref{salpha}.

The reason for choosing the gauge group SU(2) for this study is primarily its cheapness in simulations. For larger groups such an extensive study would not have been possible with the available resources. Moreover, so far no significant qualitative difference has been found in comparison to other gauge groups for the quantities studied here \cite{Maas:2010qw,Cucchieri:2007zm,Maas:2011se,Bogolubsky:2009dc,Cucchieri:2008fc,Cucchieri:2007rg,Sternbeck:2007ug,Oliveira:2008uf}. The reason for studying two, three, and four dimensions is different. On the one hand, it is possible to study a much larger parameter range in lower dimensions due to reduced numerical costs. Secondly, both lower dimensional theories have properties which make them uniquely different from four dimensions. Two dimensions appears to show a qualitatively different infrared behavior \cite{Maas:2011se,Maas:2007uv,Cucchieri:2008fc,Cucchieri:2007rg}, which will also be confirmed here. The natural question is, whether this introduces a different dependency on lattice artifacts. Three dimensions show a behavior quite similar to four dimensions \cite{Maas:2011se,Cucchieri:2008fc,Cucchieri:2007rg}. However, no renormalization is necessary in three dimensions, as all correlation functions are finite. It is therefore possible to study whether the lattice artifacts show a different behavior in the presence of renormalization. The answer appears to be mostly no, as will be discussed in detail below.

\section{Setup}\label{ssetup}

The lattice simulations have been performed in two, three, and four dimensions using the standard SU(2) Wilson action\footnote{See \cite{Gong:2008td} for similar calculations using an improved action.}, using the methods described in \cite{Cucchieri:2006tf} with a combination of over-relaxation and heat-bath sweeps. A lattice-volume-dependent and dimension-dependent number of updates have been discarded for thermalization and decorrelation. Unfortunately, gauge-fixing prevents a reliable determination of auto-correlation time for gauge-fixed quantities, as two differing Gribov copies do not have necessarily the same correlation for gauge-dependent quantities. Therefore, the numbers have been chosen rather large, ${\cal{O}}(100)$ between measurements and ${\cal{O}}(1000)$ for thermalization. Furthermore, all results have been obtained from ${\cal{O}}(10-100)$  independent runs to further reduce auto-correlation artifacts. Where possible, enough statistics was created to obtain less than 10\% statistical error for the effective ghost infrared exponent determined with the methods described in \cite{Maas:2007uv}, and defined below in \pref{effexpg}.

In a discussion of lattice artifacts it is also necessary to mention that lattice algorithms tend to get stuck towards the continuum limit in a sector of fixed topological (net-)charge \cite{DelDebbio:2002xa}. This could potentially affect the propagators, but there is evidence that they do not depend on the topological charge \cite{Maas:2011se,Maas:2008uz,Maas:unpublished}. Therefore, no measures will be taken here to avoid this problem, except for making many independent runs with long decorrelation times for each lattice setup.

To determine the lattice spacing $a$, interpolations of the results for the string tension for four dimensions from \cite{Fingberg:1992ju} and for three dimensions from \cite{Teper:1998te} have been used. In two dimensions the exact (infinite-volume) formula from \cite{Dosch:1978jt} has been used. The employed value of the string tension is $\sigma=(440$ MeV$)^2$ in all dimensions. This gives the following formulas for the lattice spacing
\bea
a_{2d}(\beta)&=&\sqrt{\frac{-\ln\frac{I_2(\beta)}{I_1(\beta)}}{\sigma}}\nn\\
a_{3d}(\beta)&=&\frac{1\text{ GeV}}{\sigma\beta}\left(1.337+\frac{0.95}{\beta}+\frac{1.1}{\beta^2}\right)\nn\\
a_{4d}(\beta)&=&\frac{1\text{ GeV}}{\sigma\beta}\left(31.85-\frac{237.1}{\beta}+\frac{574.4}{\beta}-\frac{444.3}{\beta^3}\right)\nn
\eea
\no where the formula in three dimensions is from \cite{Teper:1998te} and $I_i$ are the $i$th Bessel functions. These captured for the employed range of $\beta$, 6.75-27000 in two dimensions, 3.48-46.9 in three dimensions, and 2.19-2.861 in four dimensions, the quantity $a(\beta)$ sufficiently well. This was monitored by the dependence of the plaquette on $a$, which showed a smooth behavior\footnote{There are indications that a lattice bulk transition exists in the lower-dimensional cases in these regions of $\beta$ \cite{Burgio:2007np}. The results here did not show any impact due to the possible presence of such a transition, except in form of larger statistical fluctuations in this range of $\beta$ values.}. This range of $\beta$ corresponds to lattice spacings starting always with 0.22 fm down to 0.0033 fm in two dimensions, 0.013 fm in three dimensions, and 0.029 fm in four dimensions, spanning therefore between one and two orders of magnitude.

\begin{figure}
\includegraphics[width=\linewidth]{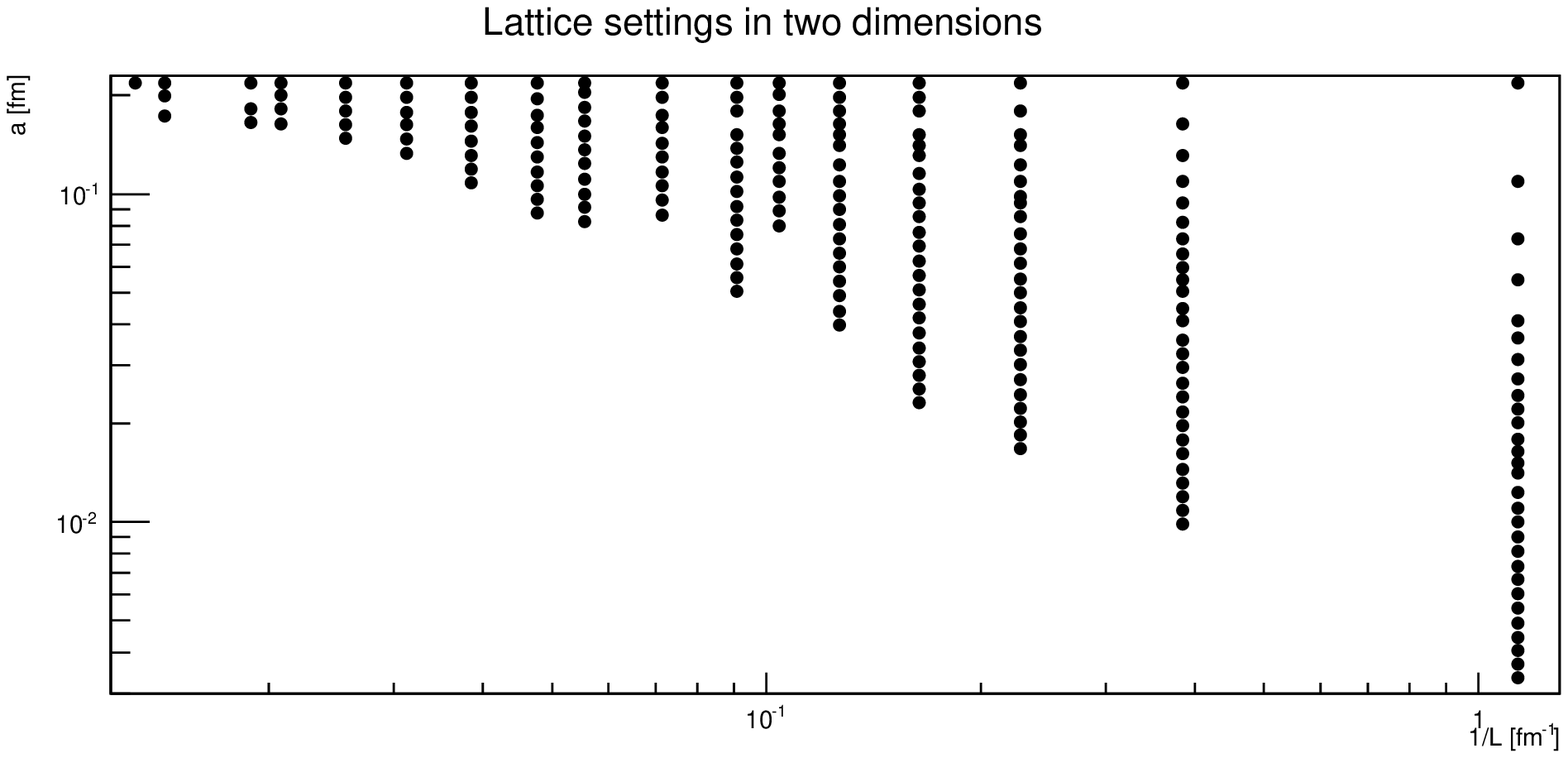}\\
\includegraphics[width=\linewidth]{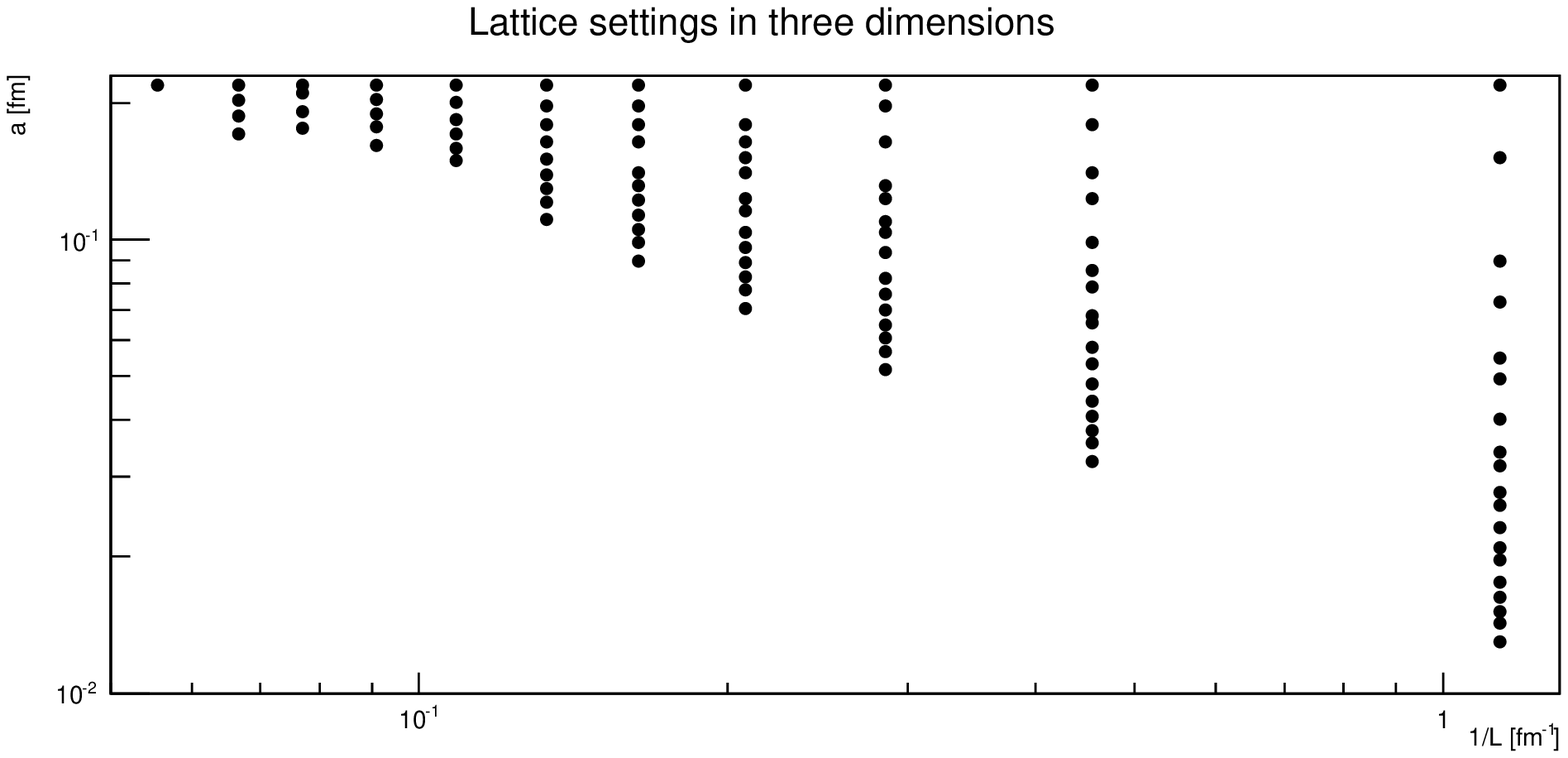}\\
\includegraphics[width=\linewidth]{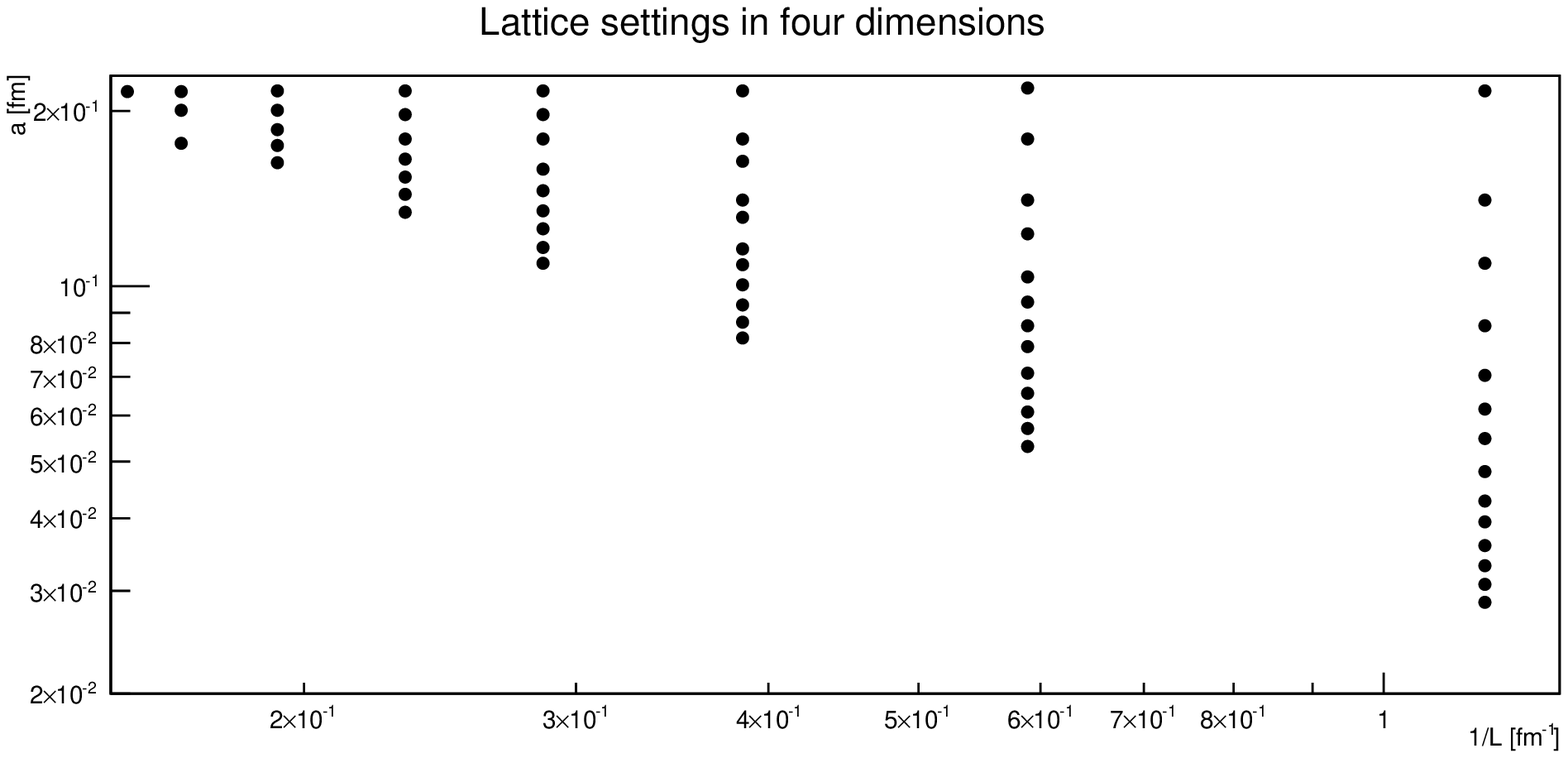}
\caption{\label{fig:va}The lattice settings employed, in physical units, for two dimensions (top panel), three dimensions (middle panel), and four dimensions (bottom panel).}
\end{figure}

The set of physical parameters for all dimensions $d$ employed for the present investigation are displayed in figure \ref{fig:va}. The symmetric lattice volumes $N^d$ range between 10-406 in two dimensions, 8-88 in three dimensions, and 6-34 in four dimensions. This corresponds to physical volumes $V=L^d=(aN)^d$ ranging between always $(0.9$ fm$)^d$ to $(53$ fm$)^2$ in two dimensions, $(15$ fm$)^3$ in three dimensions, and a rather small $(6$ fm$)^4$ in four dimensions. Thus, compared to the large volume results on coarse lattices of \cite{Sternbeck:2007ug,Bogolubsky:2007ud,Cucchieri:2007rg,Cucchieri:2008fc,Dudal:2012td}, the present investigation is somewhat restricted, especially in four dimensions.

Once the configurations had been obtained, they were fixed to the minimal Landau gauge \cite{Maas:2011se} using self-adjusting stochastic over-relaxation \cite{Cucchieri:2006tf}. This gauge-fixing procedure shows in itself an interesting behavior, as the efficiency is highly configuration-dependent. In fact, the maximum number of gauge-fixing sweeps is usually one to two orders of magnitude larger than the average, while the minimum number is usually at least a factor of two or more smaller than the average one. Also, it turns out that the optimal tuning parameter is highly configuration, and probably Gribov copy, dependent. Given that the gauge-fixing itself is usually one of the most expensive operations in investigations of gauge-fixed correlation functions, it would therefore be likely a worthwhile endeavor to develop an algorithm which adapts to a configuration. However, it would then have to be ensured that this does not alter the way how Gribov copies are selected, as this would alter the non-perturbative completion of Landau gauge, at least on any finite lattice \cite{Maas:2011se}.

\begin{figure}
\includegraphics[width=\linewidth]{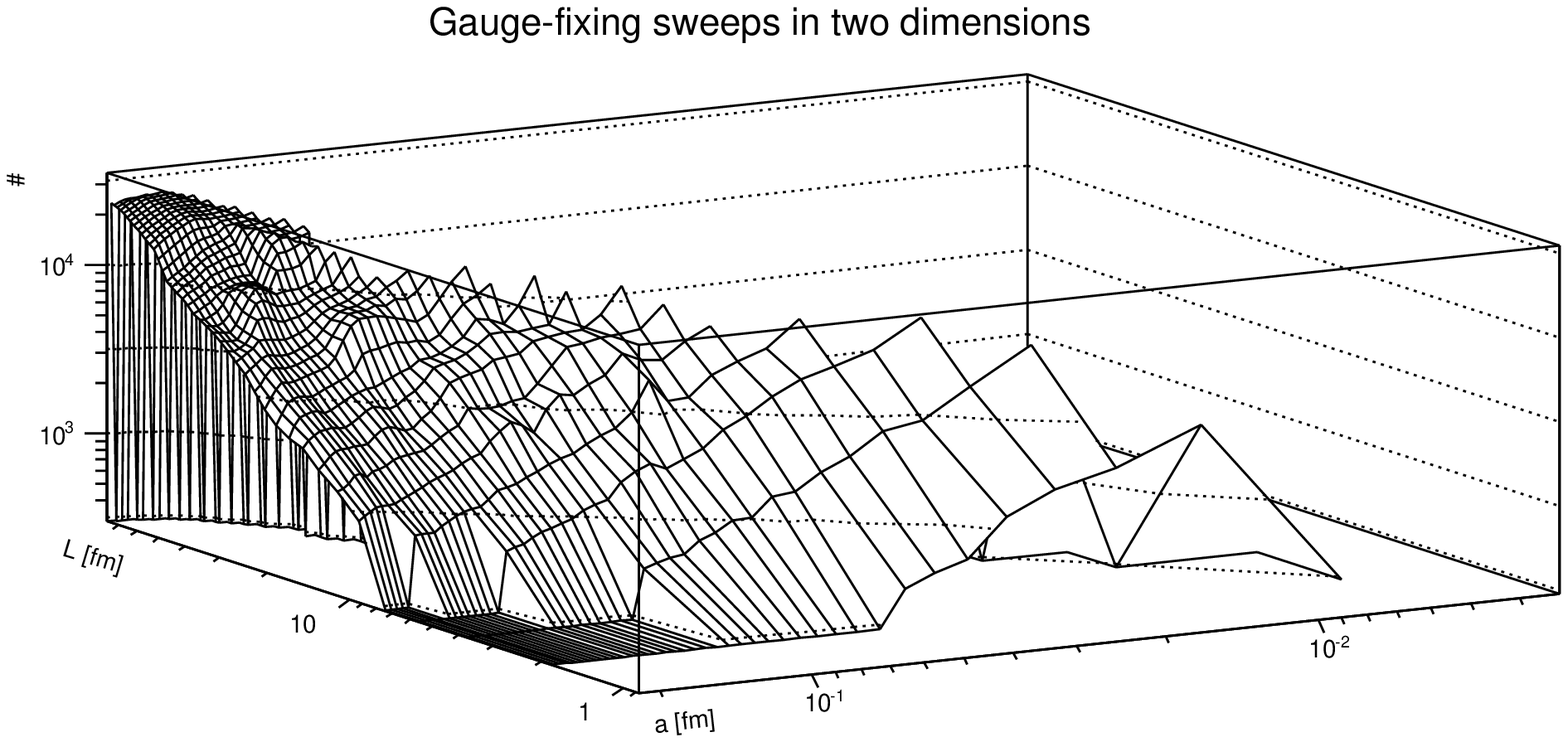}\\
\includegraphics[width=\linewidth]{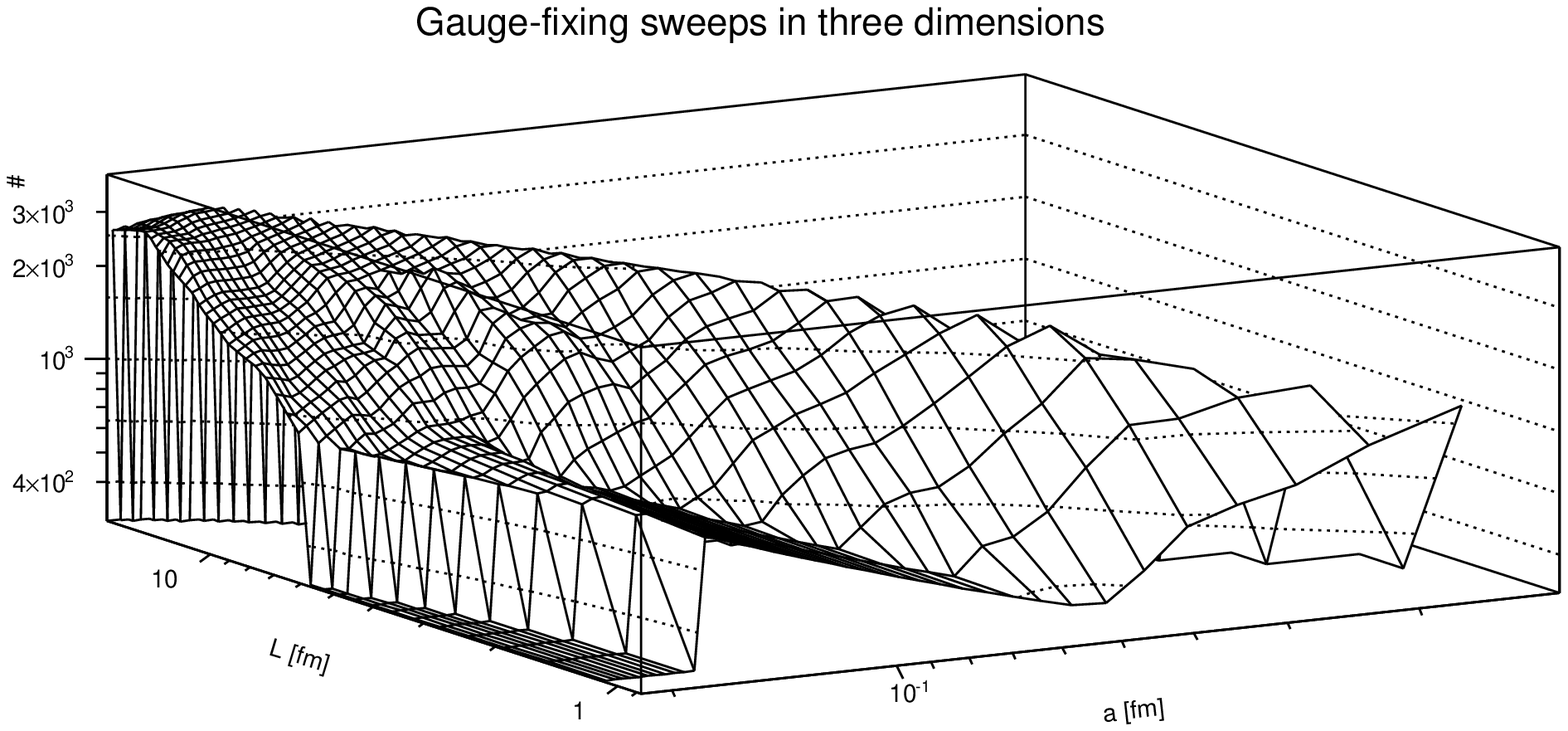}\\
\includegraphics[width=\linewidth]{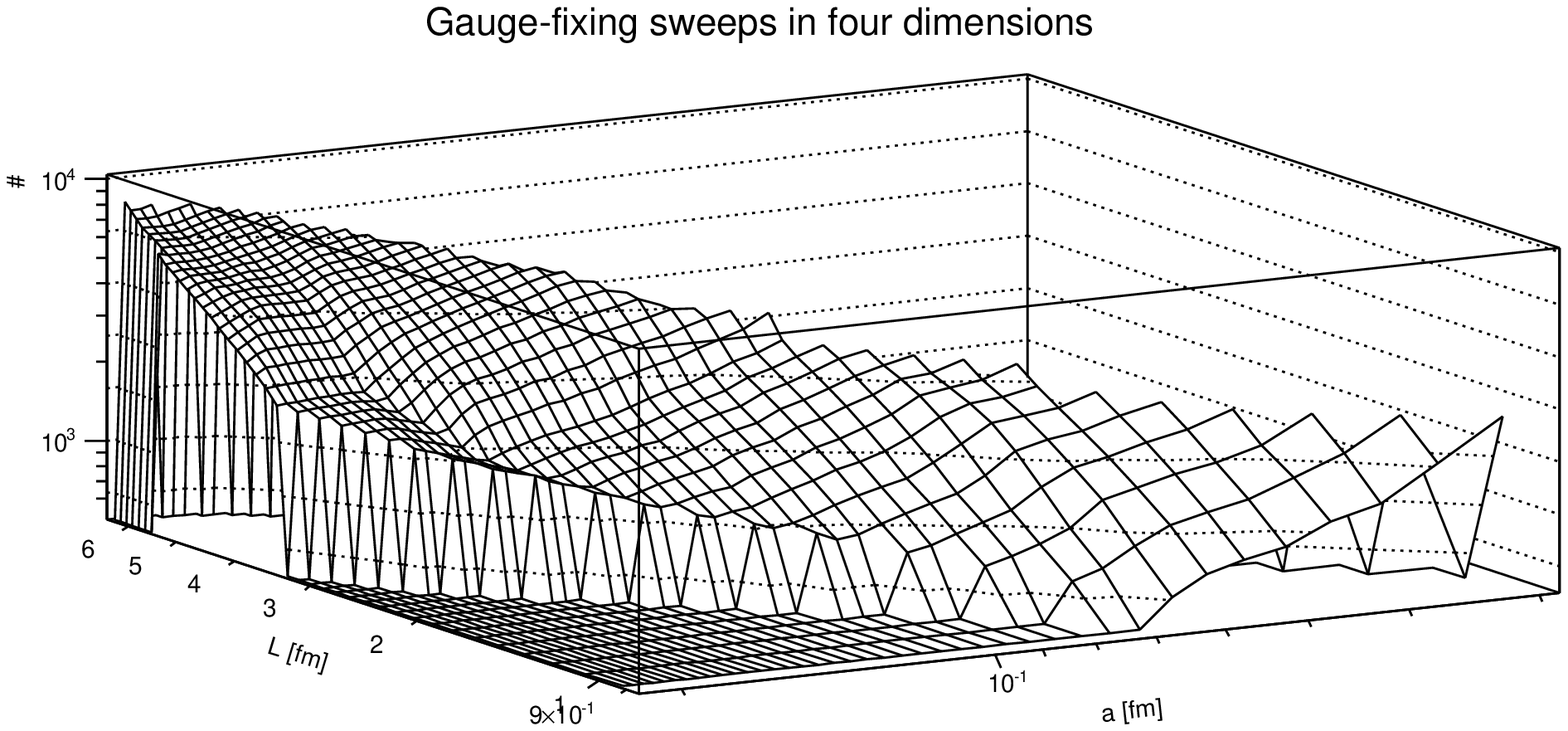}
\caption{\label{fig:sweeps}The average number of gauge-fixing sweeps against lattice extension and lattice spacing, for two dimensions (top panel), three dimensions (middle panel), and four dimensions (bottom panel).}
\end{figure}

The average performance of the gauge-fixing algorithms in general, and of the one used here in particular, has long been known to show critical slowing down at fixed discretization \cite{Cucchieri:1995pn}. This has been investigated here for the much larger set of different lattice settings. The result for the average number of gauge-fixing sweeps are shown in figure \ref{fig:sweeps}. It is seen that after some initial stage at small volumes and coarse discretizations the average number indeed behaves like a power-law, in both volume and discretization. The results are somewhat noisy, and therefore no attempt has been made to extract a critical exponent.

Finally, the gluon propagator and the ghost propagator in both position and momentum space were determined using the methods of \cite{Cucchieri:2006tf}. This implies that the standard definition of the gluon fields
\be
A_\mu^a=\frac{\sqrt{\beta}}{4ia}\tr \tau^a (U_\mu-U_\mu^+)+{\cal O}(a^2)\label{afield}
\ee
\no from the links $U_\mu$ has been used, where $\tau^a$ are the Pauli matrices. The ghost propagator has been determined using a conjugate gradient inversion on a point source of the Faddeev-Popov operator \cite{Boucaud:2005gg}. Thus, quantitatively, the results found here strictly speaking only apply to precisely these definitions of the propagators. However, in the continuum limit and for infinite statistics, these should yield the same results as any other procedure.

Throughout, since the propagators appear to be color-diagonal\footnote{This has been checked once more explicitly in the present work for all lattice settings.} \cite{Maas:2010qw}, only the color-averaged results will be presented. For the gluon also the trivial Landau-gauge Lorentz tensor structure will be removed, such that only a scalar function remains. Finally, the negative ghost propagator \cite{Suman:1995zg,Maas:2011se} will be multiplied by $-1$ to make it positive. This leaves the two positive, scalar propagators $D$ and $D_G$ for the gluon and the ghost, respectively.

\section{Renormalization}\label{srenorm}
 
In four dimensions, the propagators are formally divergent for all momenta because of the (with the cutoff logarithmically) divergent wave-function renormalization constant. Hence, it is necessary to renormalize them. This can be achieved by a multiplicative factor. For the sole purpose of displaying the individual propagators, the two renormalization constants $Z_3$ and $\tilde{Z}_3$ of the gluon and ghost, respectively, can be chosen independently. But they are not independent, and related by the condition
\be
Z_3\tilde{Z}_3^2=1\label{zrel},
\ee
\no if the finite ghost-gluon vertex renormalization constant is chosen to be one, the so-called miniMOM scheme \cite{vonSmekal:1997vx,vonSmekal:2009ae}. From this follows that the running coupling is a renormalization-group-invariant in the continuum \cite{vonSmekal:1997vx}. In principle, it would therefore be expected that using this relation would correctly renormalize both propagators when choosing an arbitrary renormalization point $\mu$. However, because both propagators are affected in a different way by lattice artifacts, particularly finite volume effects, this is no longer true on a finite lattice.

To address this constructively, the propagators will be renormalized independently at some given $\mu$ in the sections where they are discussed separately, i.\ e. sections \ref{sgp} and \ref{sghp}. For the purpose of the running coupling in section \ref{salpha}, the validity of the constraint \ref{zrel} will not be enforced, and renormalization is performed such that lattice artifacts are minimized.

In two and three dimensions, the wave-function renormalization constants in Landau gauge are finite. Hence, in lower dimensions no renormalization is necessary. However, a multiplicative tadpole correction can be used to improve the approach to the continuum limit for the propagators \cite{Bloch:2003sk,Lepage:1992xa}: By multiplying the propagators with a particular power of the plaquette expectation value $\langle P\rangle$, $-1/2$ for the gluon and $1/4$ for the ghost, respectively, this reduces discretization errors. In four dimensions, this effect is absorbed in the renormalization, but it is relevant for the lower dimensions, especially when considering $D(0)$ in section \ref{sgp}. The typical size of these corrections at a lattice spacing of ${\cal O}(0.1$ fm$)$ is of the order of 5-10\%, depending on the dimensionality. This effect therefore substantially influences the (logarithmic) running of the renormalization constants in four dimensions.

Note that these factors cancel out, as a consequence of \pref{zrel}, for the running coupling in any dimension.

\section{Center-dependence}\label{scdep}

It has been remarked early on \cite{Karsch:1994xh,Cucchieri:1997dx} that the definition of the gluon field \pref{afield} is not blind to the center charge, i.\ e.\ the values of the Polyakov loops of a configuration. If \pref{afield} is regarded as an expansion around a unit link in powers of the lattice spacing $a$, this expression cannot be true for all links if a configuration is not in the trivial center sector, i.\ e.\ with all Polyakov loops being real and positive.

This is most evident in a maximum tree-gauge \cite{Gattringer:2010zz} in a given direction with non-trivial Polyakov loop: Then, all but one link can be transformed into unity, and hence the gauge-invariant value of the Polyakov loop in the corresponding direction has to reside on this link. Hence, for this link the expansion \pref{afield} necessarily fails, if the Polyakov loop is not trivial. At the same time, this implies that this effect should become irrelevant at large extensions, as the boundary should then no longer influence the bulk. Note that this is thus a lattice artifact: The proper continuum gluon field, and thus all its finite correlation functions, is a genuine algebra element, and therefore cannot depend on anything connected to the center. Note that being in the trivial sector is thus a necessary, but not a sufficient condition, for \pref{afield} to work on all links. Even in the trivial center sector, some links can still fluctuate far away from unity. A genuine resolution of this problem would require to isolate the algebra element directly, rather only up to lattice artifacts, e.\ g.\ using stereographic projection or a logarithmic definition \cite{Ilgenfritz:2010gu,vonSmekal:2007ns}. However, in Landau gauge no such simple argument can be made, and hence it is unclear whether possible artifacts arise due to the use of \pref{afield}, rather than such a more sophisticated determination of the gauge fields.

Nonetheless, this subtlety is usually ignored in finite-volume calculations and/or assumed to be irrelevant, using \pref{afield} indiscriminately. If this assumption would not be correct, this would represent an additional systematic error. Here, this assumption will be tested. This will be done with the following procedure: Measure the (gauge-invariant) $d$ Polyakov loops in all directions \cite{Gattringer:2010zz}. Then fix to Landau gauge, and determine the propagators for either all Polyakov loops positive, all negative, or without selection\footnote{In fact, the correct gauge group in the full standard model is S(U(3)$\times$U(2))$\sim$SU(3)/Z$_3\times$SU(2)/Z$_2\times$U(1), and all center factors have to be divided out \cite{O'Raifeartaigh:1986vq}. Thus, in the standard-model case such a sorting is trivial, and always yields the sector with all Polyakov loops positive.}. Since this does not take into account the distribution of the local center phase over the configuration on a per-link basis, this can at best give a lower limit of the effect.

\begin{figure}
\includegraphics[width=\linewidth]{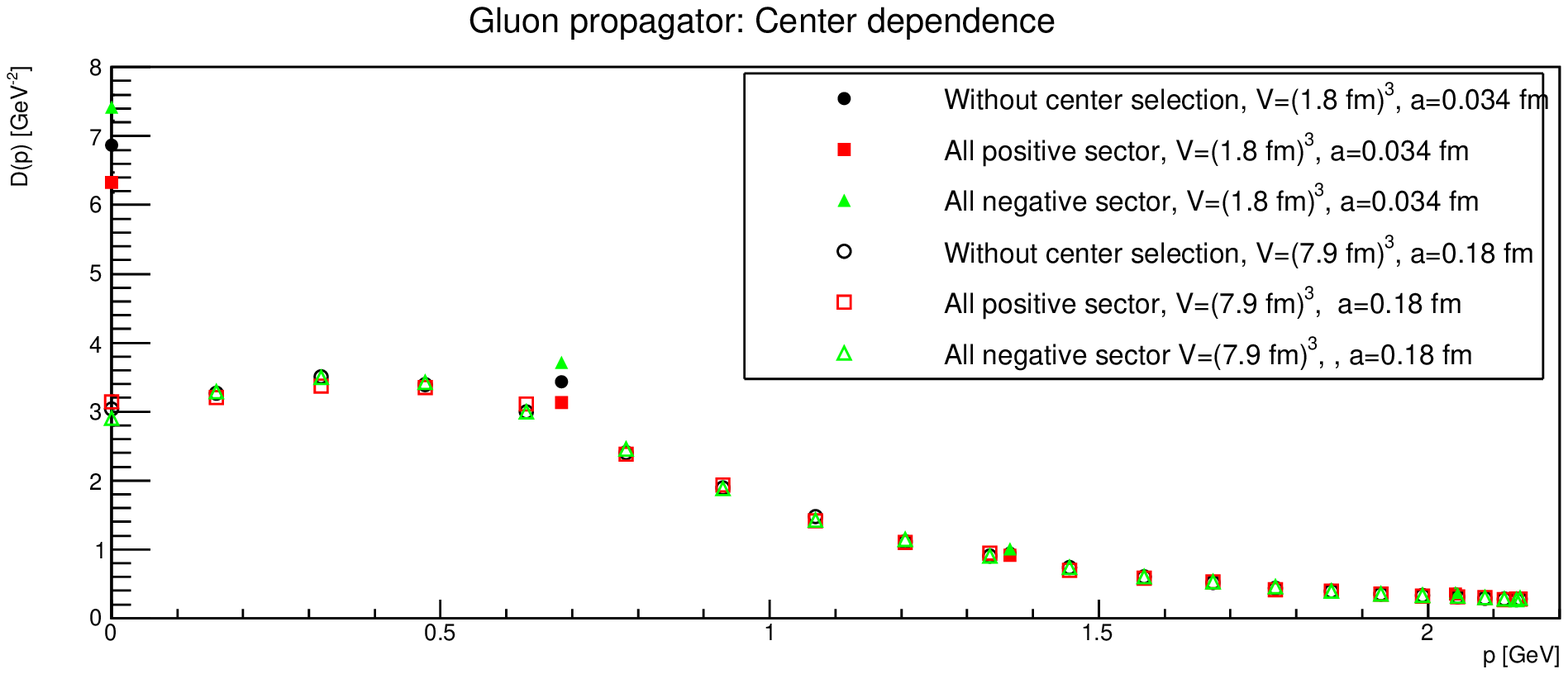}\\
\includegraphics[width=\linewidth]{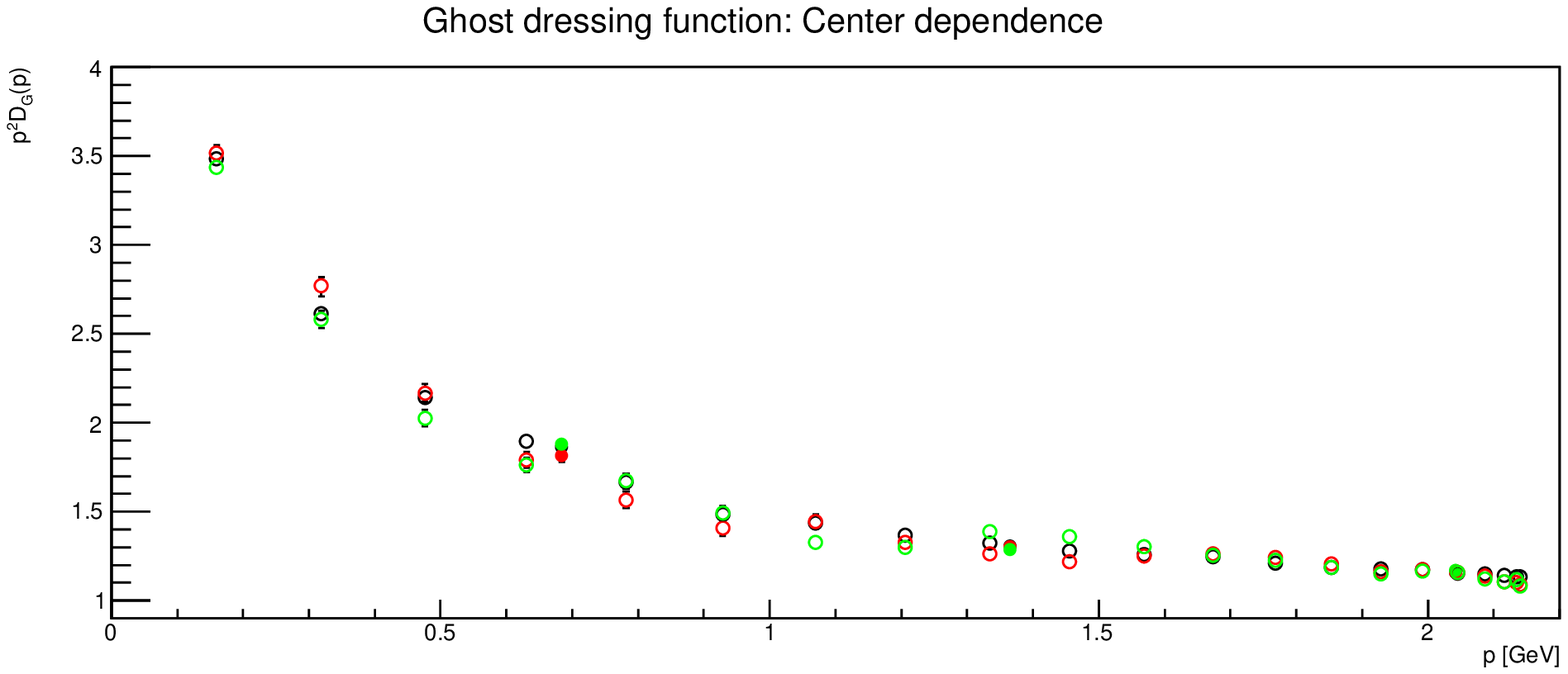}
\caption{\label{fig:center}The impact of the selection of various center sectors on the gluon propagator (top panel) and the ghost dressing function (bottom panel). All momenta along the $x$-axis. 1$\sigma$-Error bars here and hereafter are always shown, but may be smaller than the symbols. Results are from three dimensions.}
\end{figure}

The result is shown in figure \ref{fig:center}. Especially the gluon propagator differs, depending on whether all directions are in the positive, mixed, or negative sectors. It is also visible that the center-dependence not only becomes irrelevant at large physical volumes, as expected, but also at large momenta, and that the effect on the ghost propagator is almost negligible. Investigating the volume-dependence, in two dimensions, the effect became, within statistical accuracy, irrelevant for volumes larger than $(5$ fm$)^2$, in three dimensions for volumes larger than $(4$ fm$)^3$, and in four dimensions for volumes larger than $(2.5$ fm$)^4$. At smaller volumes even very fine discretizations do not remove the effect, increasing (decreasing) the gluon propagator in the negative (positive) center sectors substantially at low momenta. It is also seen that the propagator in the positive center sector is, as expected, closest to the one into which the propagators in the different sectors merge for sufficiently large volumes.

This also implies that at finite temperature along the compactified direction, care is necessary. As a consequence, there always the propagator in the positive sector is calculated, as there the links are, at least on the average, closest to one \cite{Karsch:1994xh}. In fact, above the phase transition, clusters of the same local center phase fill up essentially the whole space above the phase transition  \cite{Gattringer:2010ug,Dirnberger:2012gn}.

At zero temperature, this may hence be considered an esoteric and irrelevant effect. However, heat-bath algorithms are not very efficient in changing center sectors, if the volume is small. Thus, if the statistics is too small, or results are only from one or few runs, a considerable bias may exist, which could alter the results, as seen in figure \ref{fig:center}. In the present investigation, simulations at small volumes have been started from a  cold configuration, such that the majority of configurations is in the trivial center sector. However, in general it has to be ensured that no artifacts due to this problem arise in small-volume simulations.

Note that the observation has been made that the propagators also depend on the center aspects, if during gauge-fixing non-periodic gauge transformations are included, which essentially are an algebra-valued gauge transformation in combination with a center transformation, locking both independent \cite{O'Raifeartaigh:1986vq} symmetries \cite{Cucchieri:1997dx,Bogolubsky:2007bw}. Since the gauge-fixing is performed for the links, rather than for the algebra fields itself, this will change the gauge conditions on the level of Gribov copies, though not at the level of the perturbative Landau-gauge condition $\pdm A_\mu^a=0$, by construction \cite{Cucchieri:1997dx,Bogolubsky:2007bw}. Since this is a boundary effect, it would naively be expected that this effect should diminish with increasing volume. However, this is a question beyond the scope of this investigation.

\section{Rotational symmetry}\label{srot}

One of the trade-offs made for performing lattice simulations is to give up rotational symmetry at finite lattice spacing. As a consequence, momenta in different directions are not necessarily equivalent anymore. Naively, it would be expected that this only affects the large momenta behavior, as this is most felt when probing distances where the lattice spacing is becoming of similar sizes as the employed energies. However, because the Landau gauge condition is equivalent to minimizing the per-configuration momentum integral of the gluon propagator \cite{Maas:2008ri}, gauge-fixing can potentially mix infrared, intermediate, and ultraviolet degrees of freedom, and therefore also an effect at low momenta cannot be excluded. In fact, throughout the rest of the paper this will be seen repeatedly. Such effects have been been investigated in the past, see e.\ g.\ \cite{Cucchieri:1997dx,Maas:2007uv,Boucaud:2008gn,Sternbeck:2010xu,vonSmekal:2009ae,Bloch:2003sk,Cucchieri:2006tf}, and are found to be most pronounced when investigating the dressing functions $Z(p)=p^2D(p)$ and $G(p)=p^2D_G(p)$.

\begin{figure}
\includegraphics[width=\linewidth]{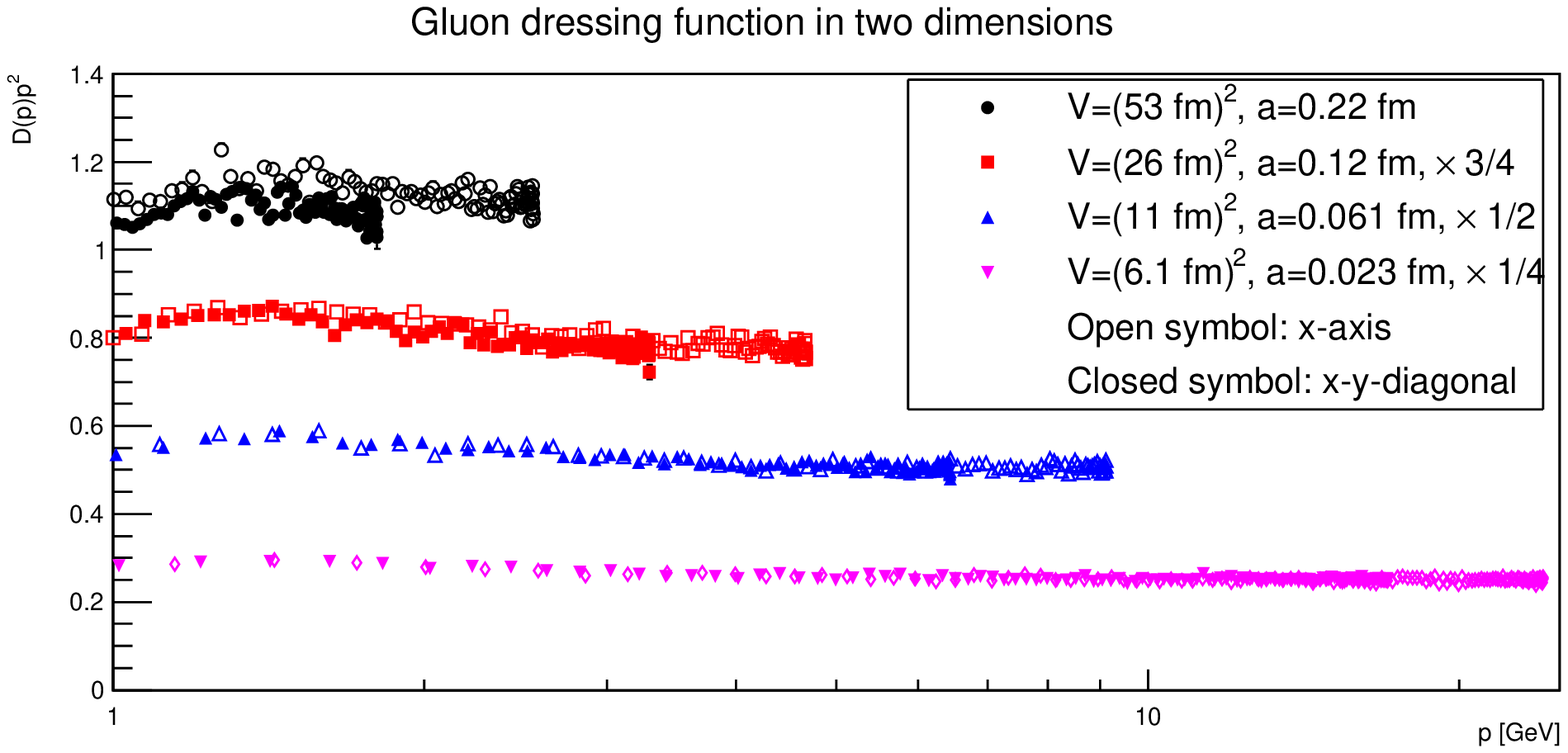}\\
\includegraphics[width=\linewidth]{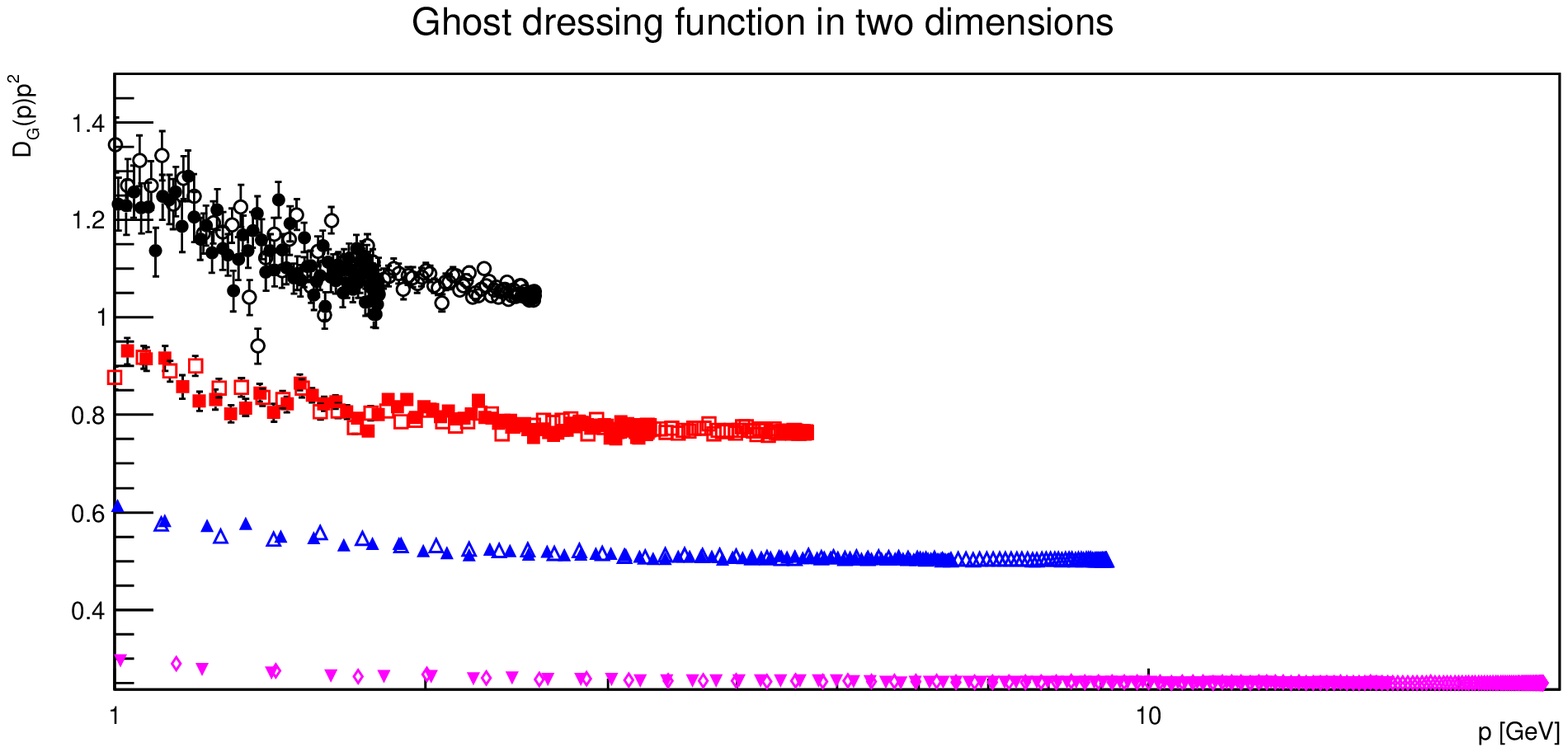}
\caption{\label{fig:rot2d}The impact of the violation of rotational symmetry in two dimensions. The top panel shows the gluon dressing function, and the bottom panel the ghost dressing function. Closed symbols are along the $x$-axis, and open symbols are along the $xy$-axis. Note that for better visibility the results for different lattice spacings have been rescaled by constant factors.}
\end{figure}

\begin{figure}
\includegraphics[width=\linewidth]{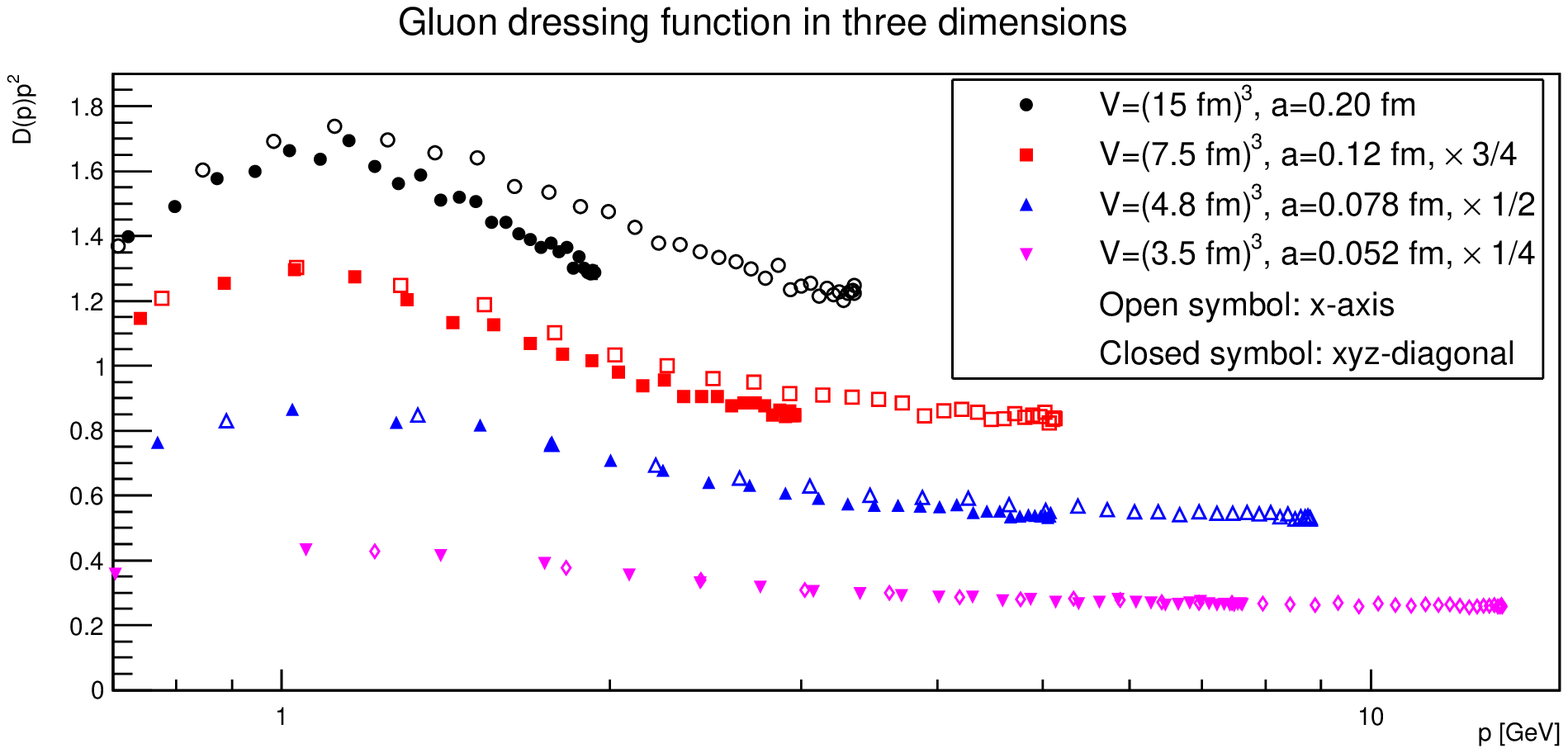}\\
\includegraphics[width=\linewidth]{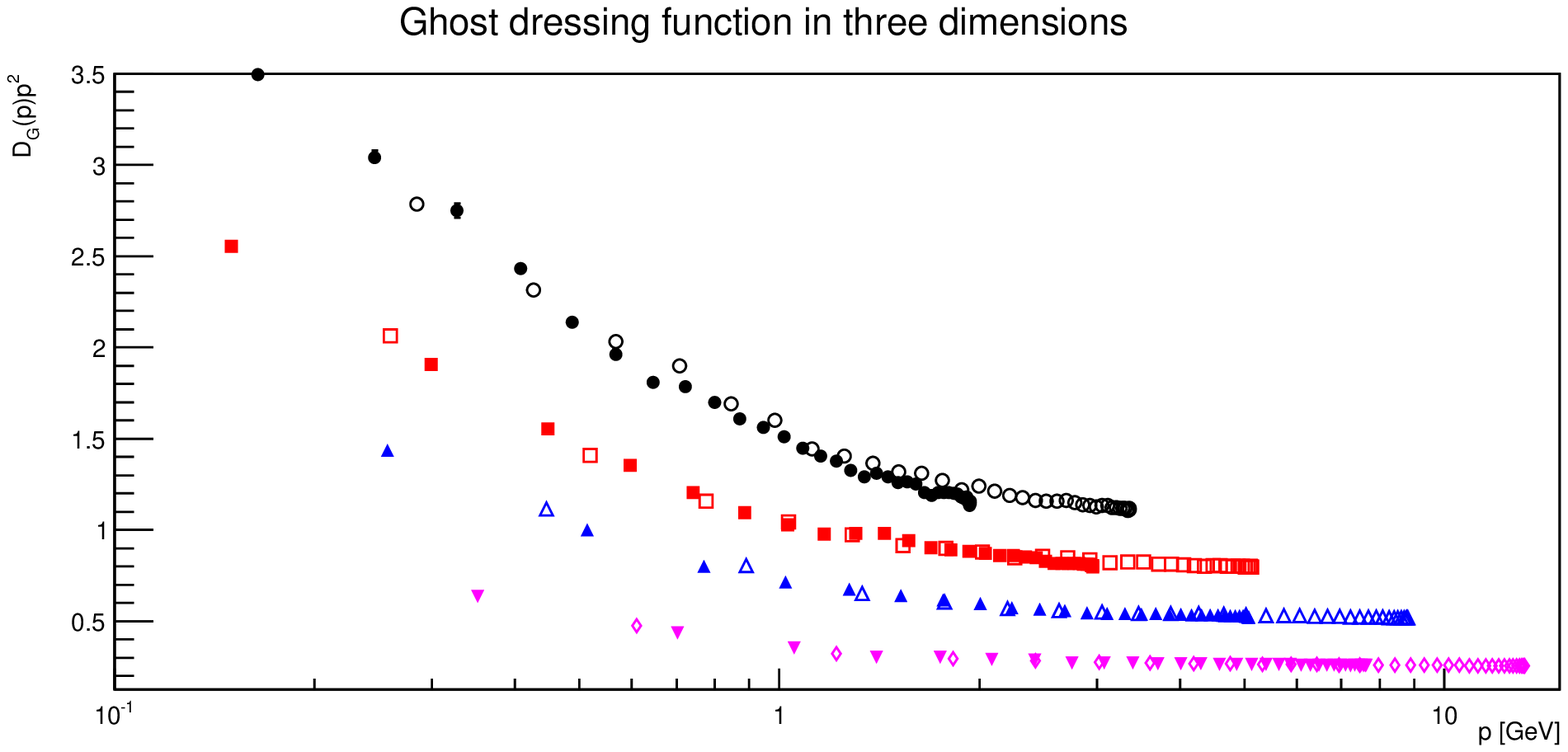}
\caption{\label{fig:rot3d}The impact of the violation of rotational symmetry in three dimensions. The top panel shows the gluon dressing function, and the bottom panel the ghost dressing function. Closed symbols are along the $x$-axis, and open symbols are along the $xyz$-axis. Note that for better visibility the results for different lattice spacings have been rescaled by constant factors.}
\end{figure}

\begin{figure}
\includegraphics[width=\linewidth]{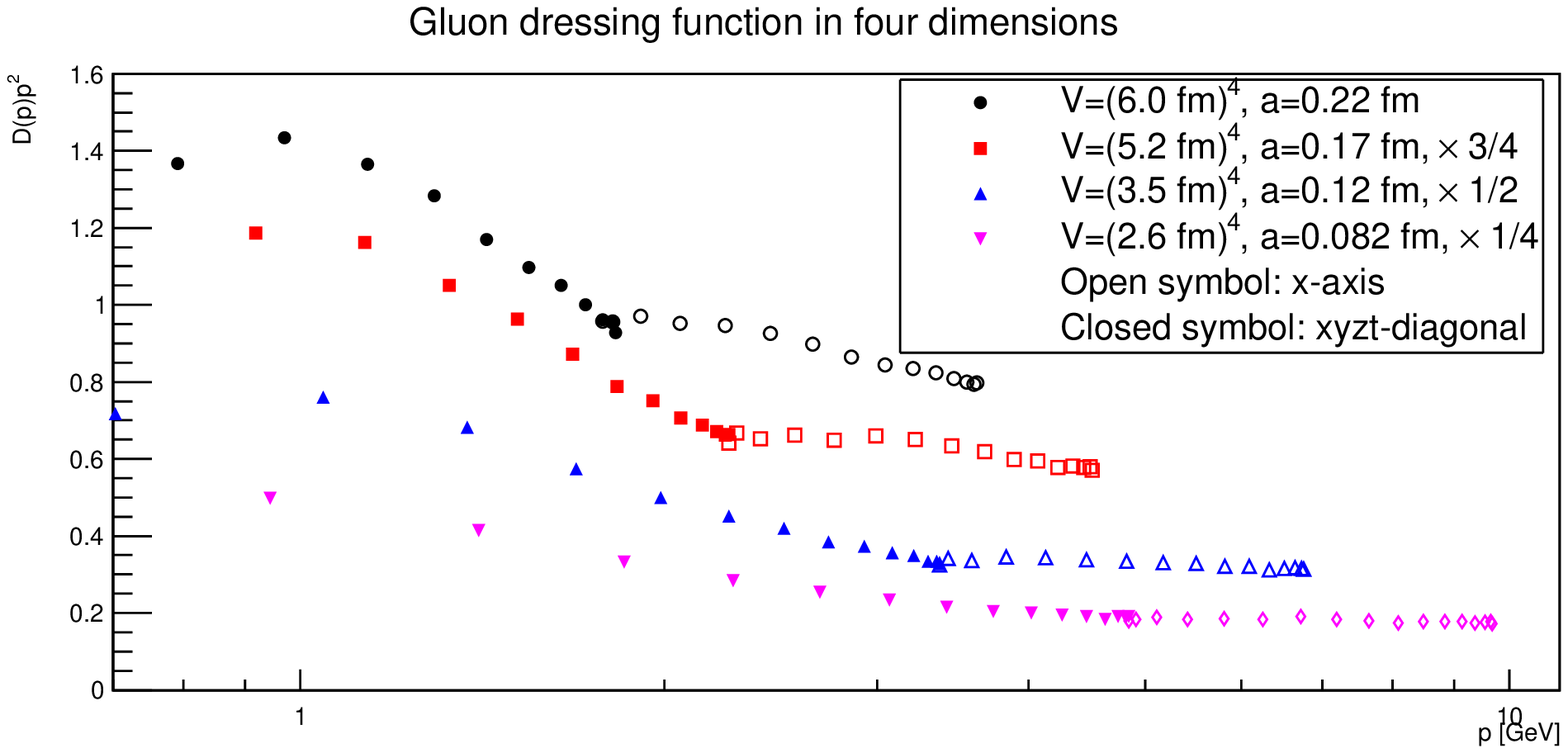}\\
\includegraphics[width=\linewidth]{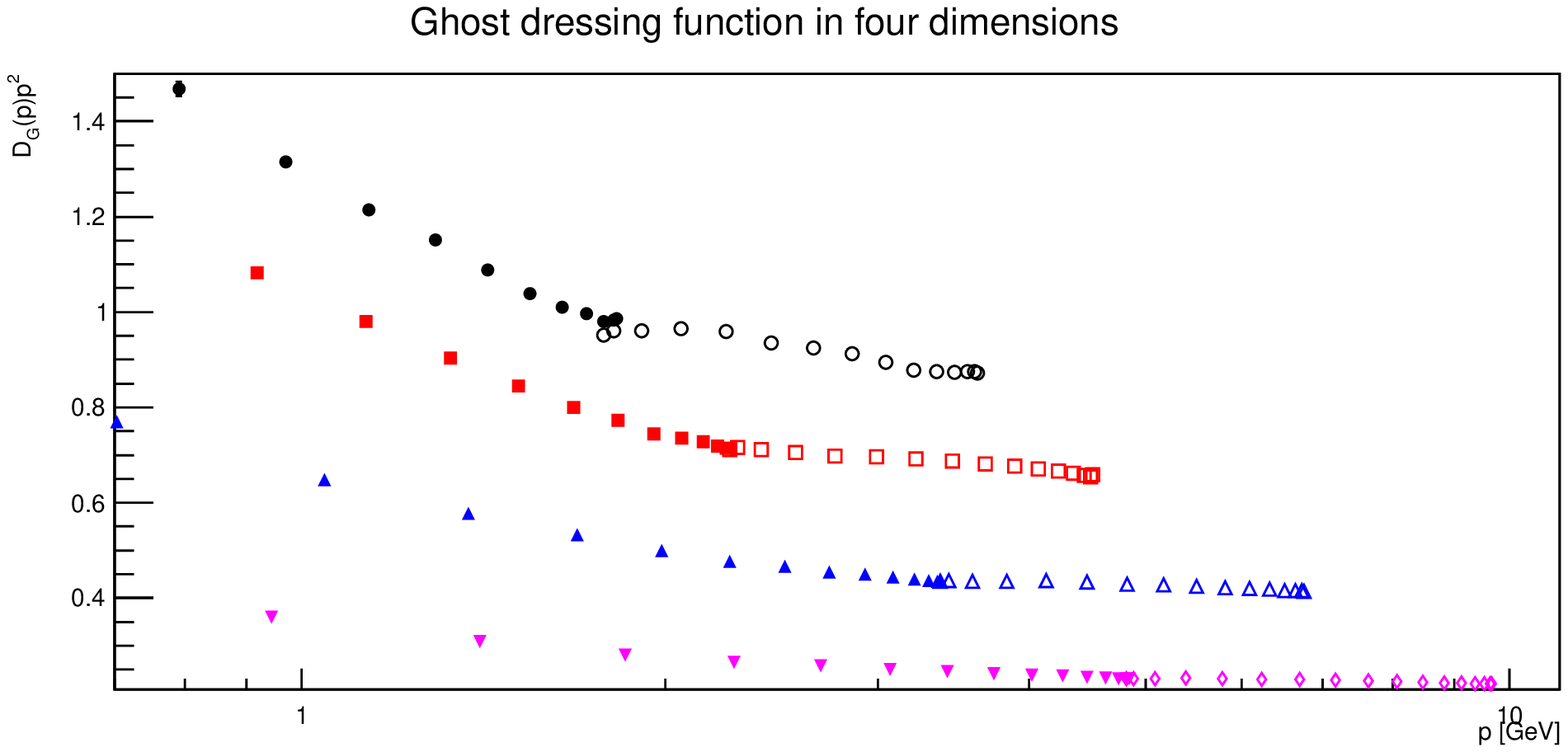}
\caption{\label{fig:rot4d}The impact of the violation of rotational symmetry in four dimensions. The top panel shows the gluon dressing function, and the bottom panel the ghost dressing function. Closed symbols are along the $x$-axis, and open symbols are along the $xyzt$-axis. Note that for better visibility the results for different lattice spacings have been rescaled by constant factors, and the results have been renormalized at 2 GeV.}
\end{figure}

In a sense, the most extreme differing cases are those where the momenta are either along an axis, which will be chosen here to be conventionally the $x$-axis, or along the space-time diagonal. The impact of choosing these different momenta are shown for some example lattice setups in figure \ref{fig:rot2d}-\ref{fig:rot4d}. Comparing also space or area diagonals, for dimensionalities were this is possible, indeed confirms that the largest difference is for the two axes chosen.

The results show that for the gluon propagator the consequences of violation of rotational invariance are strong, and become stronger the higher the momenta. Only above $a^{-1}\gtrapprox 3$ GeV, or $a\lessapprox 0.06$ fm, the effects become so small as that rotational symmetry is effectively restored. It is, to some extent, expected that the effects become stronger the higher the dimension, as the effective length of the largest diagonal of an elementary lattice cube increases, and thus the distance over which the lattice substructure can be felt becomes larger.

The effects for the ghost propagator are less severe, probably because it is a scalar particle. However, some of the remaining effects may be due to the choice of a point source, and may differ for a plain wave source \cite{Cucchieri:2006tf}. Nonetheless, again at large momenta the diagonal direction gives the best continuum prediction, but here already a lattice spacing of $(2$ GeV$)^{-1}$/0.1 fm seems to be sufficient to restore effectively rotational symmetry.

The finer the discretization, the more the result along the $x$-axis tends to the one along the diagonal. Therefore, investigations at large momenta should use the diagonal momentum direction. At small momenta, however, no distinct difference is found, as expected. Thus there momenta along the $x$-axis, which permit to go to smaller momenta, are preferable. In the following, high and low momentum behavior will therefore be discussed separately, using the momentum direction more appropriate.

Of course, an alternative is to select only momenta which are least affected by the violation of rotational symmetry, e.\ g.\ by a cylinder cut \cite{Sternbeck:2005tk} or more advanced methods \cite{Sternbeck:2012qs,Simeth:2013ima,Blossier:2010ky}. Selecting instead those momentum directions which are least affected, however, increases the effective range of momenta accessible on a given lattice, and therefore makes more use of the available computing power to analyze the extremes of momenta.

\section{Gluon propagator}\label{sgp}

\subsection{Ultraviolet}\label{sgpuv}

The ultraviolet behavior of the gluon dressing function is expected to be dominated by perturbation theory. This does not imply that there are no non-perturbative contributions. However, they are expected to be polynomially suppressed compared to the (leading) perturbative contribution. This leading perturbative contribution is given (for the present SU(2) case) to leading resummed order by \cite{Maas:2011se}
\bea
p^{-2}D_{2d}(p)^{-1}&=&1+\frac{cg^2}{p^2}\label{uv2}\\
p^{-2}D_{3d}(p)^{-1}&=&1-\frac{22g^2}{64p}\label{uv3}\\
p^{-2}D_{4d}(p)^{-1}&=&\left(\frac{33g^2}{208\pi^2}\log\left(\frac{p^2}{\mu^2}\right)+1\right)^{\frac{13}{22}}\label{uv4},
\eea
\no where $g$ is the coupling constant, dimensionful in two and three dimensions. The value of $g$ is, in principle, the one independent external parameter of the theory. Since perturbation theory fails in two dimensions due to infrared singularities, the constant of proportionality $c$ cannot be determined. In a sense, it is strictly non-perturbative, and the behavior follows just from dimensional analysis and asymptotic freedom. In three dimensions, such incurable infrared singularities appear only at higher order \cite{Jackiw:1980kv}. Thus, when they would start to mix with the non-perturbative contributions, they are strictly speaking already non-existent.

\begin{figure}
\includegraphics[width=\linewidth]{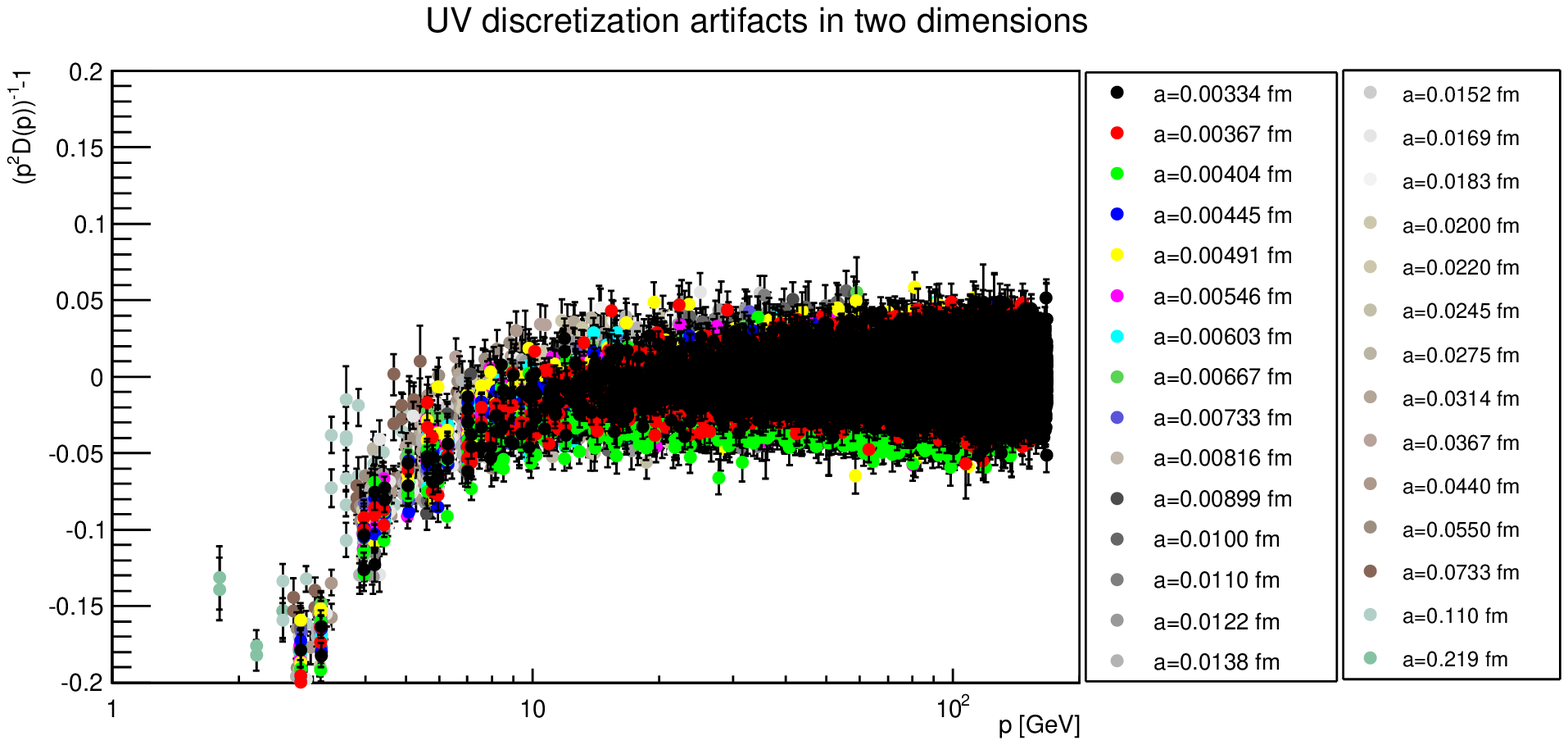}\\
\includegraphics[width=\linewidth]{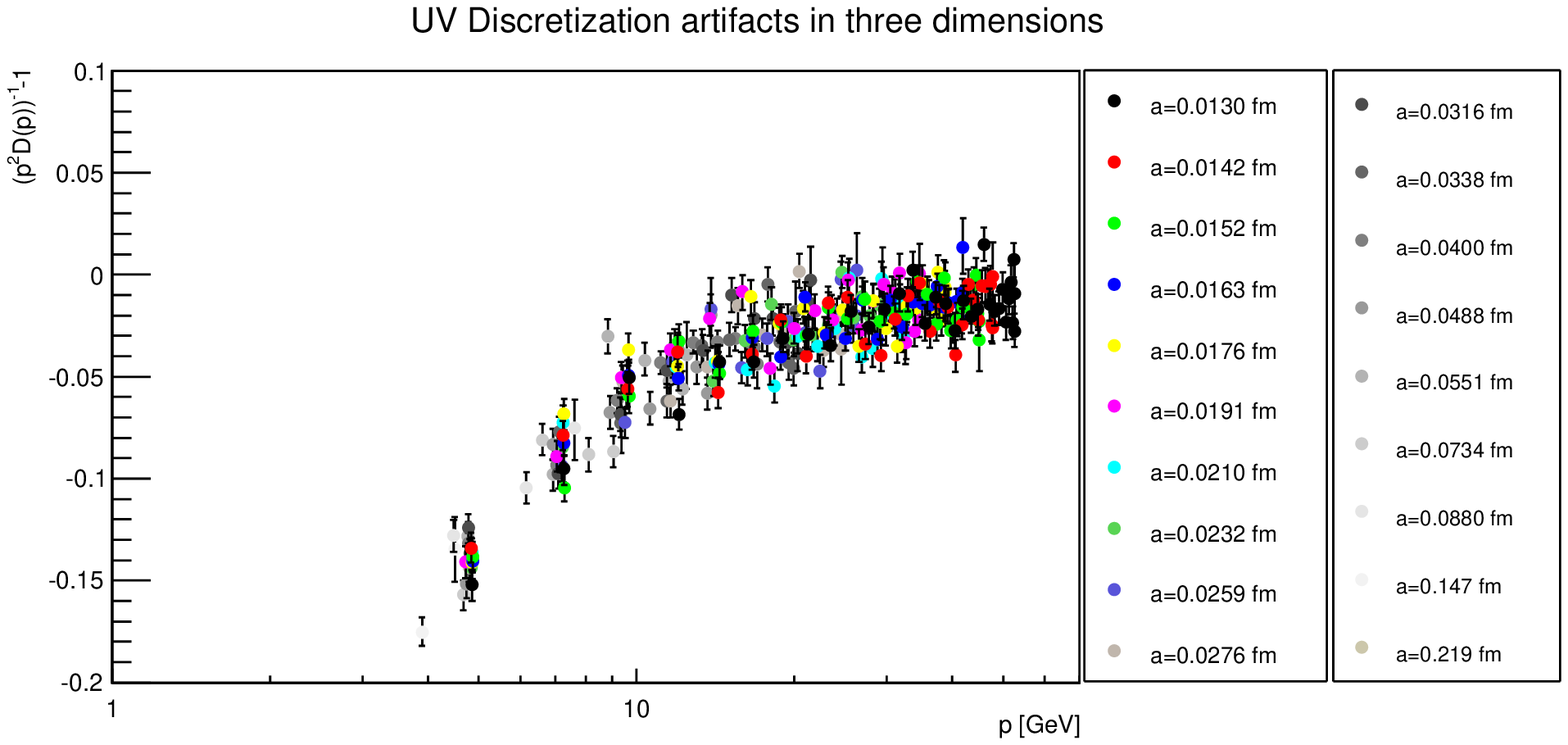}\\
\includegraphics[width=\linewidth]{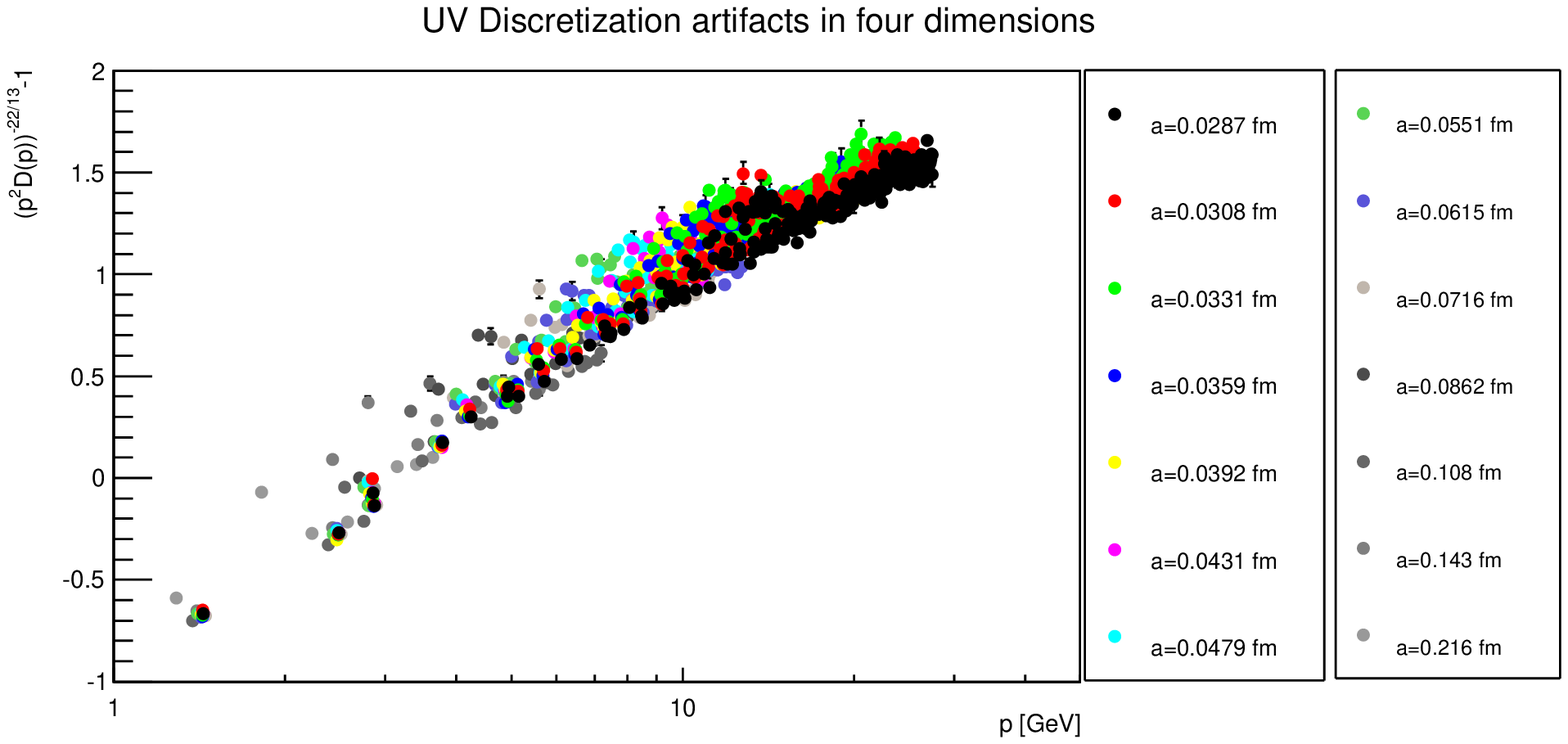}
\caption{\label{fig:gpa}The dependence of the gluon asymptotics, as isolated from \prefr{uv2}{uv4}, for the smallest volume and the different discretizations, in two (top panel), three (middle panel), and four (bottom panel) dimensions.}
\end{figure}

In addition, it is also here where the largest impact of discretization artifacts is expected. This is investigated in figure \ref{fig:gpa}. What is shown are the ultraviolet asymptotics, i.\ e.\ the leading order non-trivial contributions. In two and three dimensions, these are the polynomial second terms of \prefr{uv2}{uv3}, which are left after subtracting the tree-level 1. In four dimensions, the leading logarithm is isolated, which will give only the linear behavior seen, if the propagator has already assumed its one-loop resumed behavior.

Surprisingly, the discretization effect in theses asymptotics is very small in two and three dimensions, while it is more substantial in four dimensions. This maybe due to the absence of renormalization and the trivial manifestation of asymptotic freedom in the lower dimensions. In any case, the effects are small, already at discretizations of the order of 2-3 GeV.

\begin{figure}
\includegraphics[width=\linewidth]{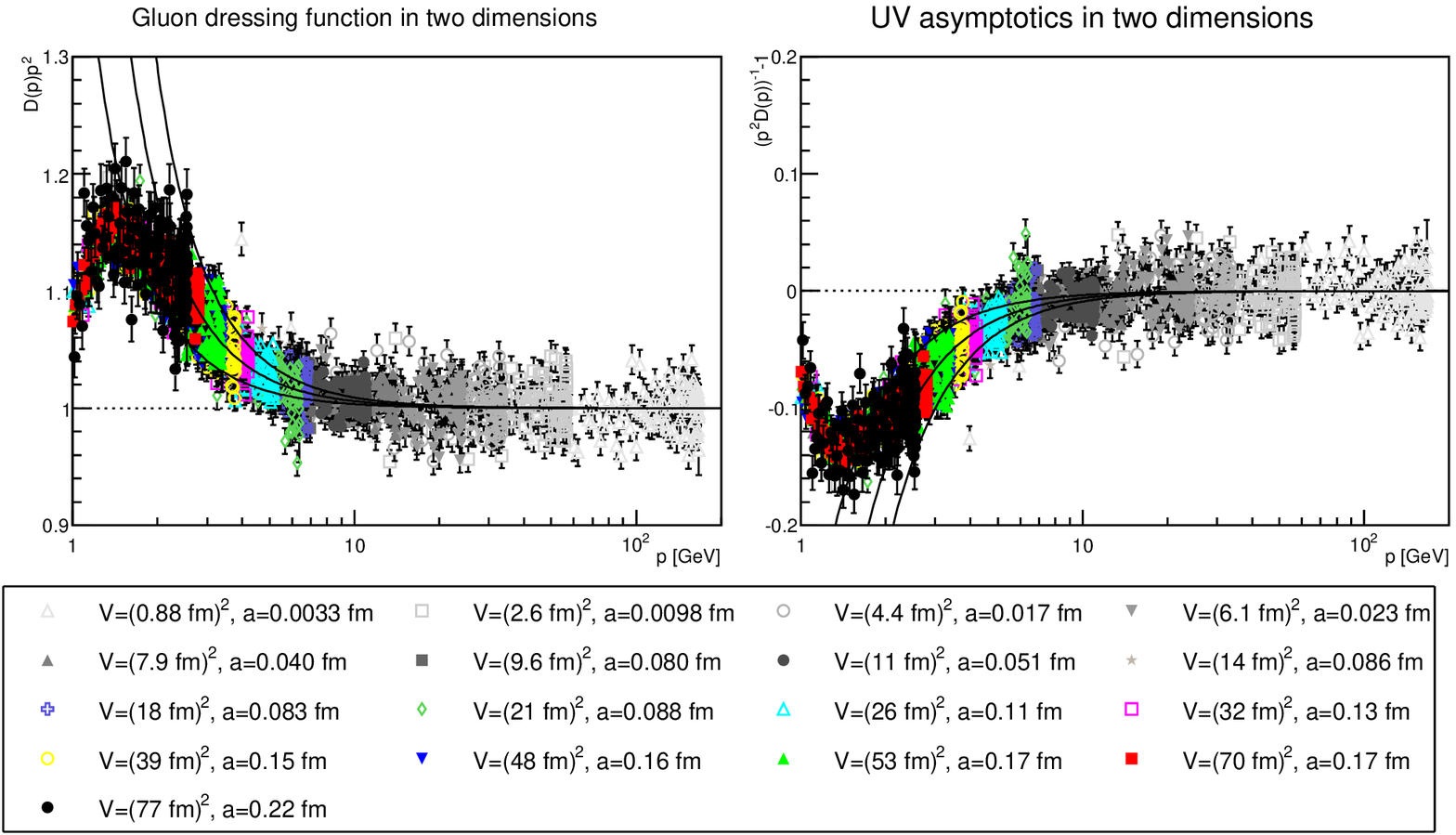}
\caption{\label{fig:gpuv2}The gluon dressing function in two dimensions at large momenta along the space-time diagonal, compared to the leading-order behavior \pref{uv2} for $cg^2=-0.60^{40}_{-25}$. On the right-hand side, the leading asymptotic has been isolated.}
\end{figure}

\begin{figure}
\includegraphics[width=\linewidth]{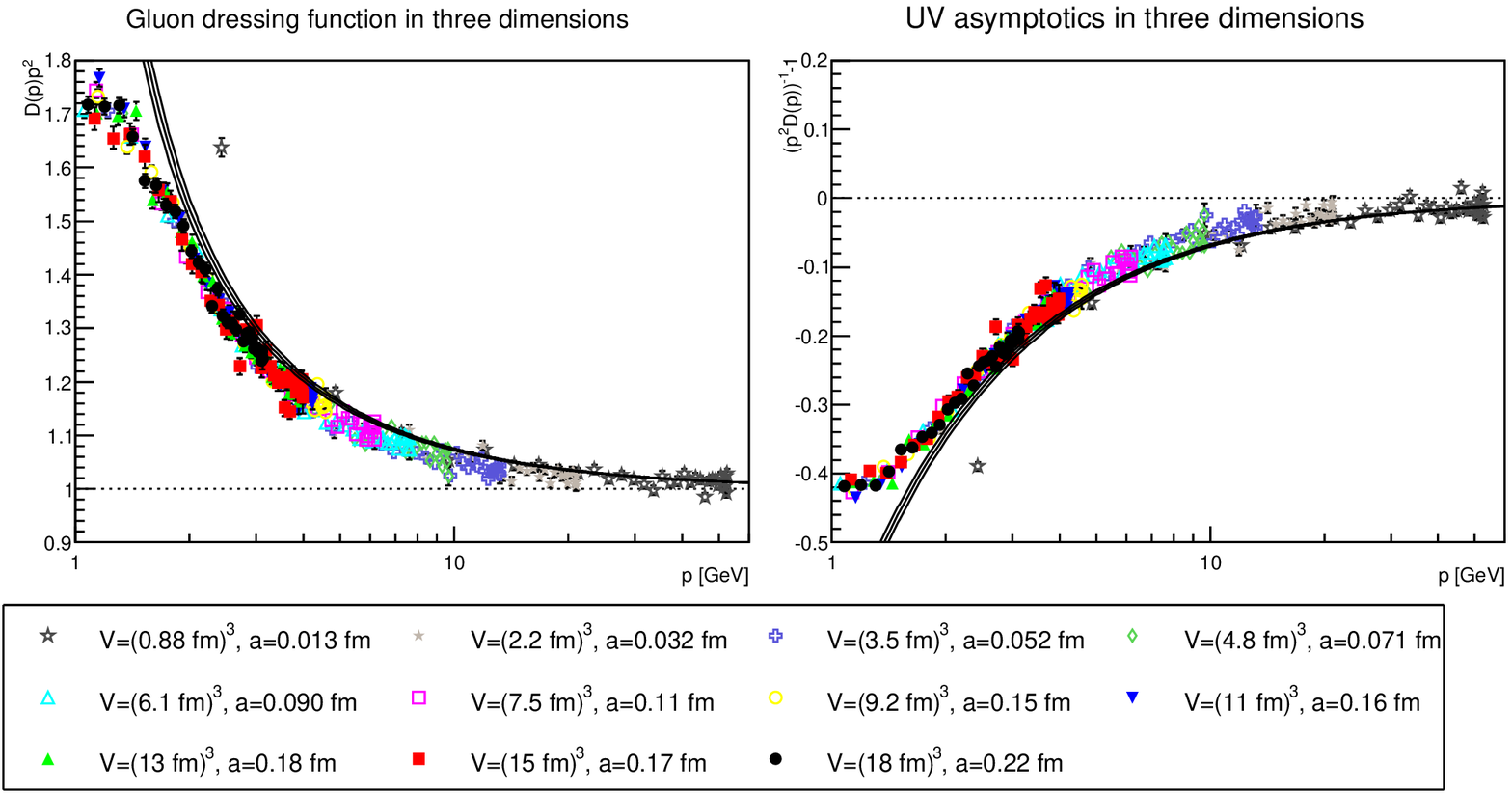}\\
\caption{\label{fig:gpuv3}The gluon dressing function in three dimensions at large momenta along the space-time diagonal, compared to the leading-order behavior \pref{uv3} for $g^2=2.05(5)$. On the right-hand side, the leading asymptotic has been isolated.}
\end{figure}

\begin{figure}
\includegraphics[width=\linewidth]{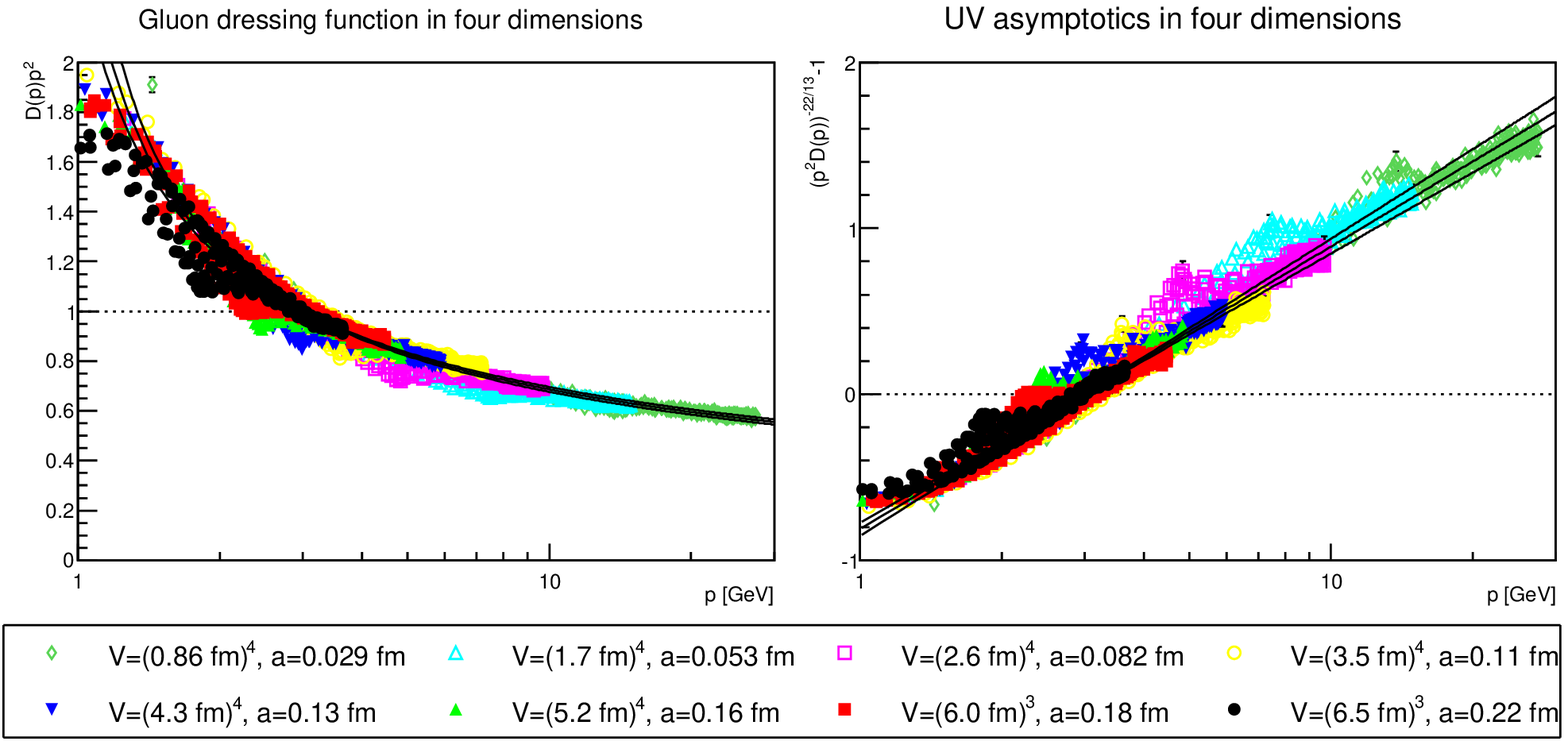}\\
\caption{\label{fig:gpuv4}The gluon dressing function in four dimensions at large momenta along all the space-diagonals for all possible energy values, compared to the leading-order behavior \pref{uv4} for $g=4.8(1)$. On the right-hand side, the leading asymptotic has been isolated. The dressing function was renormalized at $\mu=3$ GeV.}
\end{figure}

The results for the propagators are shown in figures \ref{fig:gpuv2}-\ref{fig:gpuv4} for momenta along the space-time diagonal. In all cases, the behavior is approaching the leading-order perturbative one already at 1.5-3 GeV of momenta, for $cg^2=-0.60^{+40}_{-25}$ GeV$^2$ in two dimensions, $g^2=2.05(5)$ GeV in three dimensions, and $g(\mu)=4.8(1)$ ($\alpha=g^2/(4\pi)=1.8(1)$) in four dimensions\footnote{Note that the coupling constants are not really independent. By fixing $a(\beta)$, it is in principle possible to determine the scheme-dependent value of $g$. However, in practice this is too complicated for the present illustrative purpose, and therefore skipped.}. The error bands are, to some extent, arbitrary, as the deviation could also stem from high-order or non-perturbative contributions. Hence, these values are really rather illustrative, to show for which range of parameters the results are, more or less, compatible with leading-order behavior at the given level of statistics. In three and four dimensions, the perturbative curve uses the same value of $g^2$ as for the ghost case below. In three dimensions, if one permits different values for the gluon and the ghost, the coincidence with the data could be improved. This either indicates significant sub-leading corrections or polynomial non-perturbative corrections of the same order $g^2/p$, though older continuum studies do not show strong indications of such a behavior \cite{Maas:2004se}.

This implies that essentially all non-perturbative information is contained in the behavior below 2 GeV, which will be exclusively investigated below. The resummation effects in four dimensions, in comparison to lower dimensions, and the presence of an anomalous exponent are also very clearly visible.

As is visible in figures \ref{fig:gpuv2}-\ref{fig:gpuv4}, there is almost no impact of the different physical volumes. The discretization errors at large momenta are mainly dominated by the violation of rotational symmetry, where the variation between different momentum axes give an estimate for the systematic uncertainties. However, in the case of the space-time-diagonal, there is almost no variation with discretization, at least within the available statistics. This can again be seen in figure \ref{fig:gpa}, where for the smallest volume the different discretizations for the leading ultraviolet behavior are shown.

\subsection{Infrared}

\begin{figure}
\includegraphics[width=\linewidth]{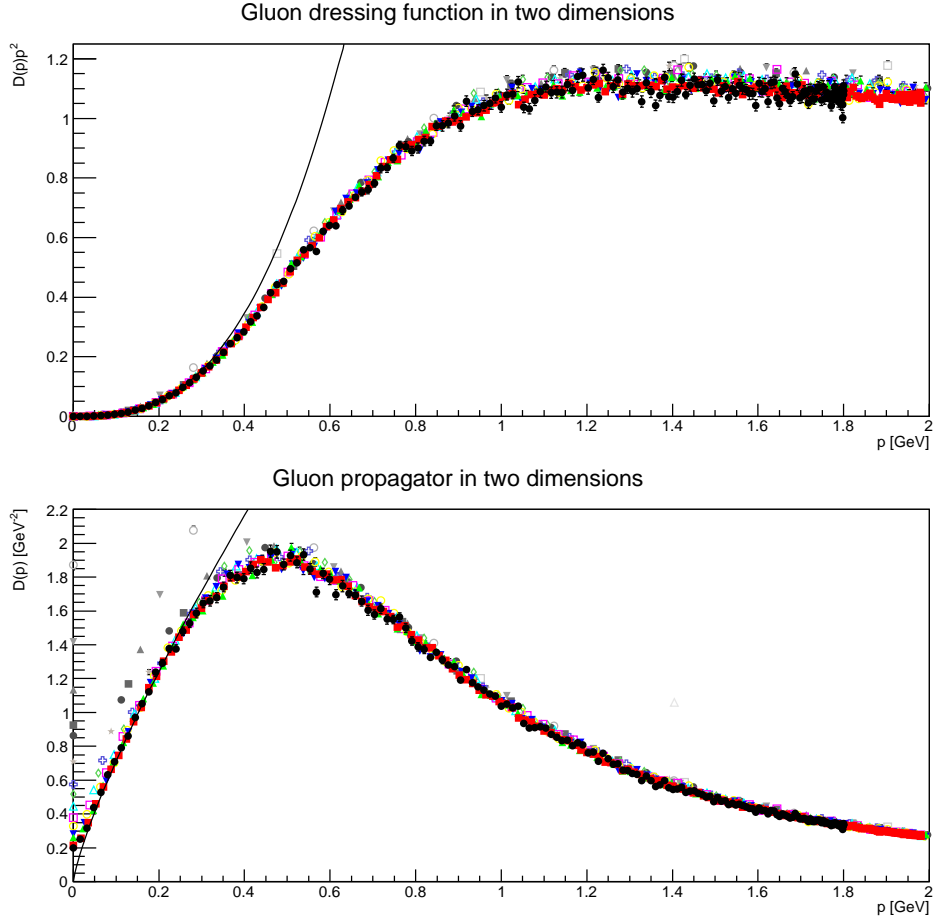}
\caption{\label{fig:gpir2d}The gluon dressing function (top panel) and propagator (bottom panel) at small momenta along the $x$-axis in two dimensions. The symbols have the same meaning as in figure \ref{fig:gpuv2}. The function shown is $4.5p^{2.8}$ and $4.5p^{0.8}$ for the dressing function and propagator, respectively.}
\end{figure}

\begin{figure}
\includegraphics[width=\linewidth]{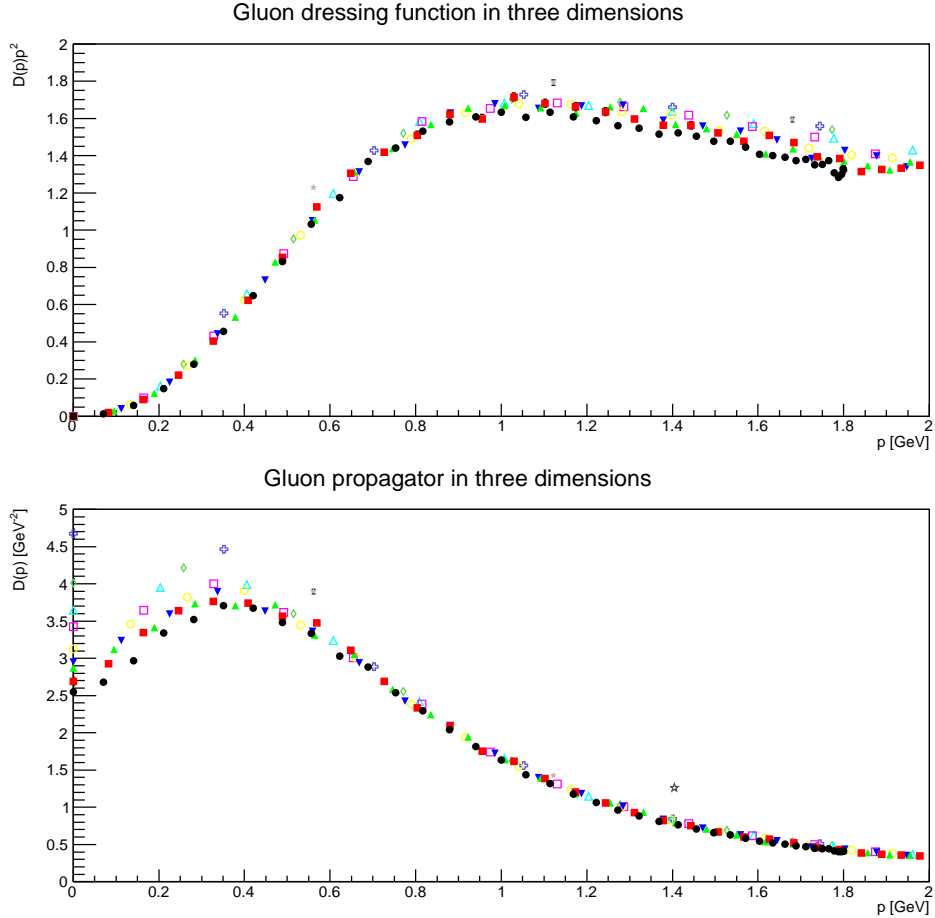}
\caption{\label{fig:gpir3d}The gluon dressing function (top panel) and propagator (bottom panel) at small momenta along the $x$-axis in three dimensions. The symbols have the same meaning as in figure \ref{fig:gpuv3}.}
\end{figure}

\begin{figure}
\includegraphics[width=\linewidth]{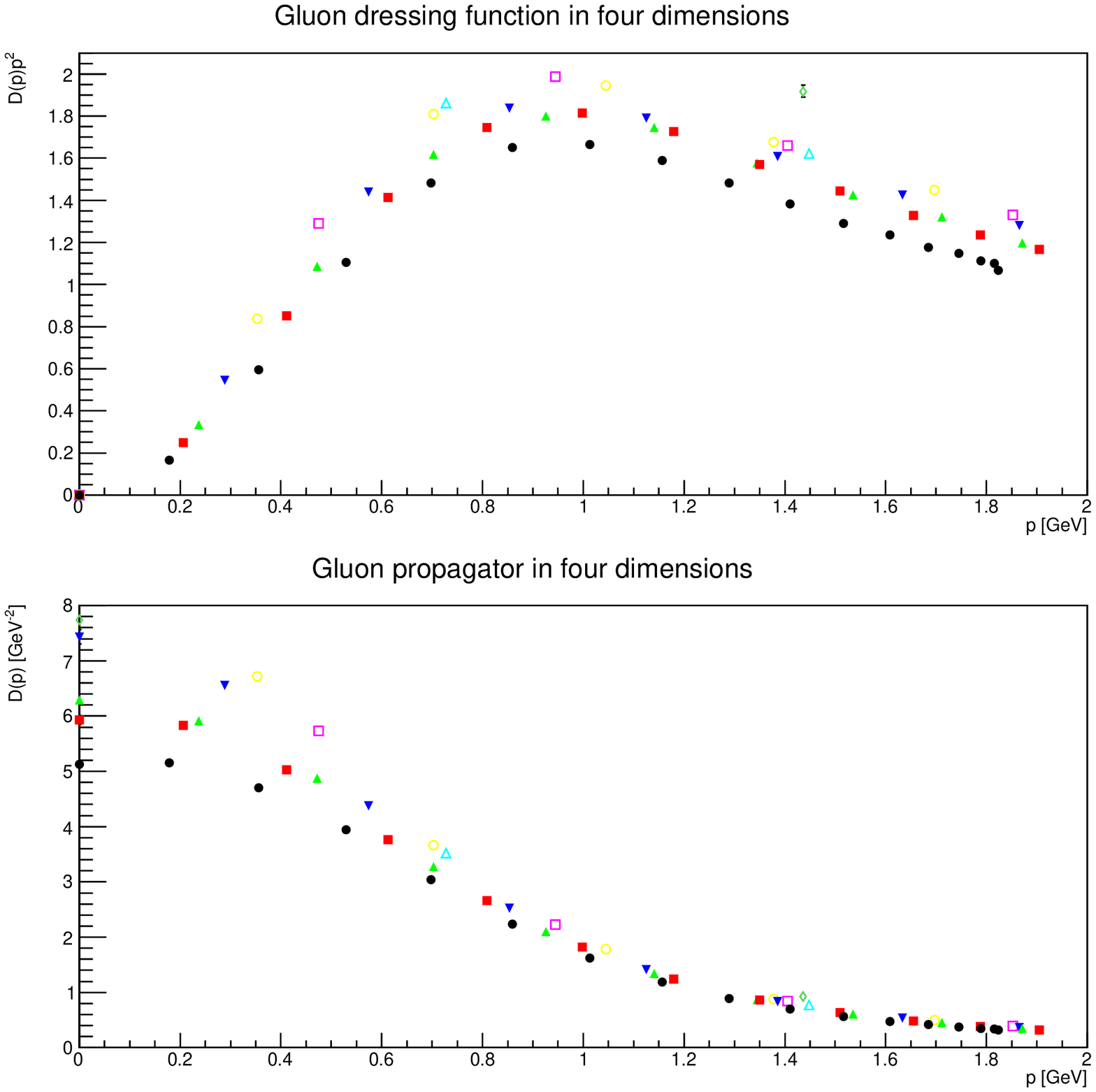}
\caption{\label{fig:gpir4d}The gluon dressing function (top panel) and propagator (bottom panel) at small momenta along the $x$-axis in four dimensions. The symbols have the same meaning as in figure \ref{fig:gpuv4}. The renormalization was done as in figure \ref{fig:gpuv4}.}
\end{figure}

The behavior of the gluon propagator and dressing function below 2 GeV for different volumes are shown in figure \ref{fig:gpir2d}-\ref{fig:gpir4d}. The results are in agreement with previous investigations, see e.\ g.\ \cite{Maas:2011se} for a list, i.\ e.\ the propagator is infrared vanishing like a power-law in two dimensions, and infrared finite in higher dimensions. It is very visible that even for rather large volumes (above (6 fm)$^d$) substantial finite volume effects still remain, especially in two dimensions. Nonetheless, in two and three dimensions clearly a maximum of the gluon propagator arises, while there is no statistically convincing hint of a maximum in four dimensions.

To assess better the development, two quantities are of particular interest. One is the gluon propagator at zero momentum, the other is the volume-dependent effective exponent \cite{Fischer:2007pf}, determined in the same way as in \cite{Maas:2007uv} from the ansatz
\be
D(p)\stackrel{p\ll\Lambda_\text{YM}}{\sim}p^{2t}\label{effexpz}.
\ee
\no The formal prescription to fit the effective exponent is to discard the two points at the lowest non-vanishing momentum. Then the next five momentum values were used to fit a power-law. To obtain errors, the steepest and shallowest curve consistent with a 1$\sigma$-confidence interval was determined as well. In three and four dimensions, this exponent appears to approach zero for sufficiently large volumes, while it is often expected to approach a value of 0.4 in two dimensions, based on the arguments in \cite{Zwanziger:2001kw,Lerche:2002ep}.

\begin{figure}
\includegraphics[width=\linewidth]{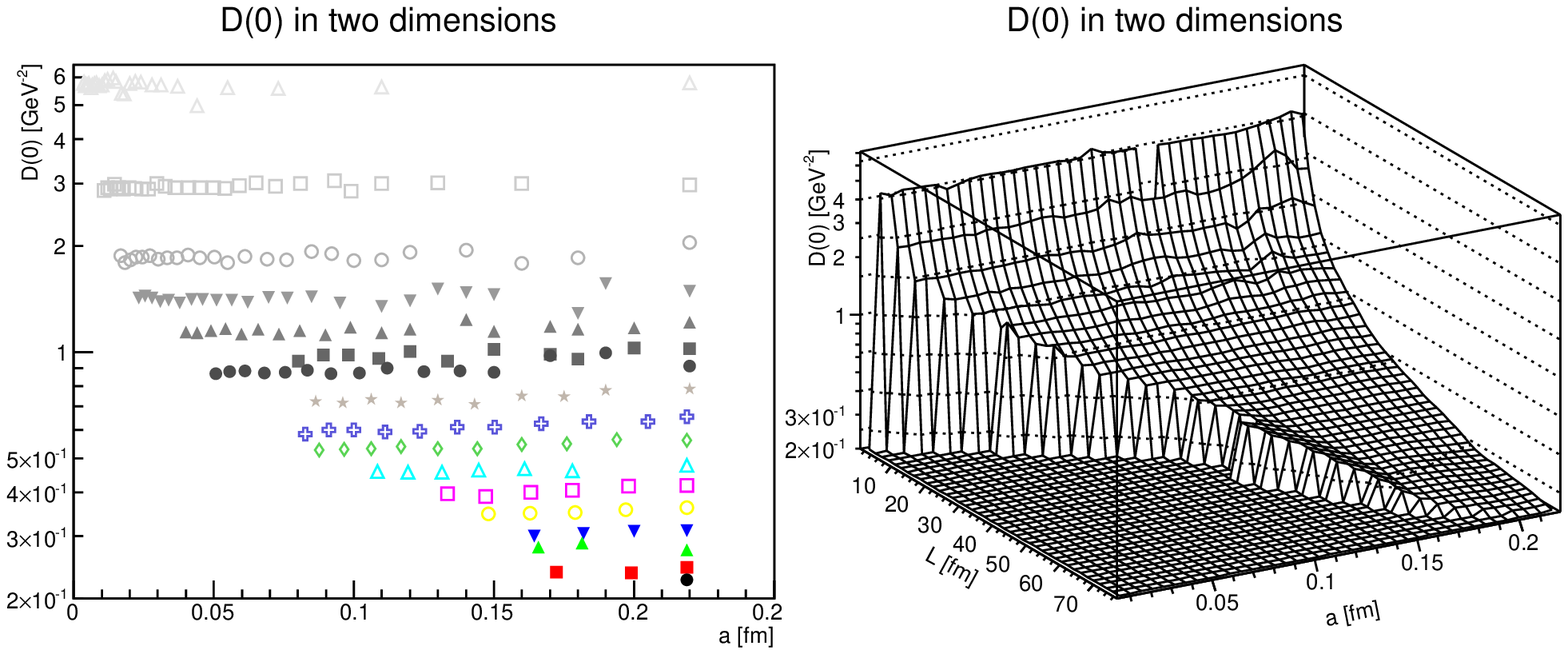}\\
\includegraphics[width=\linewidth]{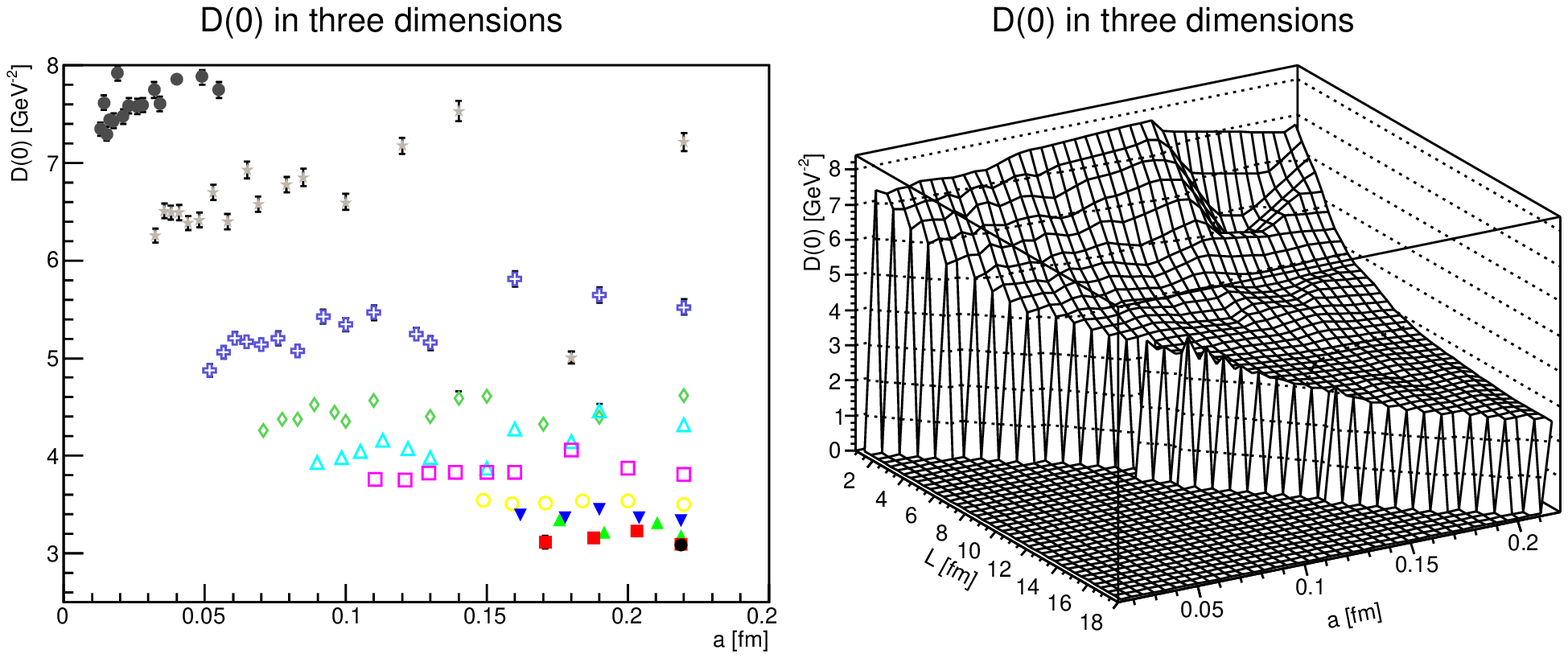}\\
\includegraphics[width=\linewidth]{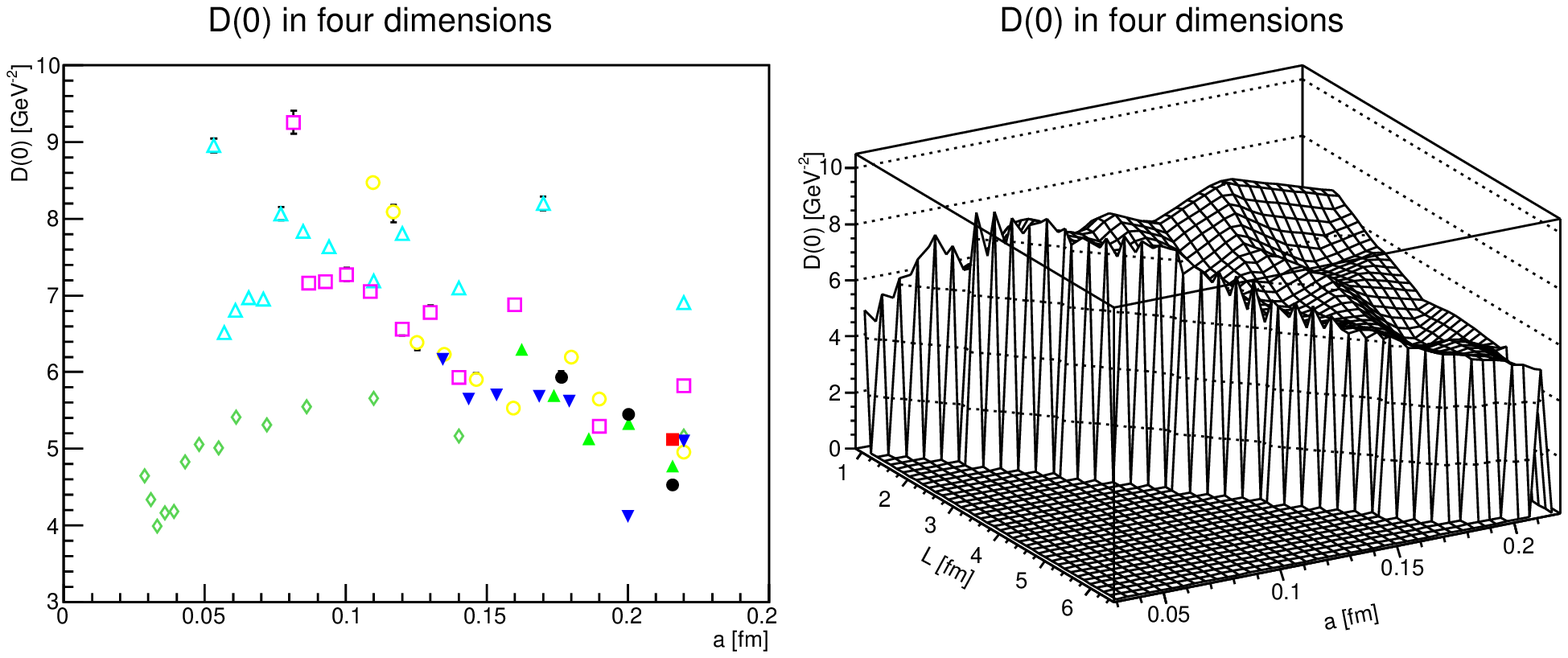}
\caption{\label{fig:d0}The dependence of the gluon propagator at zero momentum as a function of lattice spacing and extension in two (top panel), three (middle panel), and four (bottom panel) dimensions. The results in two and three dimensions have been tadpole-corrected \cite{Bloch:2003sk,Lepage:1992xa}, and in four dimensions a renormalization with $\mu=3$ GeV has been performed. The symbols indicate the same volumes as in figures \ref{fig:gpuv2}-\ref{fig:gpuv4}.}
\end{figure}

\begin{figure}
\includegraphics[width=\linewidth]{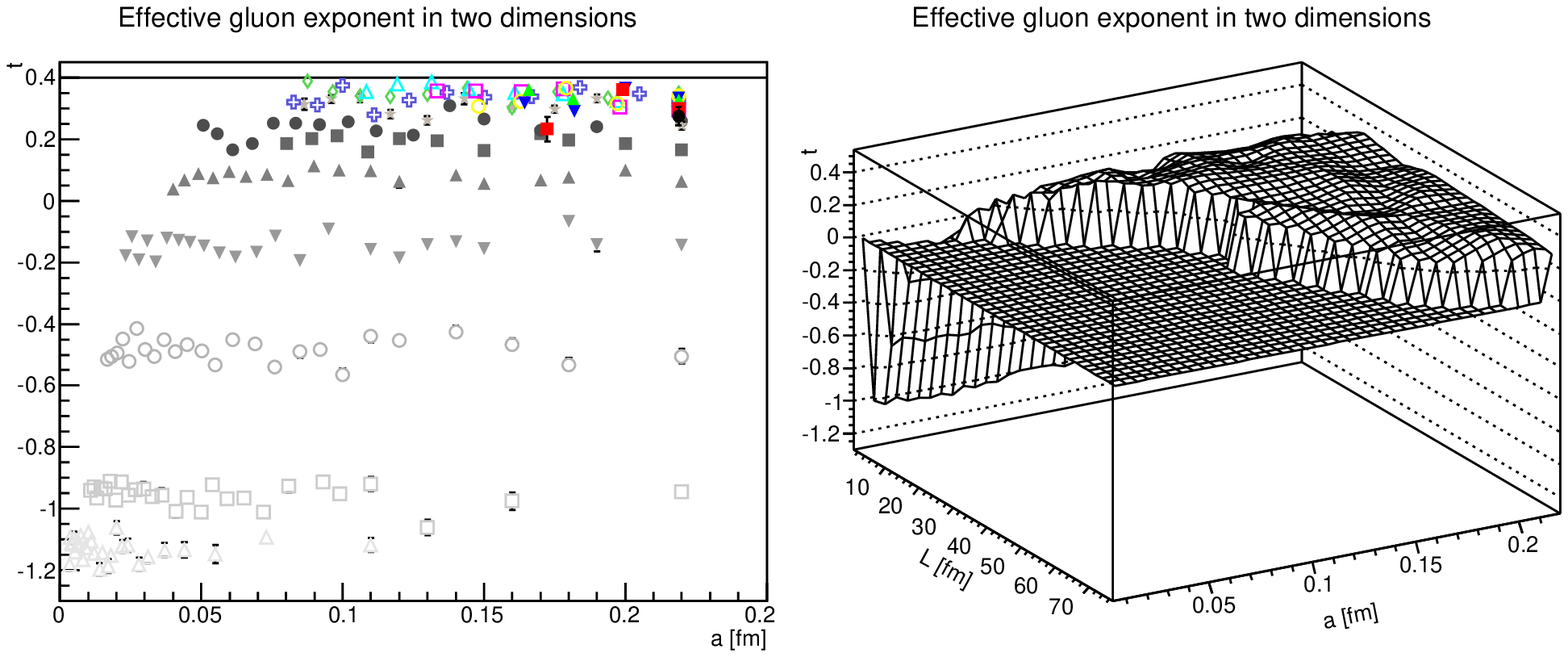}\\
\includegraphics[width=\linewidth]{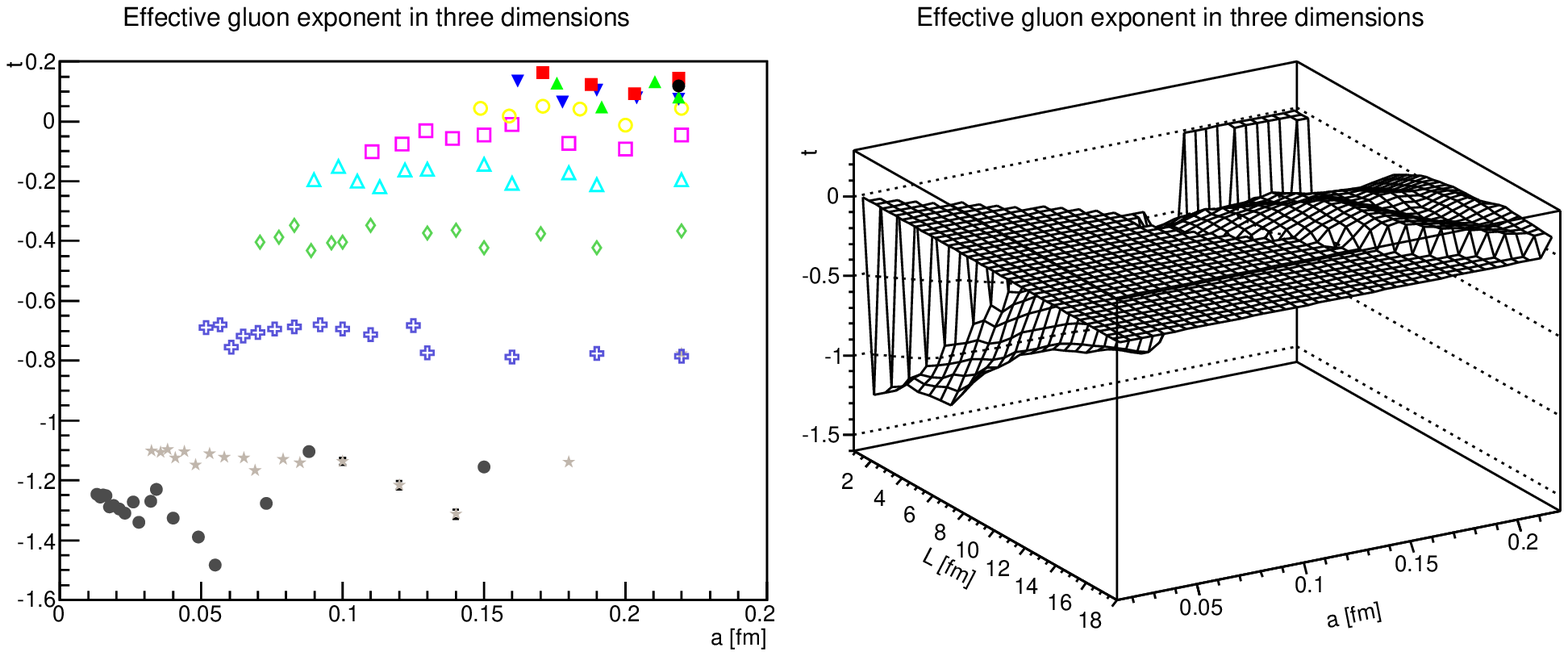}\\
\includegraphics[width=\linewidth]{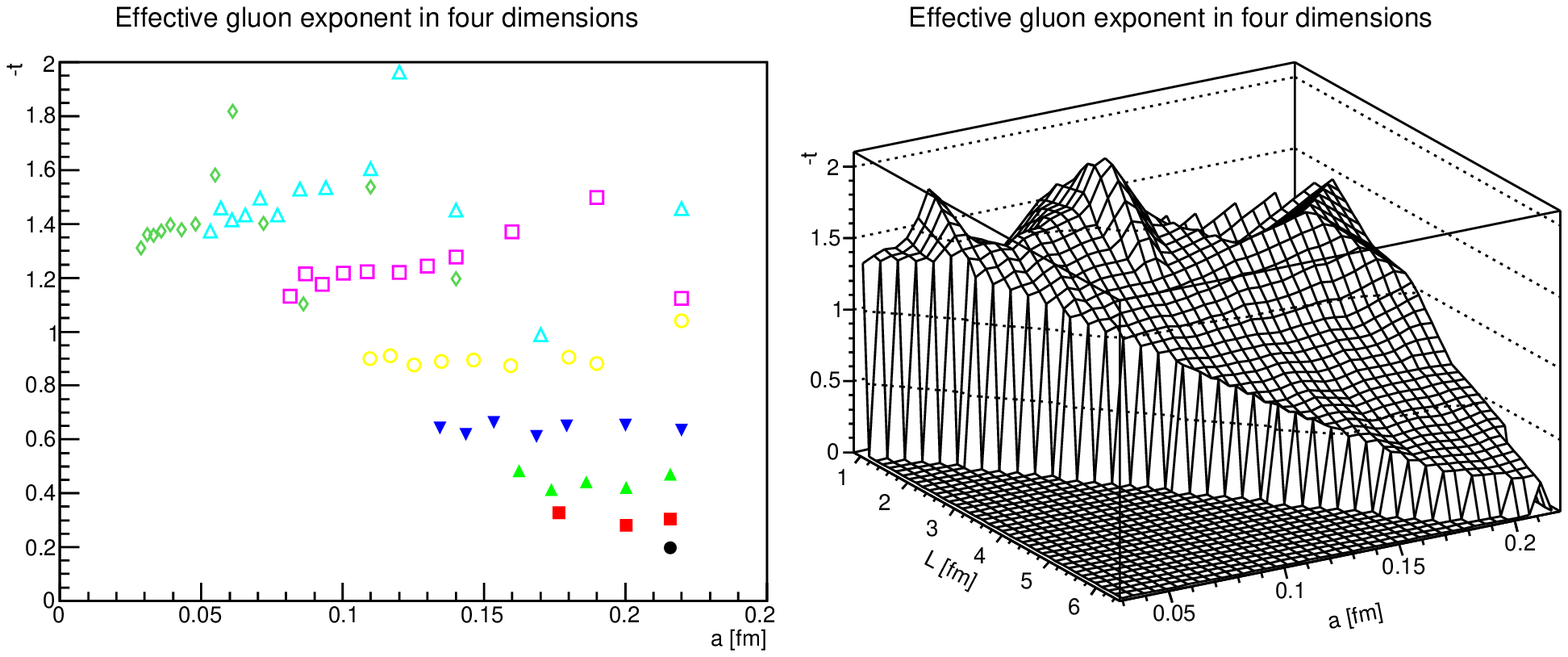}
\caption{\label{fig:gpex}The dependence of the effective gluon exponent \pref{effexpz} as a function of lattice spacing and extension in two (top panel), three (middle panel), and four (bottom panel) dimensions. The line in two dimensions is the expected value of $t=0.4$. Note the inverted scale in four dimensions for better visibility. The symbols indicate the same volumes as in figures \ref{fig:gpuv2}-\ref{fig:gpuv4}.}
\end{figure}

The gluon propagator at zero momentum, including in lower dimensions the tadpole correction \cite{Bloch:2003sk,Lepage:1992xa}, is shown in figure \ref{fig:d0}, while the effective exponent is shown in figure \ref{fig:gpex}.

For the gluon propagator at zero momentum, the behavior is rather smooth, and, when including tadpole corrections, mostly insensitive to the discretization. This also implies that without tadpole correction discretization effects have a sizable impact. It is visible how the value in two dimensions decays at sufficiently large volume like a power-law \cite{Fischer:2007pf,Maas:2007uv}, while it moves towards a constant in higher dimensions, i.\ e.\ how it decays much slower. In all cases, the infinite-volume limit has not yet been reached. The rather strong fluctuations in four dimensions is mostly due to systematic errors, like the breaking of rotational invariance, which affects the determination of the renormalization constant. There is also the quite interesting result that in four dimensions the gluon propagator at zero momentum first increases with volume, before it decreases. It is this effect which lead to the conjecture of an infrared divergent gluon propagator in very early lattice studies of the gluon propagator, when only very small volumes were available \cite{Mandula:1987rh}.

Concerning the effective exponent, the behavior in four dimensions is the one expected from previous studies: It approaches zero for large volumes. It does so from below, due to the absence of a distinct maximum. Except for some fluctuations, which come from discretization artifacts when identifying the points to extract the effective exponent \cite{Maas:2007uv}, the exponent in four dimensions is essentially independent of the lattice spacing. The situation in three dimensions is somewhat different. Due to the presence of the maximum, it overshoots the eventual value of zero, and then approaches it from above. There is a small dependence on the lattice spacing, which decreases the effective exponent further towards zero when approaching the continuum limit. Finally, in two dimensions, the exponent appears to be non-zero. There appears to be only little dependence on the volume at large volumes, though essentially none on the discretization. It does not appear to become zero at very large volumes, as investigations on extremely large lattices \cite{Cucchieri:2011ig} and continuum arguments \cite{Dudal:2012td,Huber:2012td} suggest. However, neither does it appear to reach the expected value of 0.4, and stays just a little bit smaller. To emphasize this effect, and a corresponding one for the ghost propagator, see also below figure \ref{fig:exv} in section \ref{sghpir}.

This will have some consequences to be seen below in section \ref{salpha}. This is a curious result, and somewhat unexpected. In previous extractions \cite{Maas:2007uv,Cucchieri:2011ig} of this quantity, the statistical and systematic uncertainties were too large to see this effect, and it can, of course, not be excluded that at much larger volumes and finer discretizations the exponent once more rises towards the expected value.

\subsection{Schwinger function}

The analytic structure of the gluon propagator has been a subject of much interest, as it should explain why gluons cannot be observed as free particles \cite{Alkofer:2003jj,Cucchieri:2004mf,Maas:2011se,Vandersickel:2012tg}. To find its full spectral function a solution in real time would be necessary, a task which has so far seriously only been approached with continuum methods \cite{Strauss:2012as}. There are indirect possibilities to obtain the spectral function also from the lattice data \cite{Haag:1992hx,Dudal:2010wn,Dudal:2013yva}, but those require either a statistical or systematic precision currently not available.

However, it is possible to infer such information also indirectly. One possibility is, of course, to fit it with functions of known analytic structure, see e.\ g.\ \cite{Cucchieri:2011ig}. However, this requires a prejudice on the analytic structure, and because with a finite number of points no unique statement for a fit can be made, there are always substantial systematic errors involved. An alternative is the Schwinger function \cite{Maas:2011se,Alkofer:2003jj}
\be
\Delta(t)=\frac{1}{\pi}\int_0^\infty dp_0\cos(tp_0)D(p_0^2)=\frac{1}{a\pi}\frac{1}{N_t}\sum_{P_0=0}^{N_t-1}\cos\left(\frac{2\pi tP_0}{N_t}\right)D(P_0^2)\nn,\\
\ee
\no where the first expression is the continuum one and the second one the lattice version. This function can be determined directly. However, it is still not possible to uniquely determine the analytic structure of the gluon propagator with it. But it permits to infer general properties. E.\ g.\ positivity violation of the Schwinger function directly translates into a positivity violation of the corresponding spectral function \cite{Alkofer:2003jj,Cucchieri:2004mf,Maas:2011se}, and therefore bans a particle from the physical spectrum. Furthermore, requiring that a fit works well for both the original propagator and the Schwinger function is a non-trivial constraint \cite{Maas:2011se}, and therefore provides additional valuable information. However, as it will be found below, the Schwinger function has an exponentially decaying envelope for the gluon, and therefore statistical noise seriously limits its usefulness.

\begin{figure}
\includegraphics[width=\linewidth]{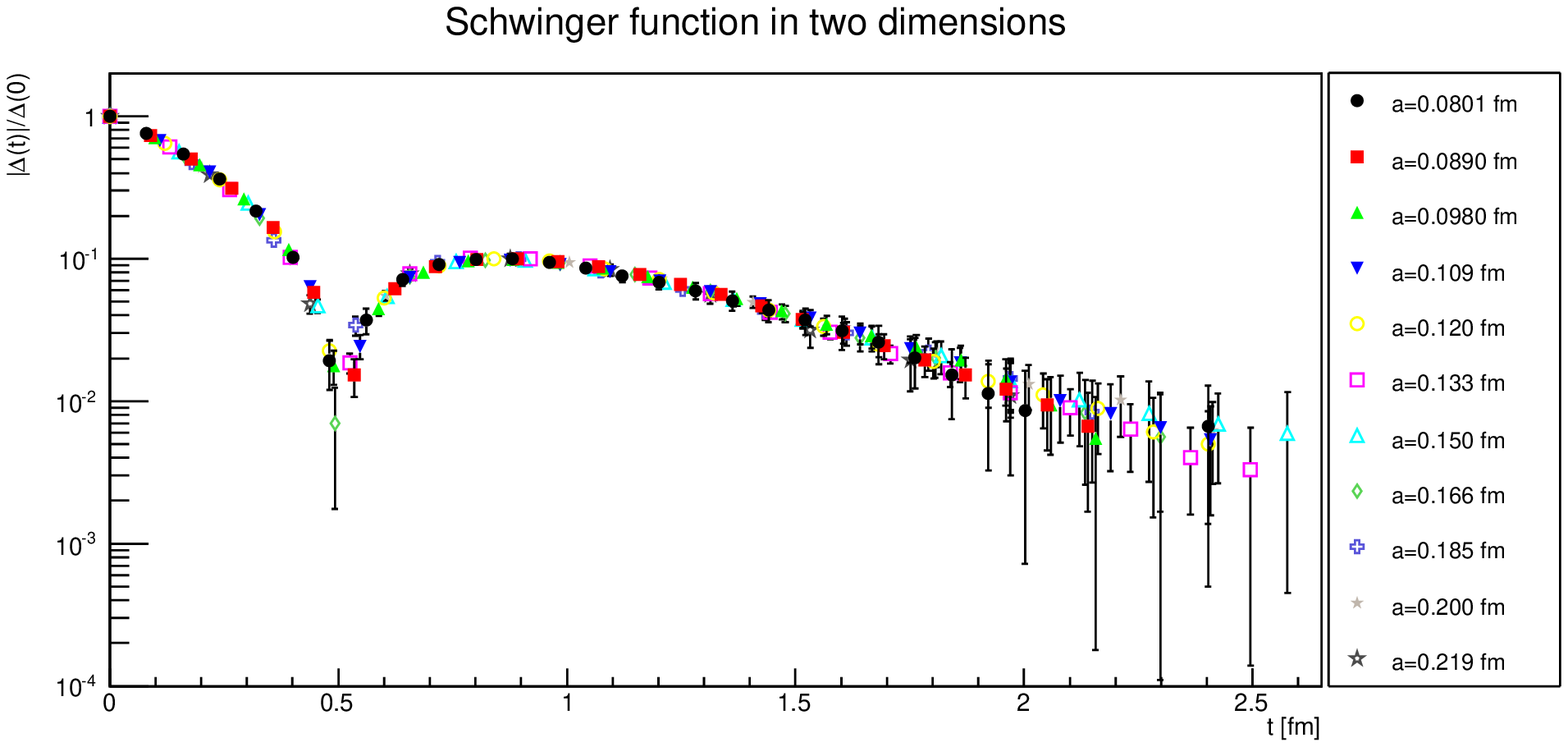}\\
\includegraphics[width=\linewidth]{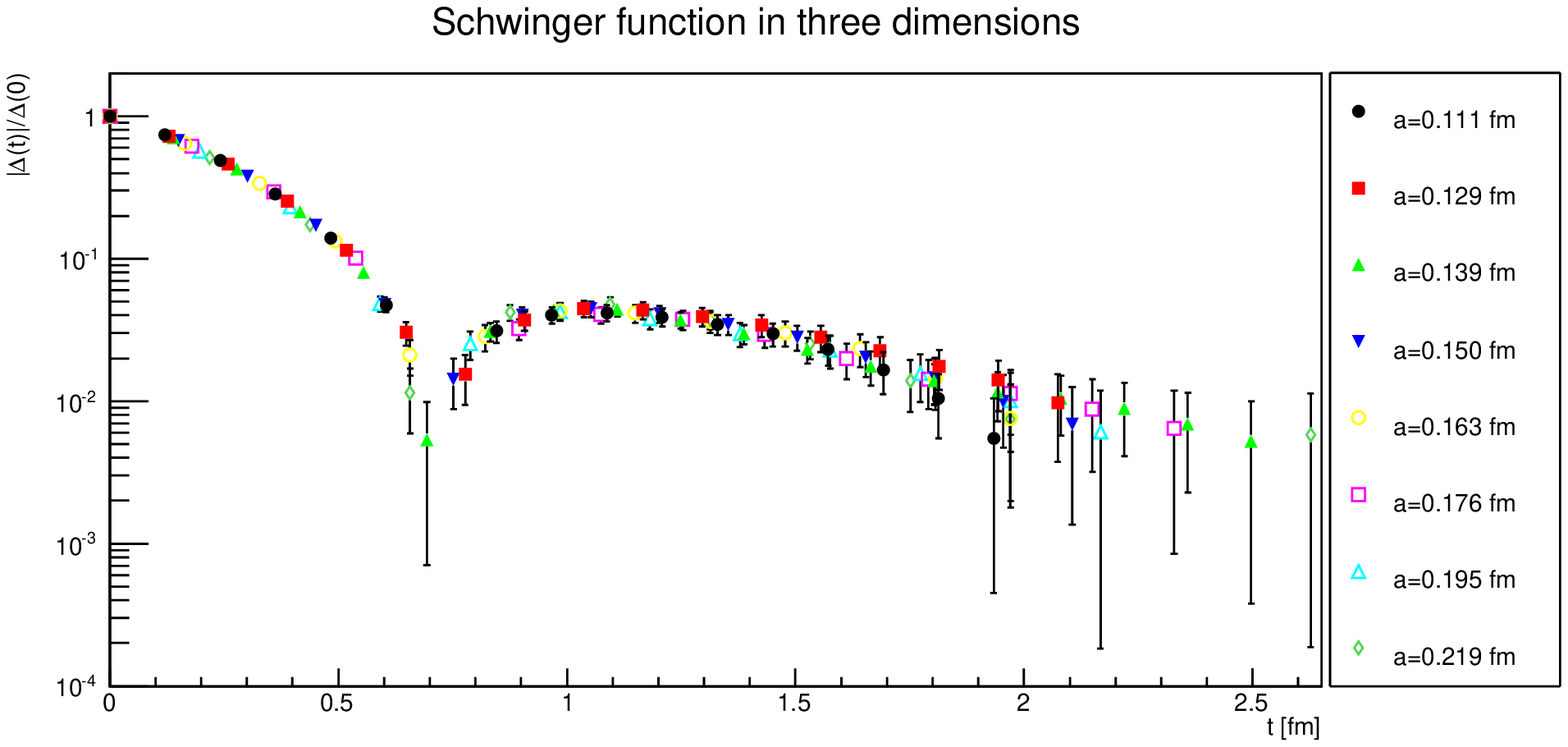}\\
\includegraphics[width=\linewidth]{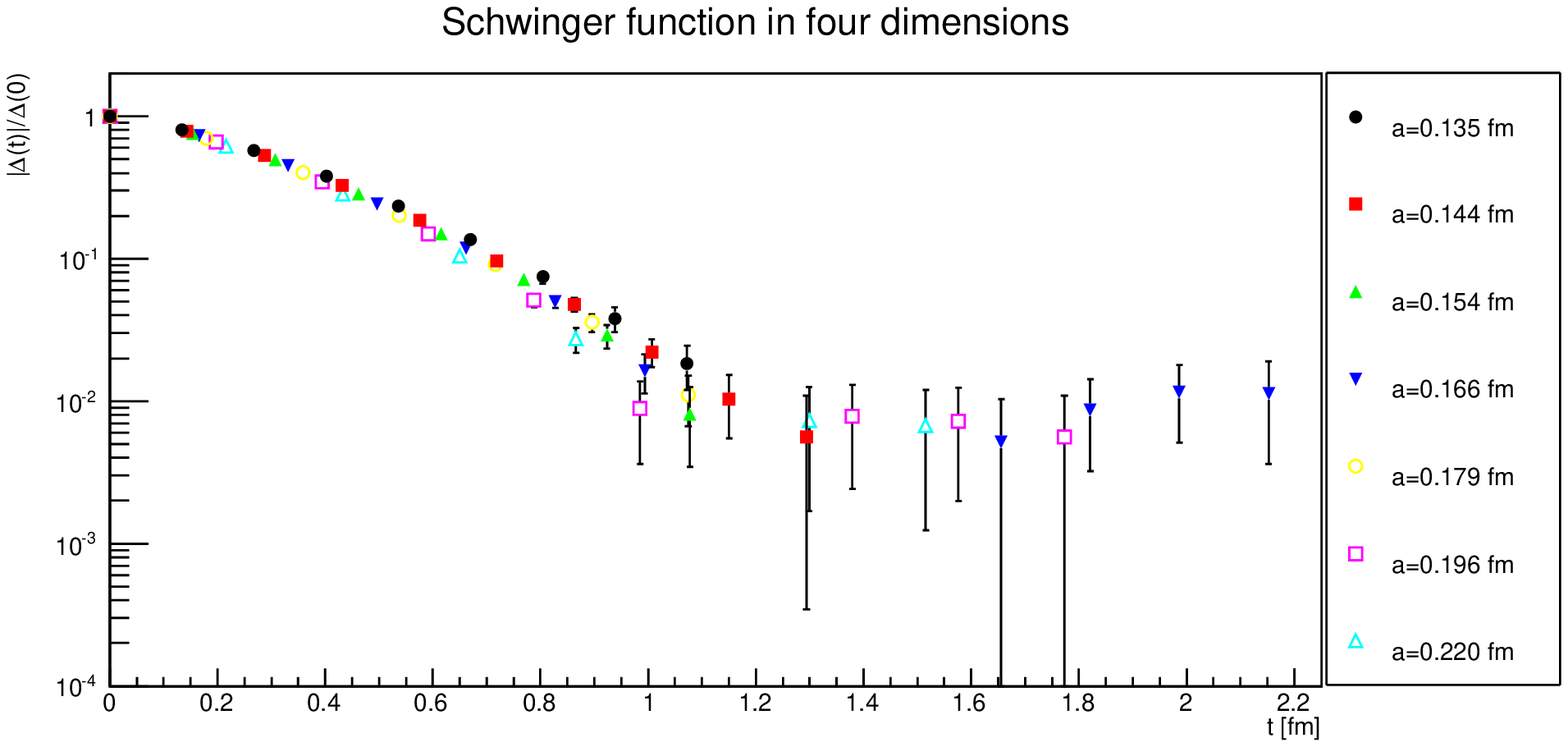}
\caption{\label{fig:gpschwingera}The dependence on the lattice spacing of the absolute value of the gluon Schwinger functions, normalized to its value at time zero. The top panel shows a fixed volume of $(9.6$ fm$)^2$ in two dimensions, the middle panel a fixed volume of $(7.5$ fm$)^3$ in three dimensions, and the bottom panel a fixed volume of $(4.3$ fm$)^4$ in four dimensions.  Points with a relative statistical error larger than 100\% have been omitted.}
\end{figure}

\begin{figure}
\includegraphics[width=\linewidth]{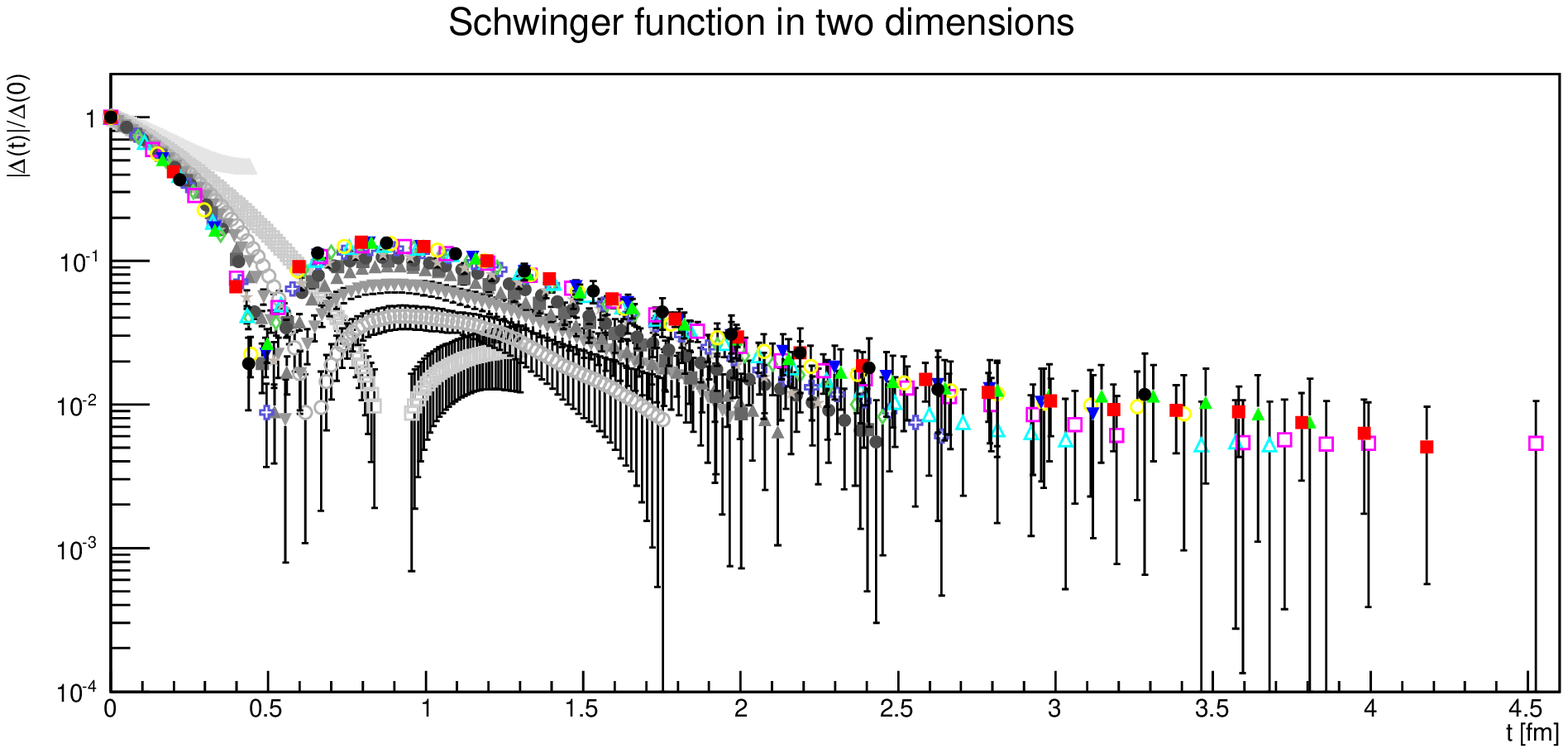}\\
\includegraphics[width=\linewidth]{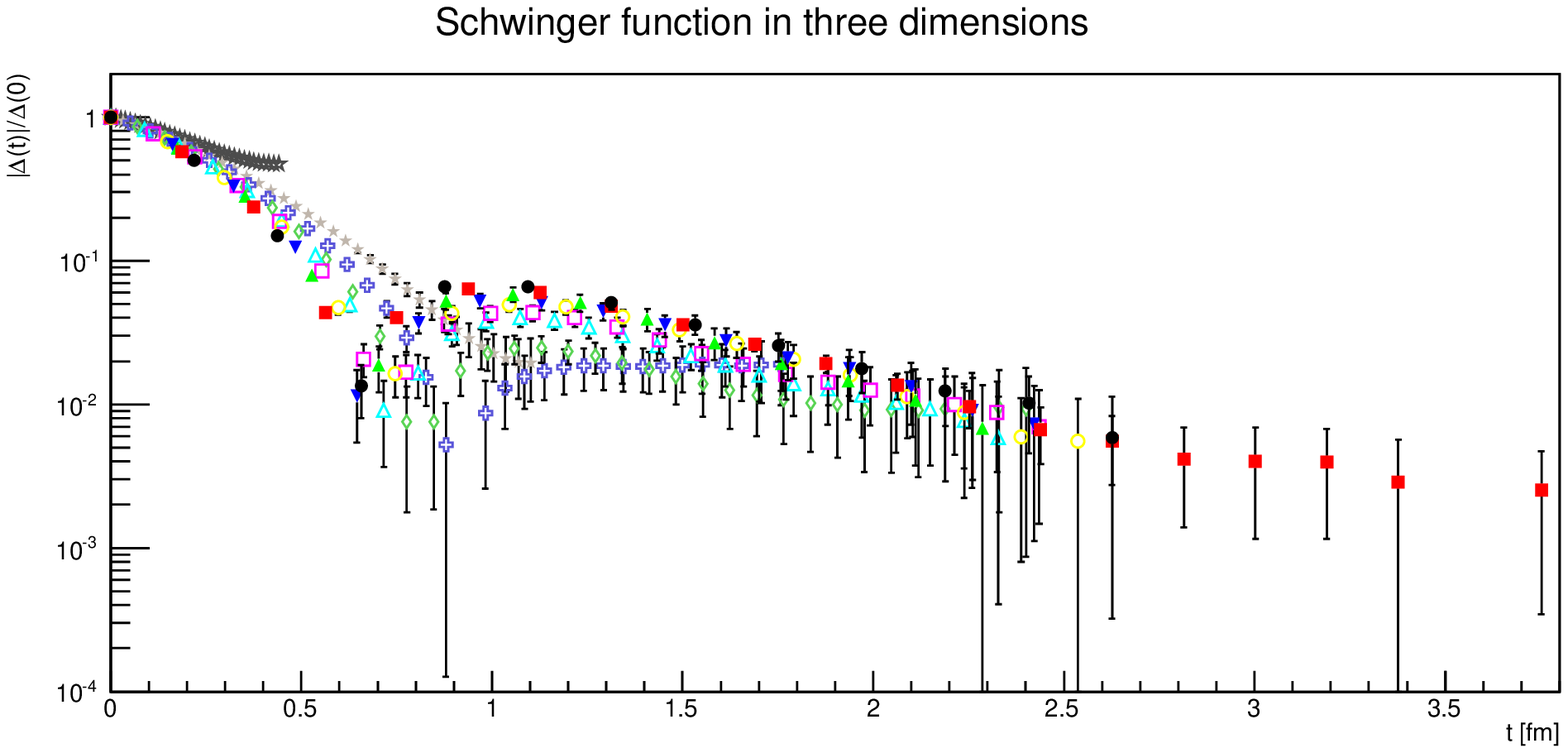}\\
\includegraphics[width=\linewidth]{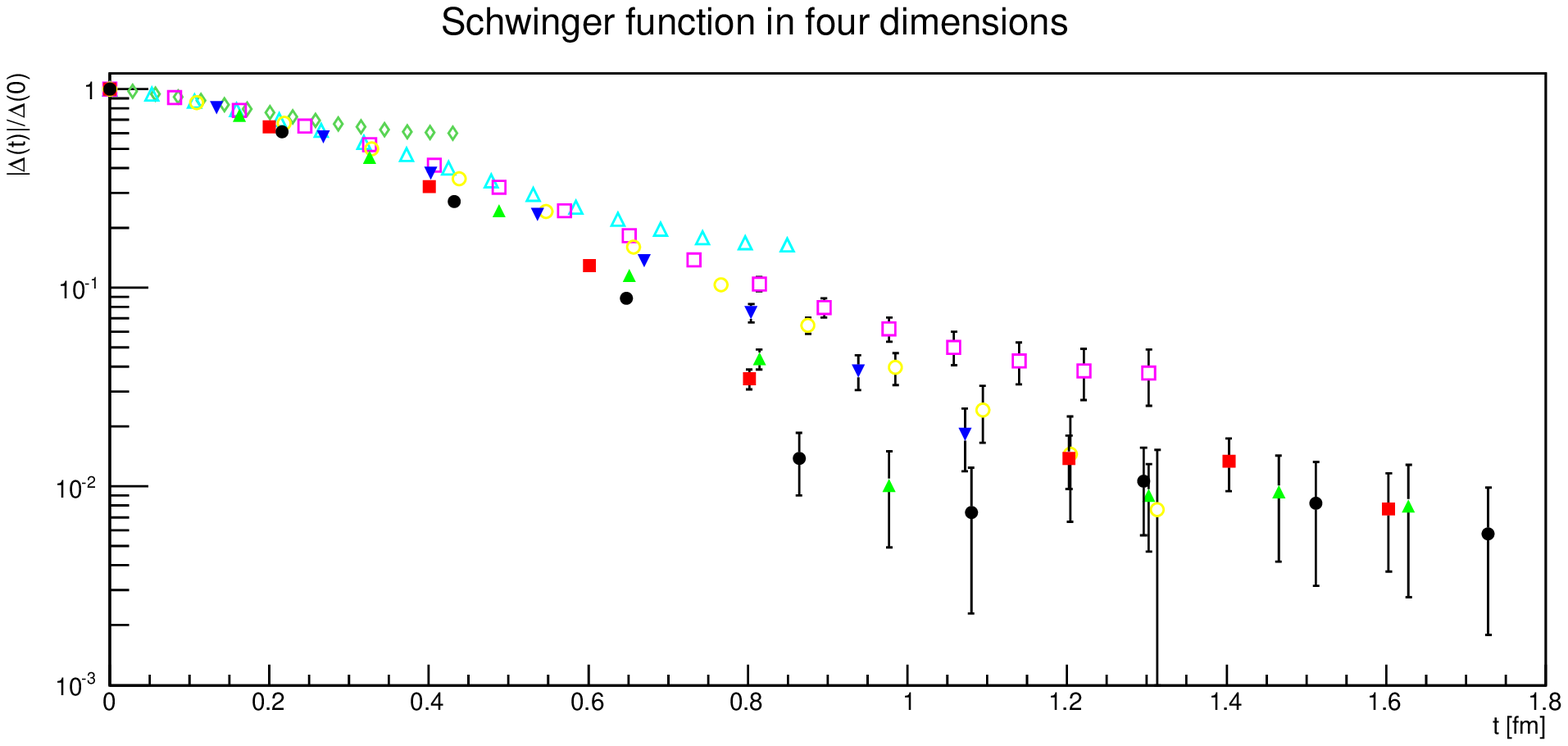}\\
\caption{\label{fig:gpschwingerv}The absolute value of the gluon Schwinger functions for different volumes, normalized to its value at time zero. The top panel shows the results in two dimensions, the middle panel in three dimensions, and the bottom panel in four dimensions. Points with a relative statistical error larger than 100\% have been omitted. Symbols have the same meaning as in figures \ref{fig:gpuv2}-\ref{fig:gpuv4}.}
\end{figure}

The Schwinger function for various lattice discretizations are shown in figure \ref{fig:gpschwingera} and for different volumes in figure \ref{fig:gpschwingerv}. First of all, for all practical purposes no dependence on the lattice spacing are observed. This leaves only the volume dependence. In two and three dimensions a positivity violation is clearly seen for volumes larger than $(5-6$ fm$)^d$, which moves from about 1 fm to smaller times with increasing volumes, and saturates at a scale of roughly half a Fermi. In four dimensions, due to the rather small volumes, the zero crossing at about 1 fm is just so observed, though it has been clearly established in studies on larger volumes \cite{Bowman:2007du}.

In two dimensions no second zero crossing is observed for at least 4-4.5 fm, which is 8-9 times the scale of the first zero crossing. In three dimensions, this is also true up to at least 3.5 fm. After this, statistical uncertainties become too large for a statement. The absence of a second zero crossing severely restricts the possibility of a double pole structure, and is much more in line with the one expected for a cut structure along the real axis \cite{Maas:2011se}, which would also agree with results from functional methods \cite{Alkofer:2003jj,Strauss:2012as}. However, it cannot be excluded that the parameters of a double-pole structure conspire, and move further zero crossings beyond the range accessible in the present calculations, see e.\ g.\ \cite{Huber:2013xb}. Nonetheless, even if this is the case, this is a major restriction for any fits with double poles.

Of course, in four dimensions, where the first zero crossing is just so observed, any further speculations are pointless.

For the ghost the determination of the Schwinger function requires a regulator due to the infrared divergence. Since this regulator dependence can affect both the long-time and positivity properties of the ghost spectral function, it is not quite clear how to extract the corresponding properties indirectly, instead of the direct approach as in \cite{Strauss:2012as}. Thus, this will not be investigated further here. Some results\footnote{See \cite{Greensite:2010tm} for similar considerations in Coulomb gauge.} for a particular regularization can be found in \cite{Maas:2011se}, which indicate that the ghost violates both, positivity and cluster decomposition. The latter is in agreement with the expectations for a confining theory \cite{Alkofer:2000wg}.

In principle, it is possible to determine a Schwinger function also for the running coupling. However, since formally the running coupling is just a three-point function in a special kinematic configuration, it is not at all clear what the physical interpretation of any structure is. Its determination will therefore be skipped here, see \cite{Maas:2011se} for an exploratory study.

\section{Ghost propagator}\label{sghp}

\subsection{Ultraviolet}

\begin{figure}
\includegraphics[width=\linewidth]{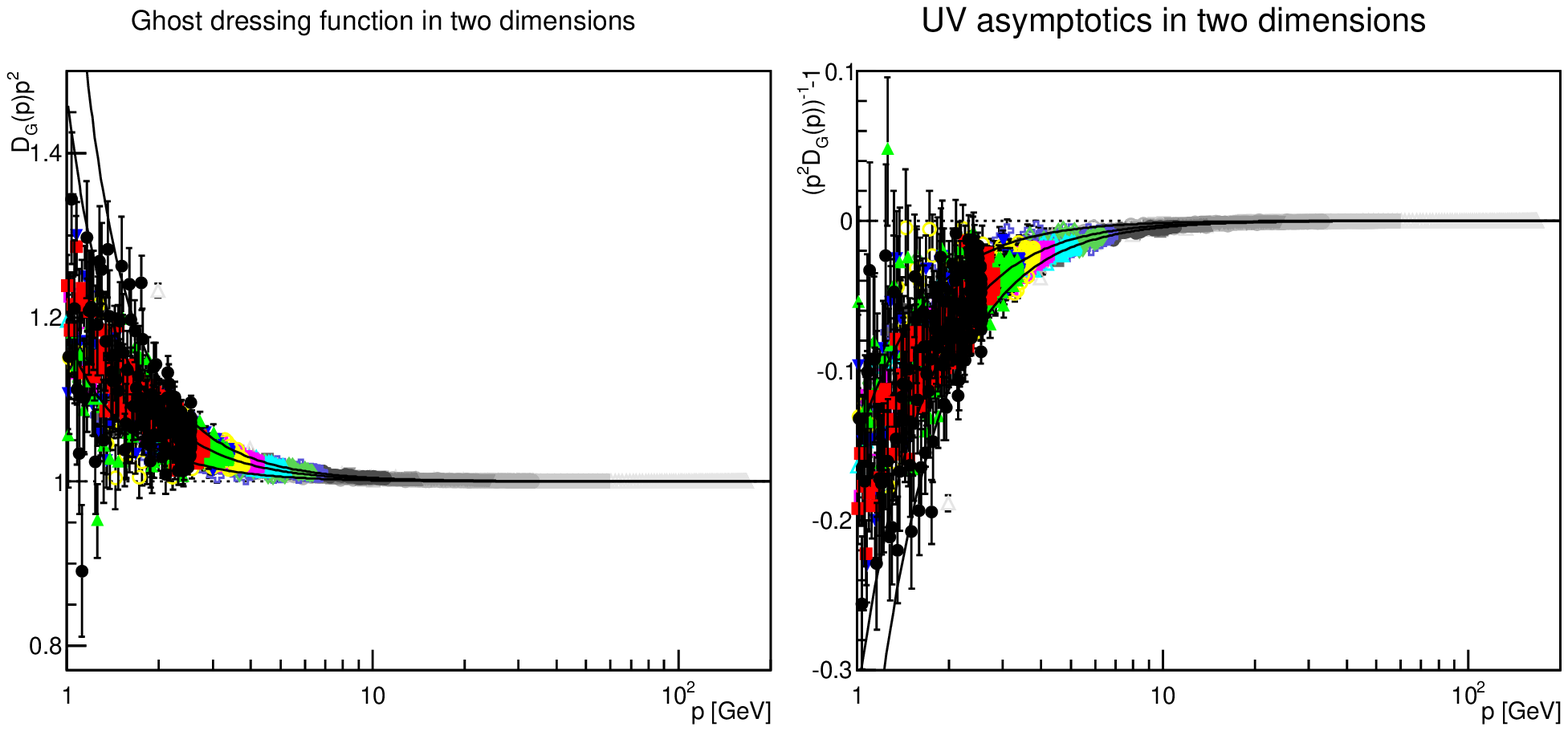}\\
\includegraphics[width=\linewidth]{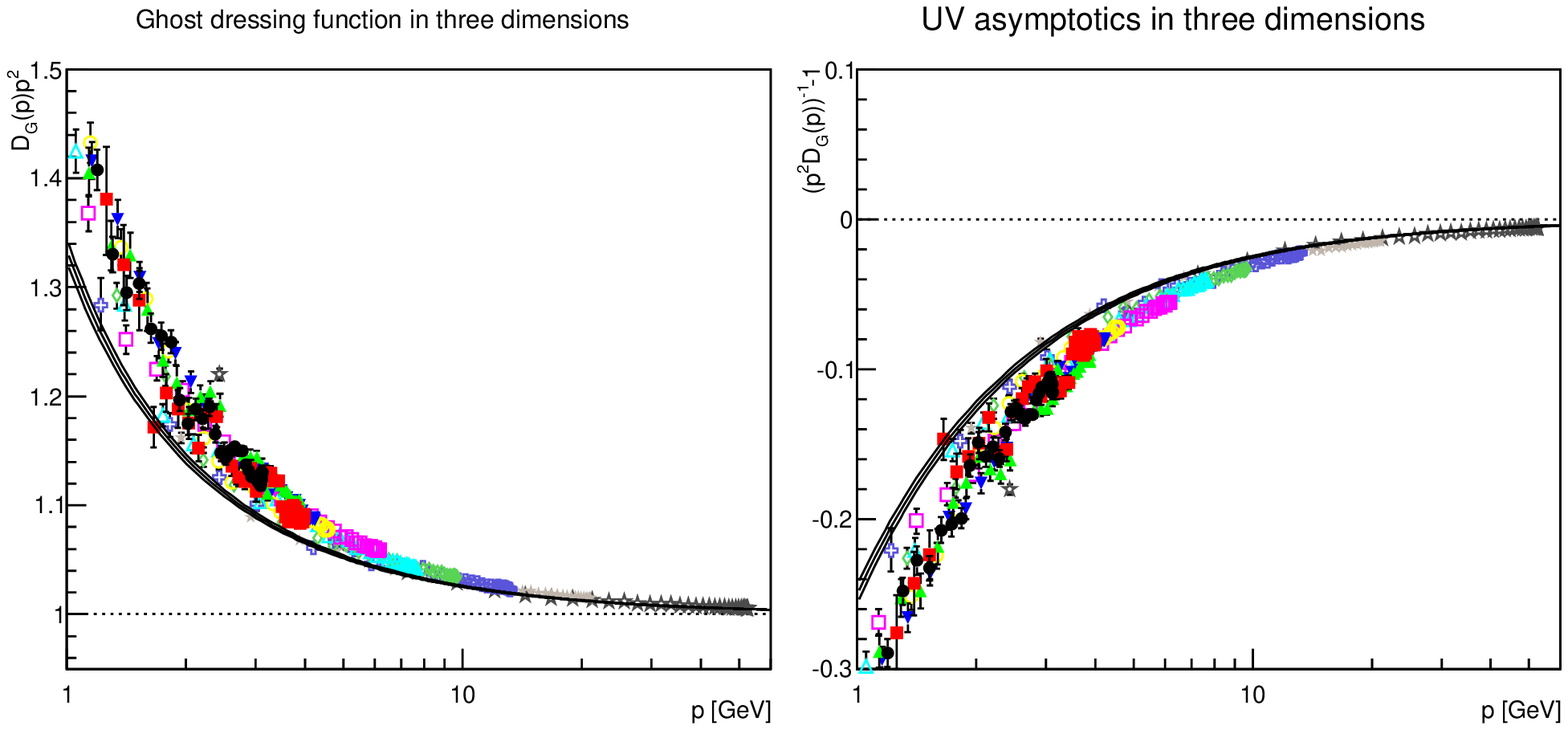}\\
\includegraphics[width=\linewidth]{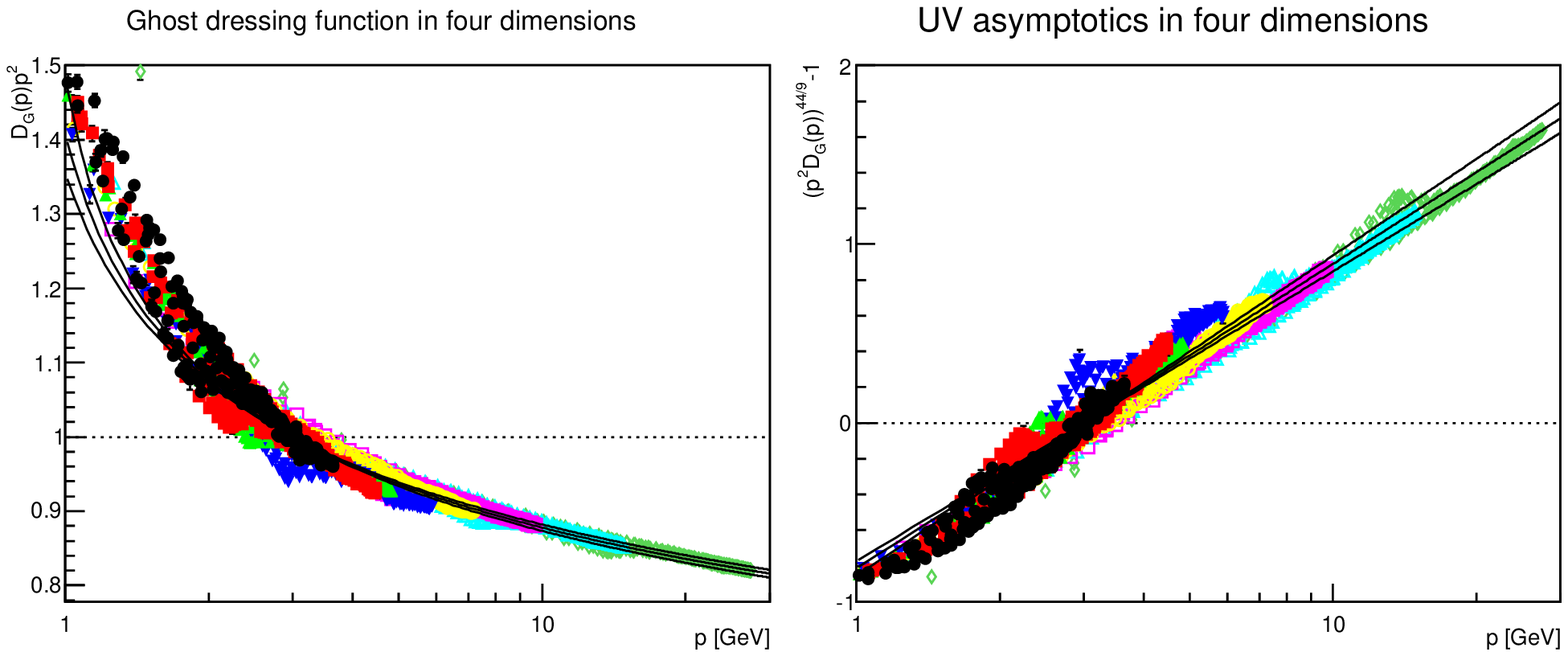}\\
\caption{\label{fig:ghpuv}The ghost dressing function at large momenta along the space-time-diagonal (in four dimensions all possible space-diagonals are shown), compared to the leading-order behavior \prefr{uvg2}{uvg4}. The values for the parameter bands are $fg^2=-0.32^{+13}_{-17}$, $g^2=2.05(5)$, and $g=4.8(1)$ in two, three, and four dimensions, respectively. On the right-hand side, the leading asymptotic has been isolated. The symbols have the same meaning as in figures \ref{fig:gpuv2}-\ref{fig:gpuv4}. Note the different scales on the right-hand side. In four dimensions the dressing function was renormalized at $\mu=3$ GeV. Top panels are two dimensions, middle panels three dimensions, and bottom panels four dimensions.}
\end{figure}

\afterpage{\clearpage}

In a very similar way as for the gluon the leading ultraviolet behavior of the ghost propagator is determined by
\bea
p^{-2}D_{G2d}(p)^{-1}&=&1+\frac{fg^2}{p^2}\label{uvg2}\\
p^{-2}D_{G3d}(p)^{-1}&=&1-\frac{g^2}{8p}\\
p^{-2}D_{G4d}(p)^{-1}&=&\left(\frac{33g^2}{208\pi^2}\log\left(\frac{p^2}{\mu^2}\right)+1\right)^{\frac{9}{44}}\label{uvg4},
\eea
\no where the value of $g$ must coincide with the one determined from the gluon propagator, though the parameter $f$ in two dimensions is, of course, independent. The results are shown in figure \ref{fig:ghpuv}. The value of $fg^2$ is found to be $fg^2=-0.32^{+13}_{-17}$. In four dimensions, the same value of $g^2$ has again been used, showing again a very good agreement to the leading-order perturbative result. The result in three dimensions is satisfactory, though not very good, for the reasons discussed before in section \ref{sgpuv}.

\begin{figure}
\includegraphics[width=\linewidth]{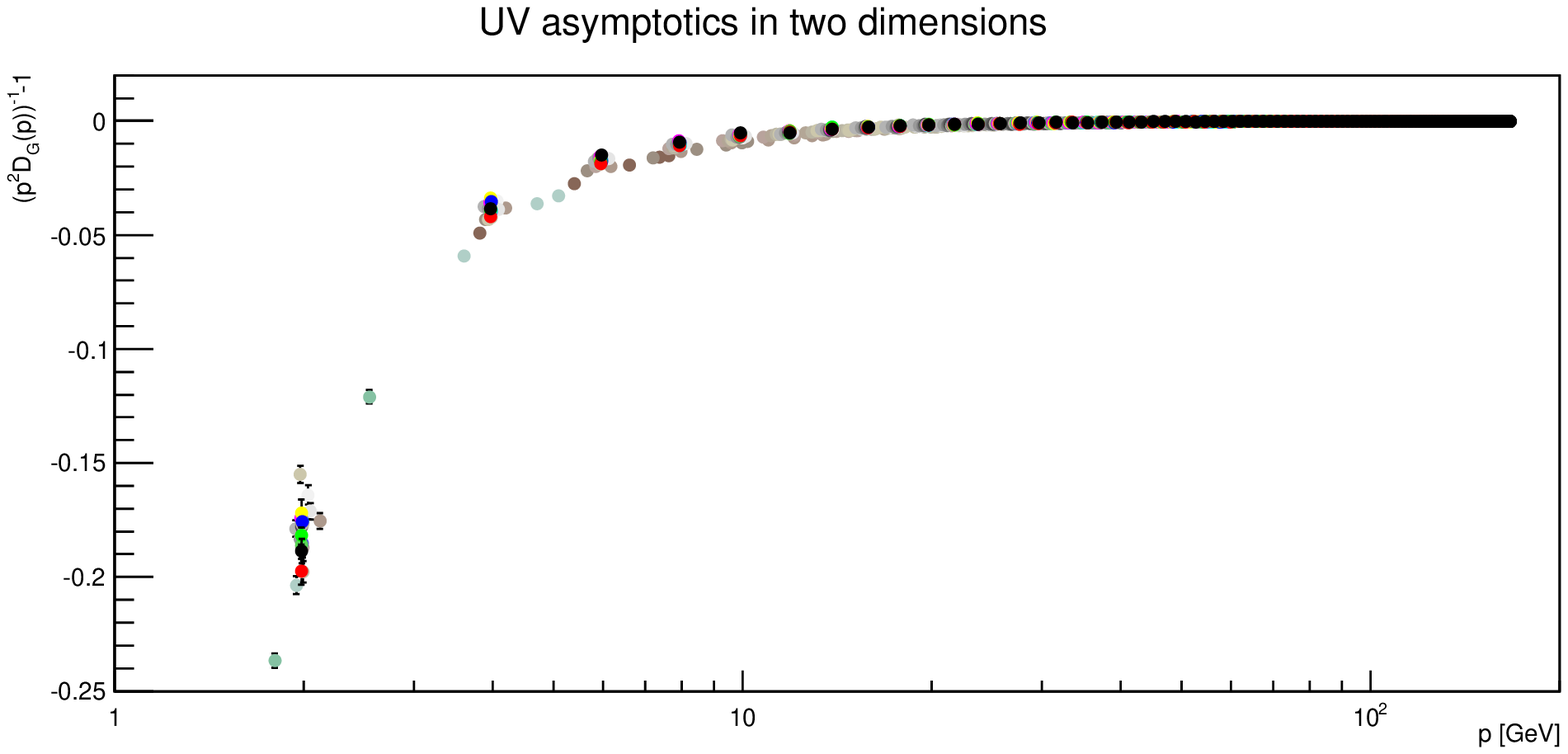}\\
\includegraphics[width=\linewidth]{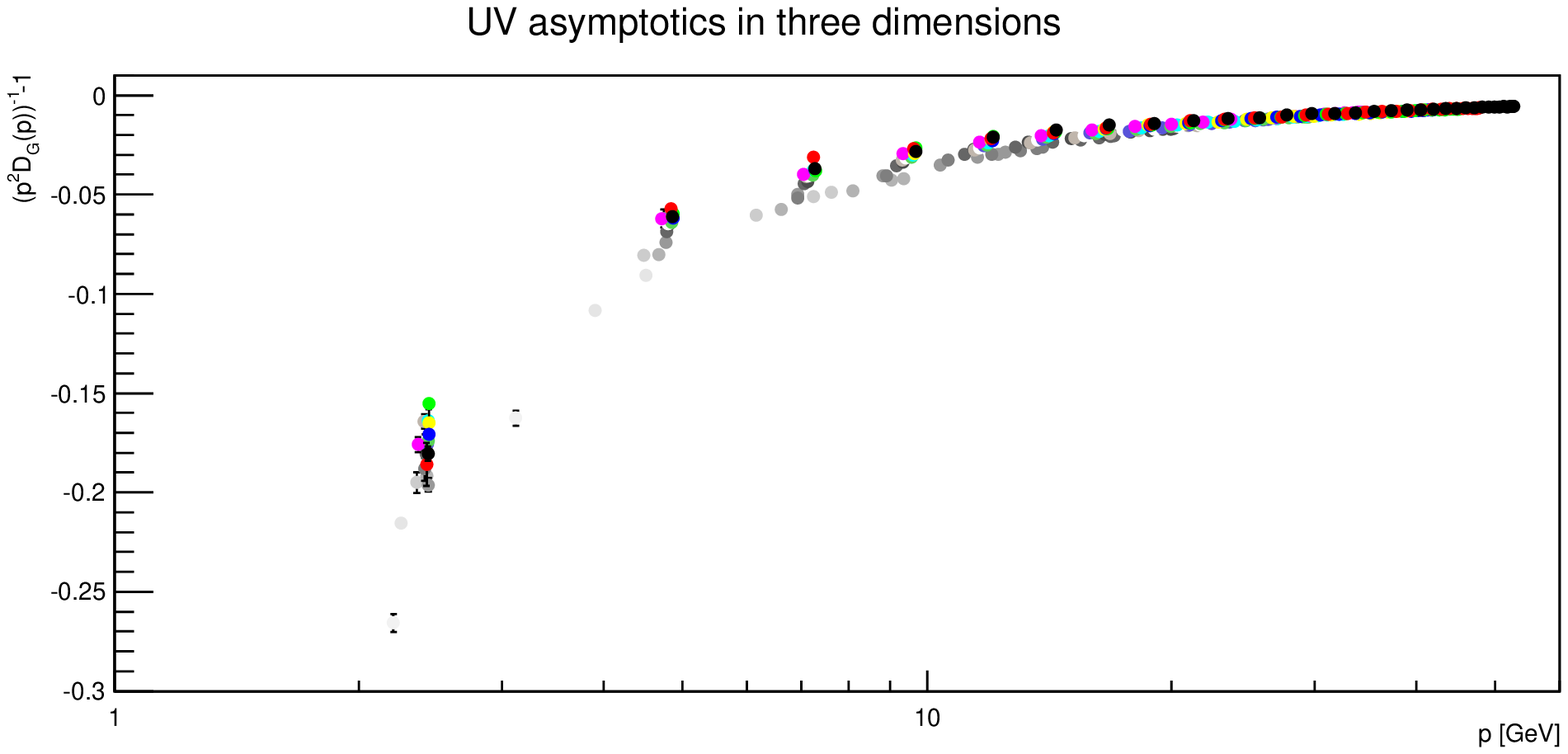}\\
\includegraphics[width=\linewidth]{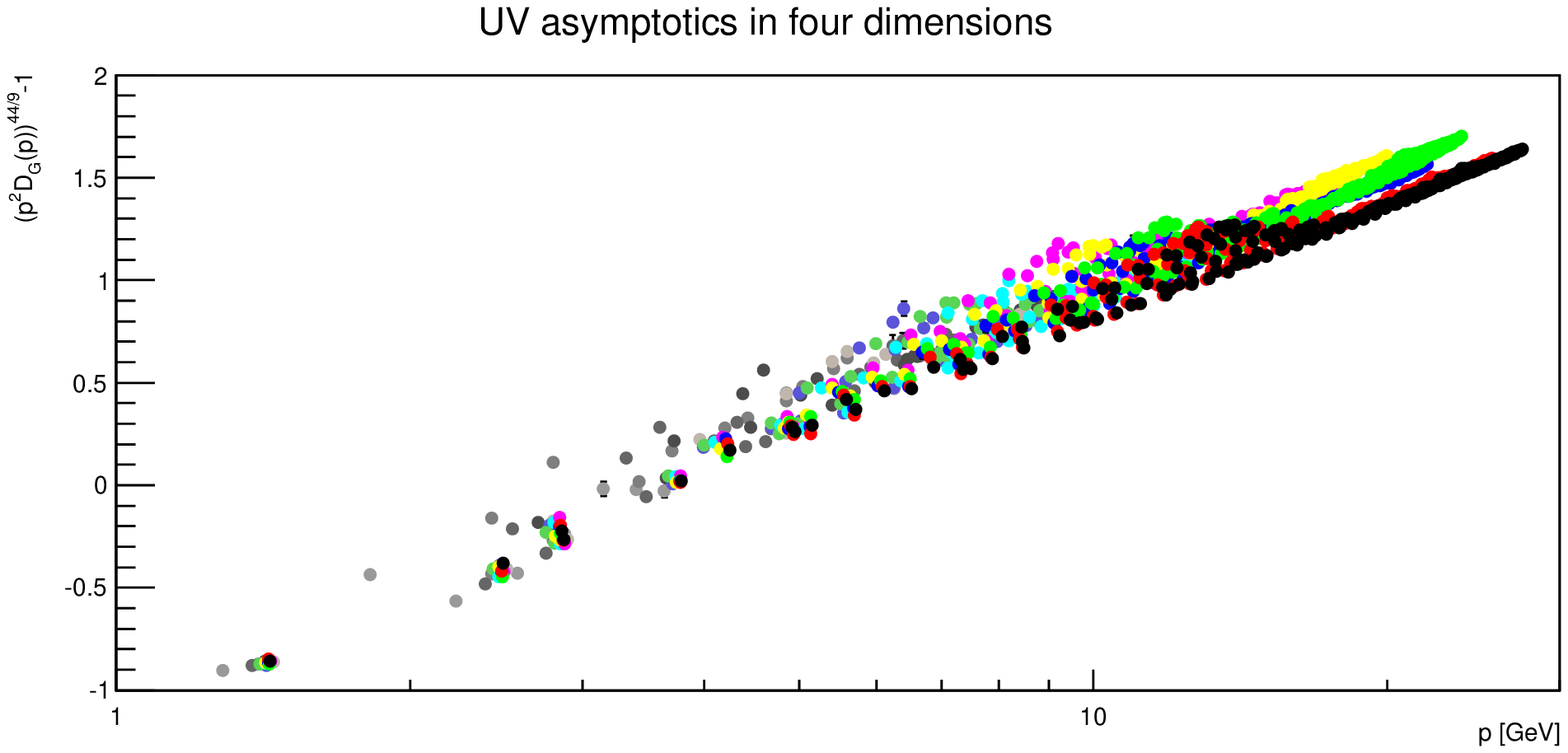}
\caption{\label{fig:ghpa}The dependence of the ghost asymptotics, as isolated from \prefr{uvg2}{uvg4} in two (top panel), three (middle panel), and four (bottom panel) dimensions. See figure \ref{fig:ghpuv} for details. The symbols have the same meaning as in figures \ref{fig:gpuv2}-\ref{fig:gpuv4}.}
\end{figure}

To investigate to which extend this result is affected by discretization artifacts, the corresponding plot from the gluon propagator in figure \ref{fig:gpa} for the ghost propagator is shown in figure \ref{fig:ghpa}. Similar to the gluon case, the importance of discretization artifacts increases with dimension, and is somewhat larger than for the gluon. Especially, essentially no discretization artifacts are seen in two dimensions, while even for the finest lattices still a systematic trend is visible in three, and in particular four dimensions. The aforementioned discrepancies in three dimensions between leading-order behavior and observed behavior may therefore also be partly due to this effect.

\subsection{Infrared}\label{sghpir}

\begin{figure}
\includegraphics[width=\linewidth]{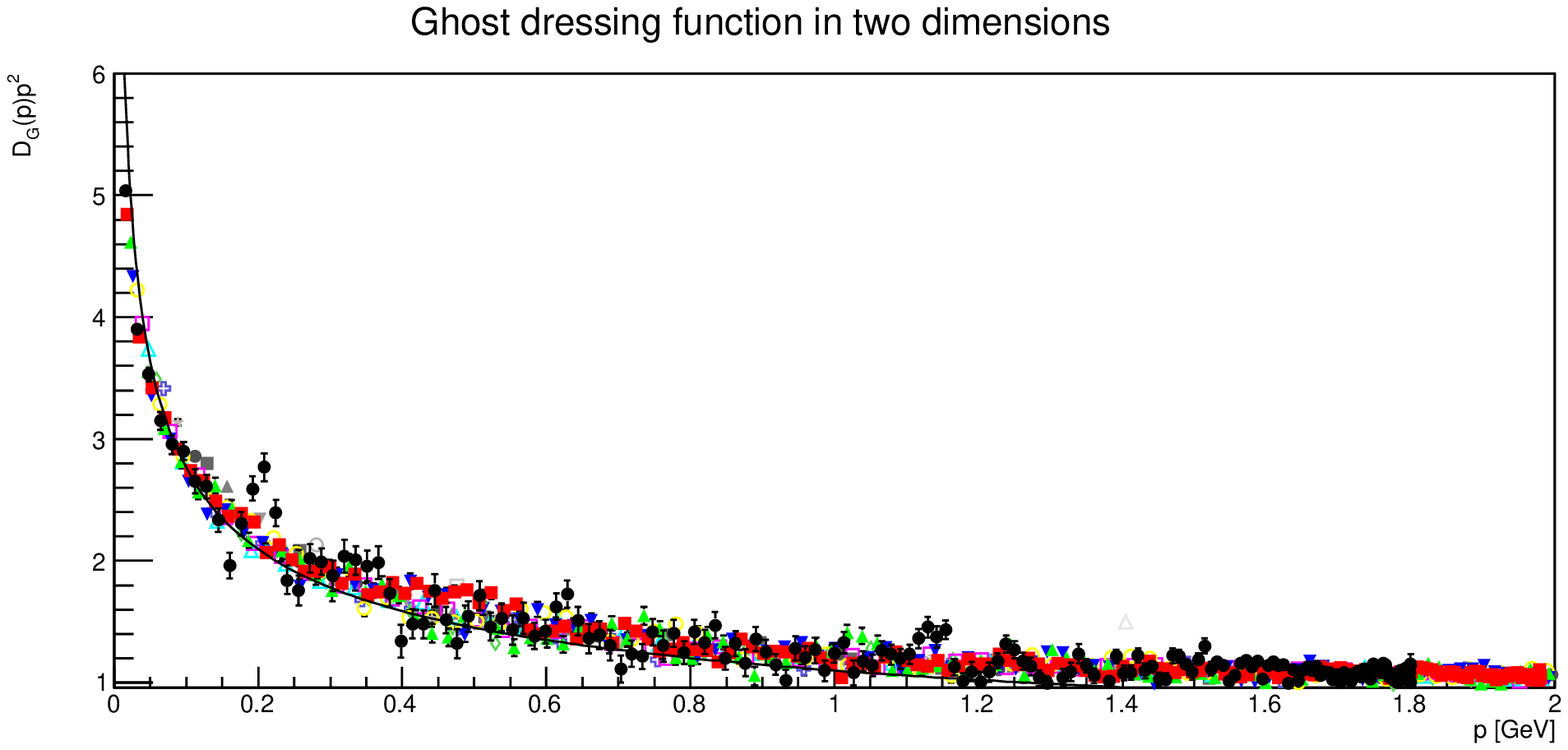}\\
\includegraphics[width=\linewidth]{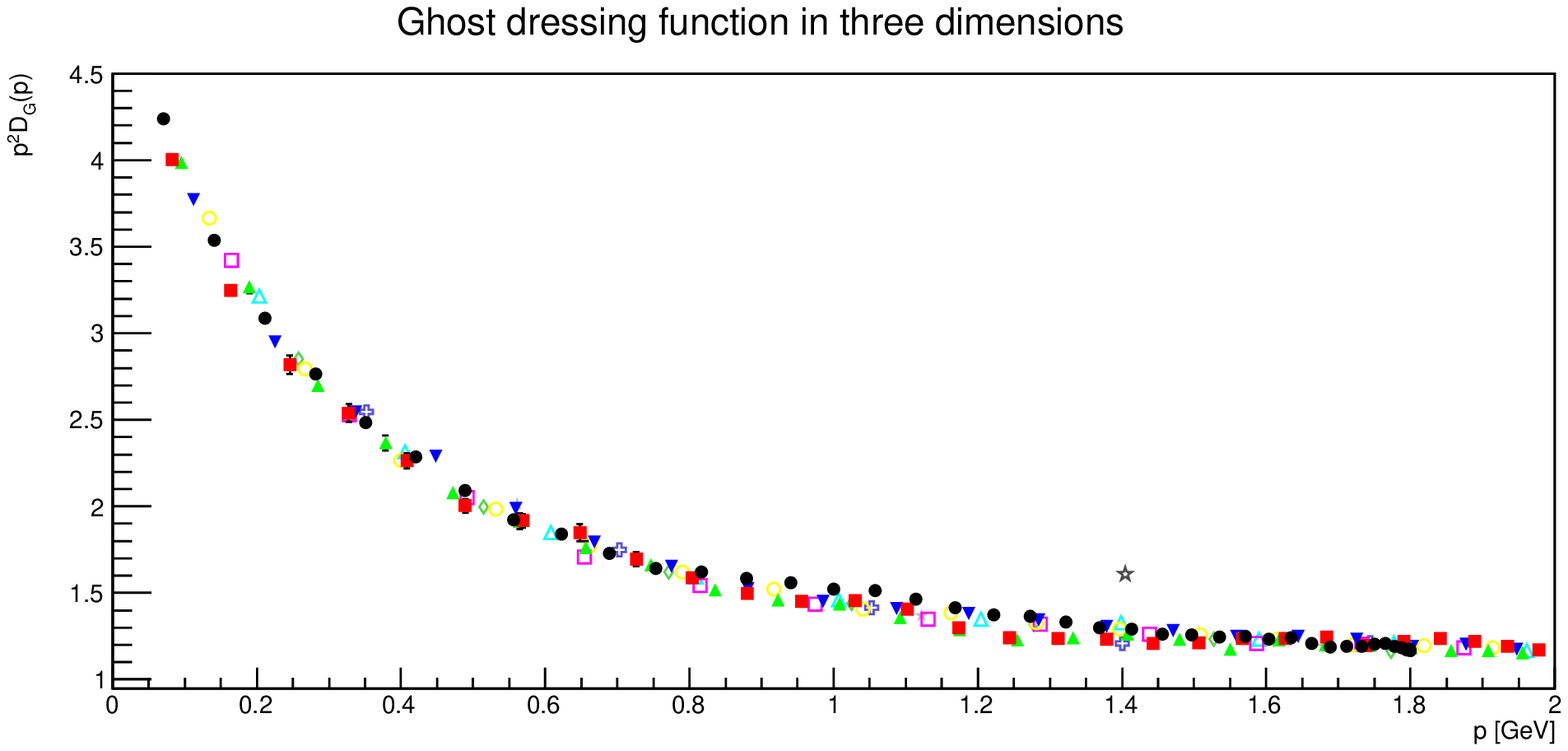}\\
\includegraphics[width=\linewidth]{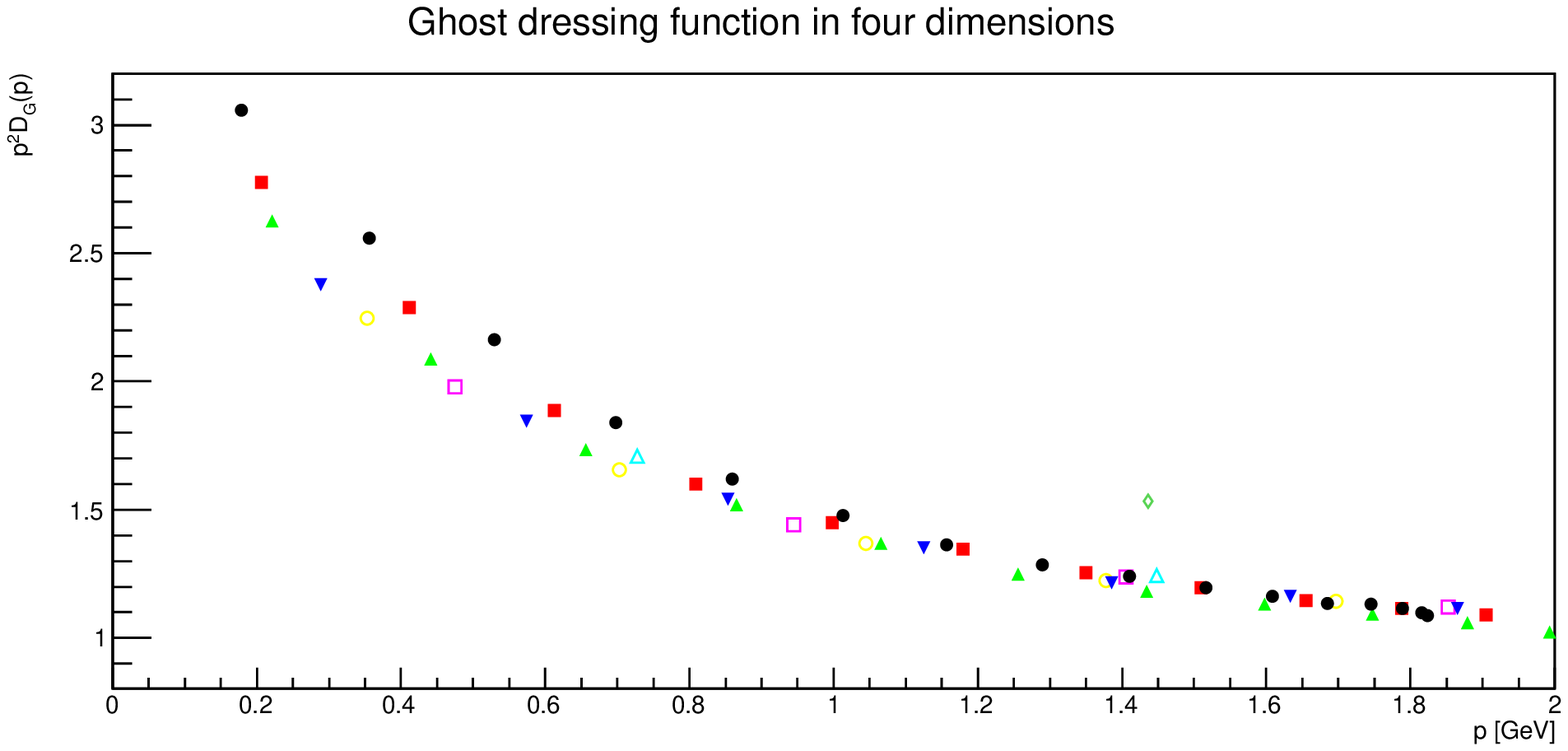}
\caption{\label{fig:ghpir}The ghost dressing function at small momenta along the $x$-axis. Symbols are as in figures \ref{fig:gpuv2}-\ref{fig:gpuv4}.  In four dimensions the dressing function was renormalized with the same renormalization factors as in figure \ref{fig:ghpuv}. The top panel shows two dimensions, the middle panel three dimensions, and the bottom panel four dimensions. The function shown for two dimensions is $1.1p^{0.4}$.}
\end{figure}

The ghost propagator is throughout essentially dominated by the trivial $1/p^2$ factor. Therefore, to study the low-momentum behavior, in figure \ref{fig:ghpir} only the ghost dressing function is shown. A significant enhancement is seen, though studies on much larger volumes \cite{Sternbeck:2007ug,Bogolubsky:2007ud,Cucchieri:2007rg,Cucchieri:2008fc} reveal that the dressing function is finite in three and four dimensions. It appears to remain divergent in two dimensions, in agreement with previous studies \cite{Cucchieri:2008fc,Maas:2007uv}.

\begin{figure}
\includegraphics[width=\linewidth]{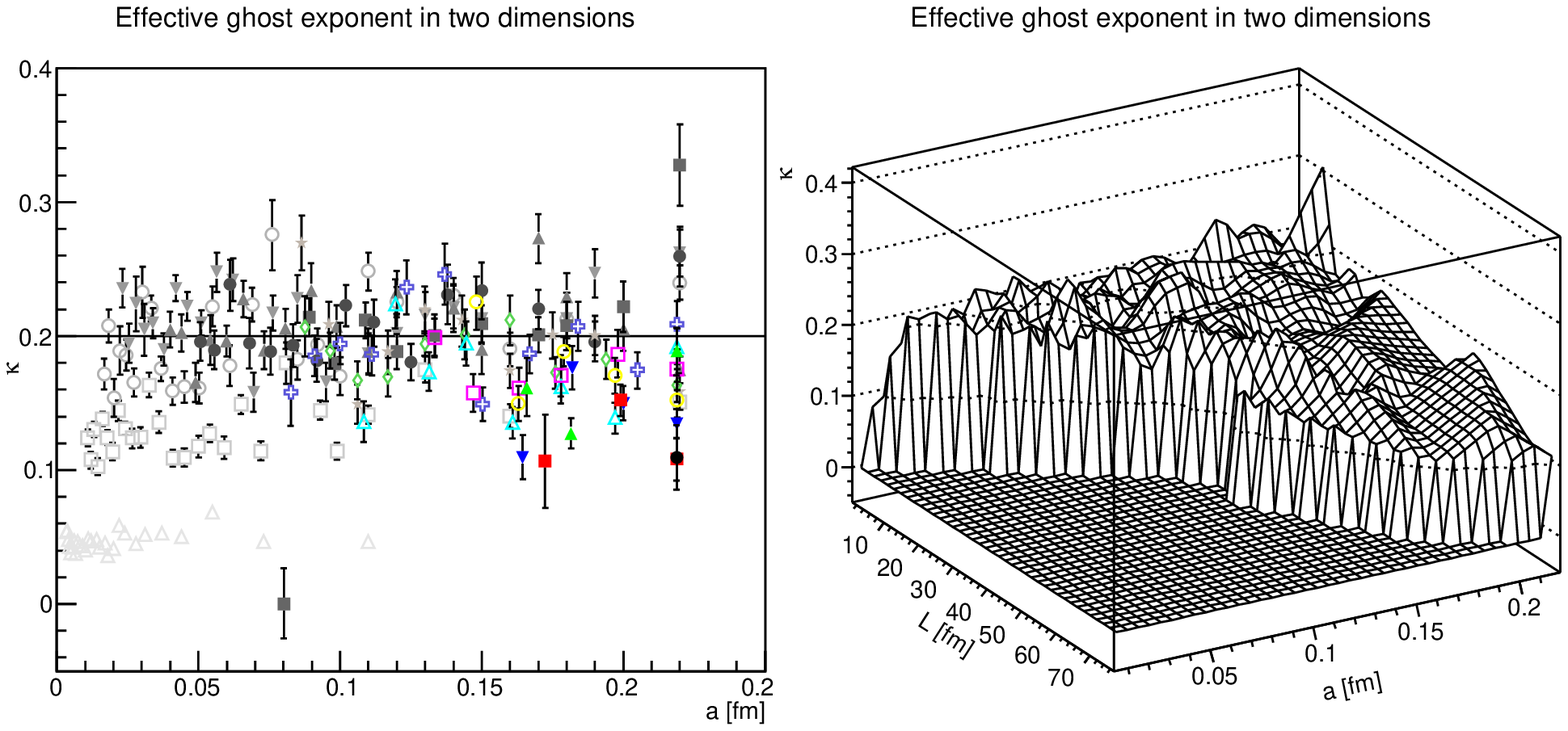}\\
\includegraphics[width=\linewidth]{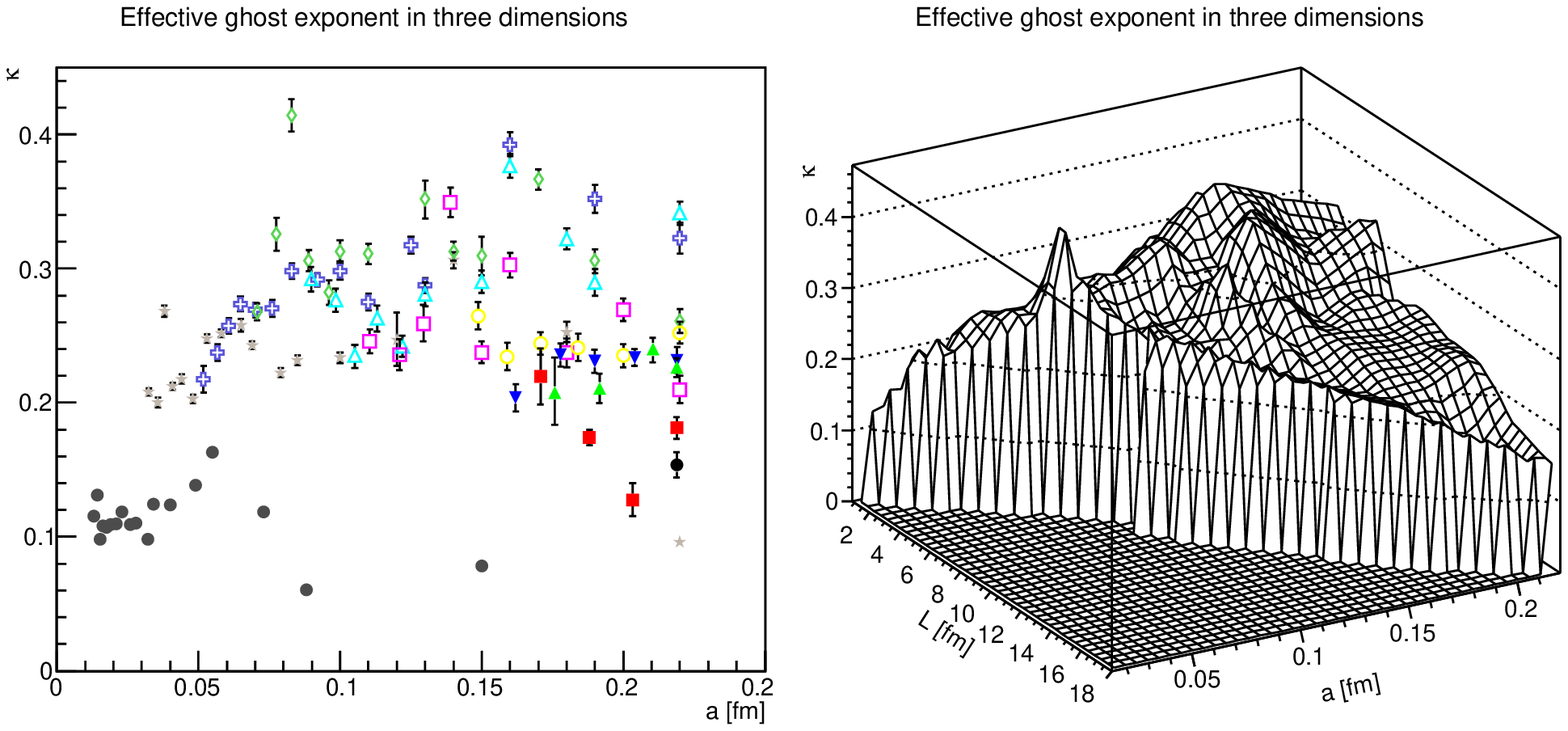}\\
\includegraphics[width=\linewidth]{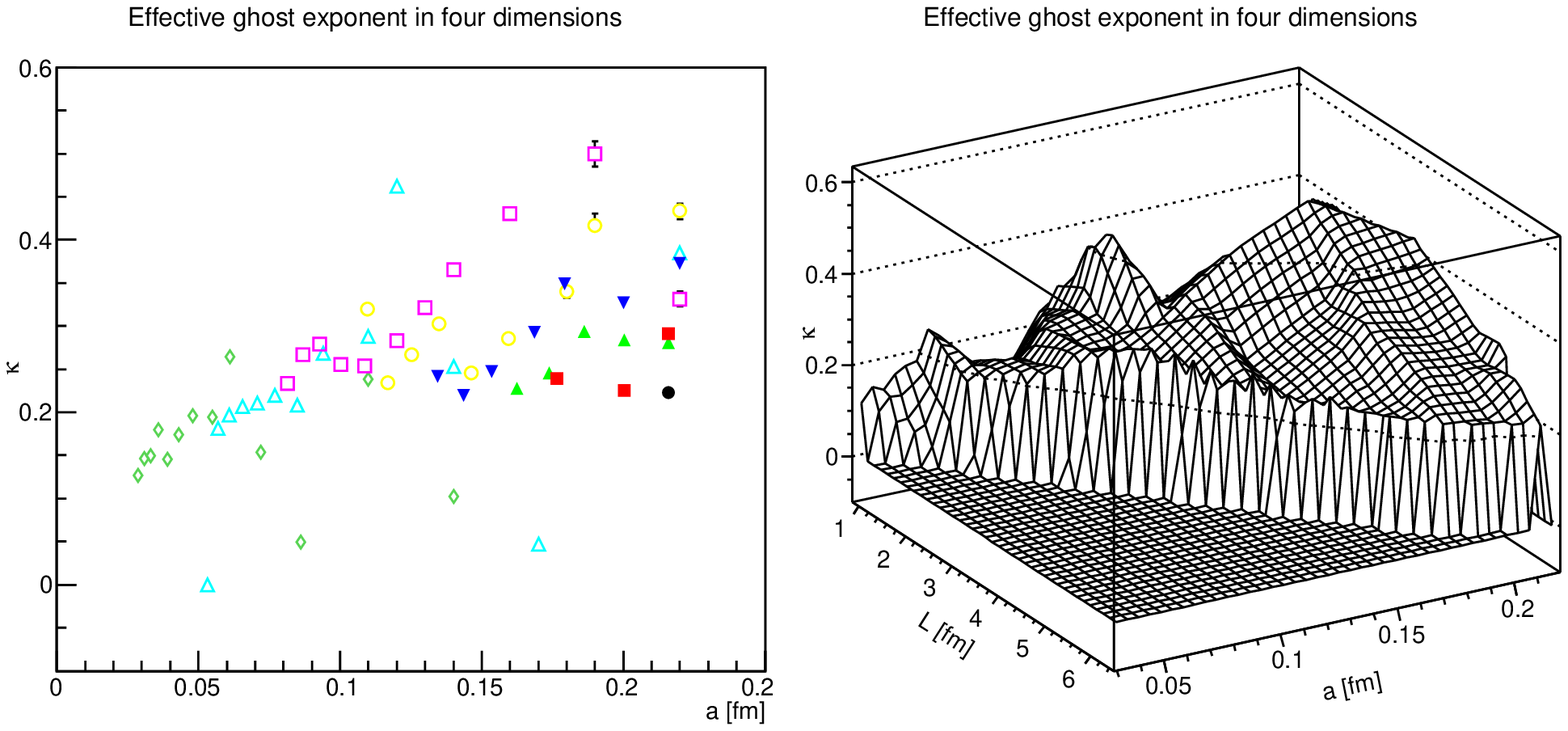}
\caption{\label{fig:ghpex}The dependence of the effective ghost exponent \pref{effexpg} as a function of lattice spacing and extension in two (top panel), three (middle panel), and four (bottom panel) dimensions. The symbols denote the same volumes as in figures \ref{fig:gpuv2}-\ref{fig:gpuv4}.}
\end{figure}

\afterpage{\clearpage}

\begin{figure}
\includegraphics[width=0.5\linewidth]{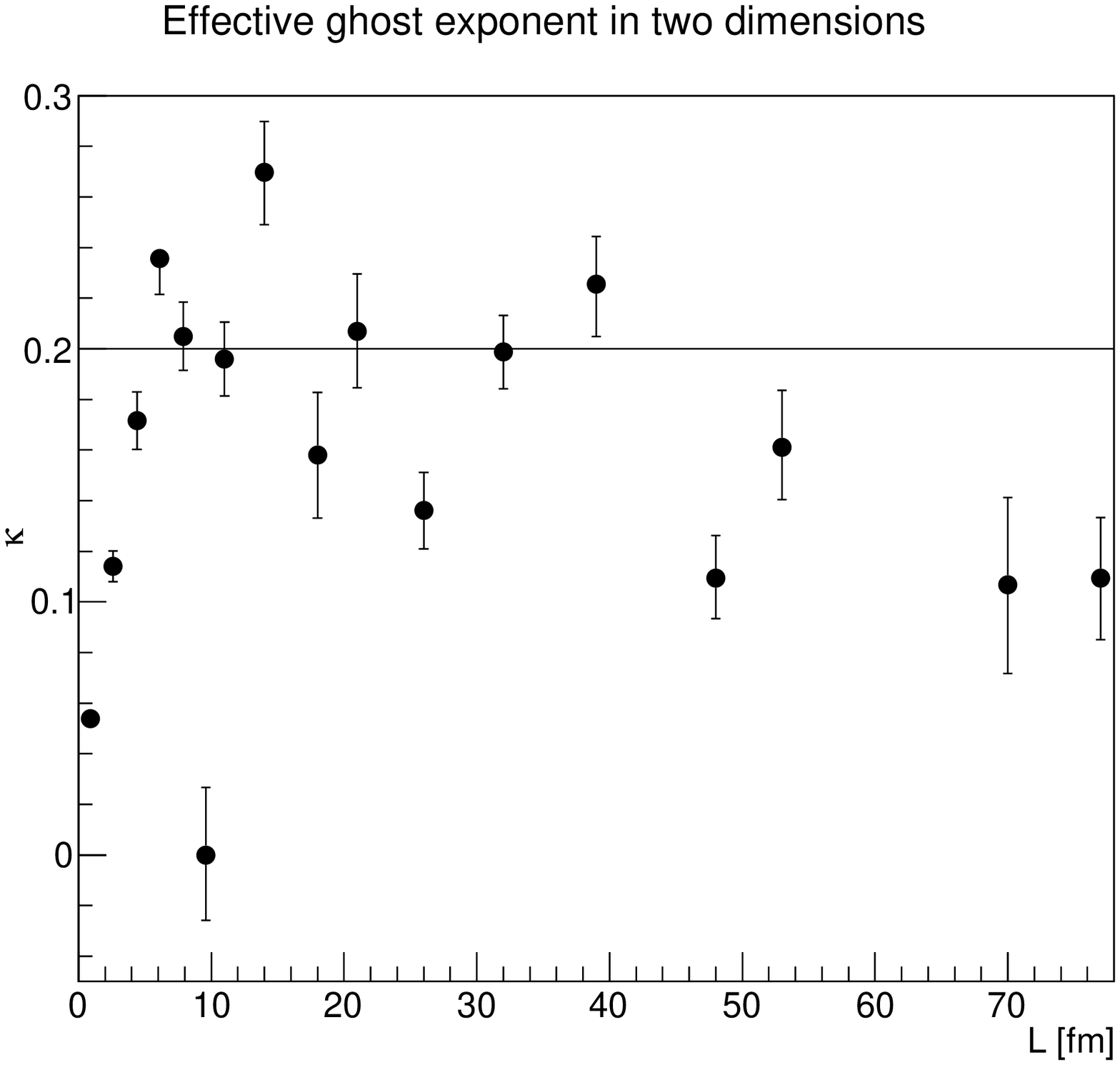}\includegraphics[width=0.5\linewidth]{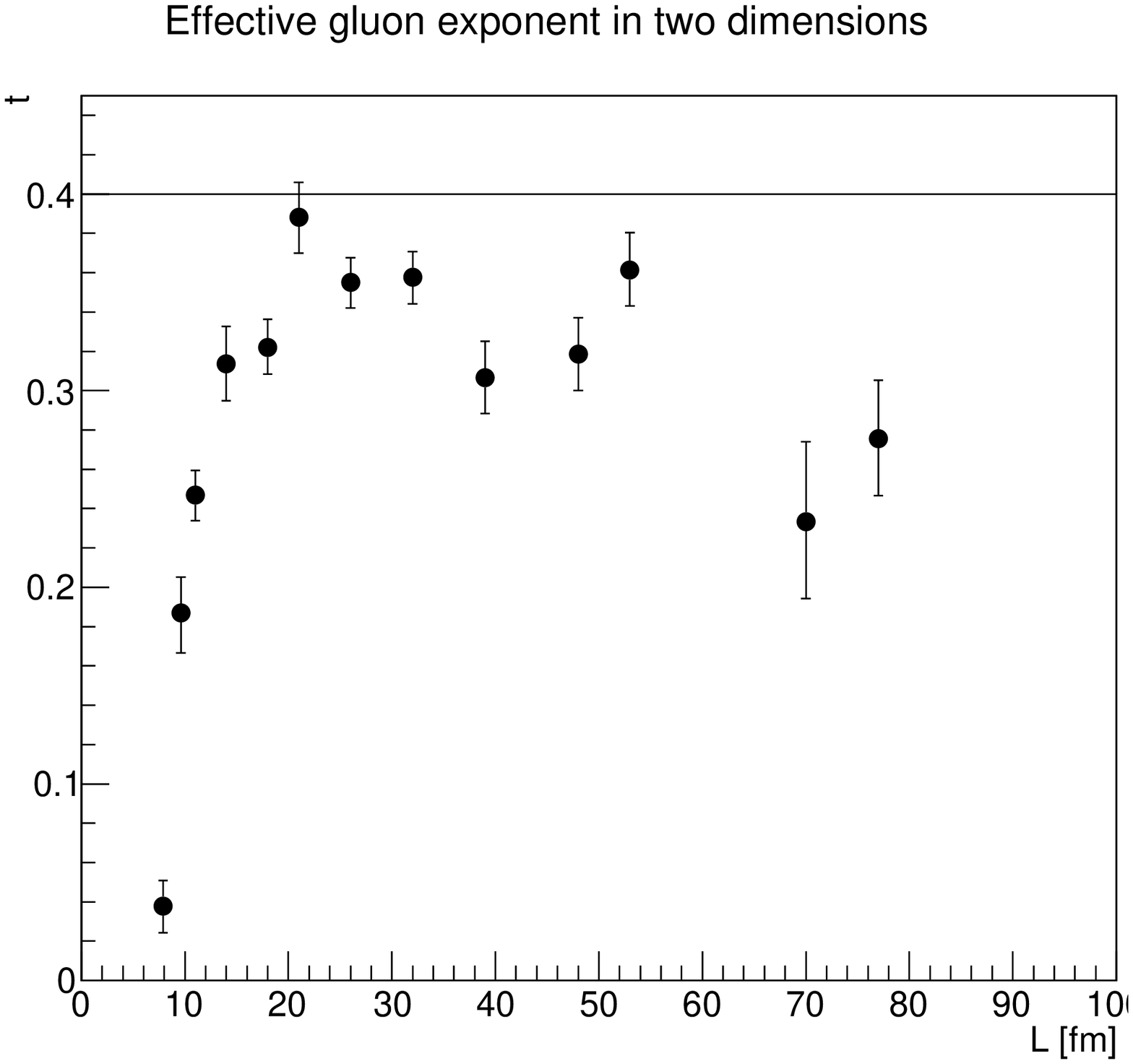}
\caption{\label{fig:exv}The volume dependence of the effective exponents for the ghost (left panel) and the gluon (right panel) in two dimensions for the respective finest discretization for every volume.}
\end{figure}

To better characterize this behavior, a fit with an effective exponent can be performed, in the same way as for the gluon \cite{Fischer:2007pf,Maas:2007uv}, using the ansatz
\be
p^2D_{G}(p)\stackrel{p\ll\Lambda_\text{YM}}{\sim}p^{2\kappa}\label{effexpg}.
\ee
\no In two dimensions the exponent $\kappa$ is expected to be 0.2 \cite{Zwanziger:2001kw,Lerche:2002ep}. The results are shown in figure \ref{fig:ghpex}. The rather strongly fluctuating results are a consequence of the larger statistical fluctuations due to the here employed point-source method. Nonetheless, the trends are clearly visible. In two dimensions, and to some extent in three dimensions, the exponent is, within errors, insensitive to the lattice spacing. In three dimensions, it slowly decreases with volume. In two dimensions, it appears to stabilize at a non-zero value, though as in the case of the gluon at a value which is not precisely coinciding with the expected one, but somewhat smaller. To emphasize this, the volume-dependence of both the gluon and the ghost exponent are shown in figure \ref{fig:exv}. Though the effect is small, at most at the few $\sigma$-level, there is a consistent trend in both cases for the exponents to be below the expected ones at large volumes. In four dimensions, the exponent decreases both with lattice spacing and volume. This is also visible in the ghost dressing function directly: A close inspection shows that the value for the dressing function at the smallest non-zero momentum slowly decreases towards a saturation value with decreasing lattice spacing.

\section{Running coupling}\label{salpha}

A particular convenient feature of Landau gauge is that the running coupling $\alpha(p^2)$, defined from the ghost-gluon vertex, can already be calculated by only using the propagators as \cite{vonSmekal:1997vx}
\be
\alpha(p)=\alpha(\mu)p^6D_G(p,\mu)^2D(p,\mu)\label{alpha},
\ee
\no as long as the propagators are renormalized as $\mu^6D_G(\mu^2,\mu^2)^2D(\mu^2,\mu^2)=1$. This has been repeatedly used in the literature to determine the characteristic scale $\Lambda$ in the presently employed minimal momentum subtraction scheme \cite{vonSmekal:2009ae}, see e.\ g.\ \cite{Sternbeck:2010xu,Boucaud:2008gn}. In lower dimensions, $\alpha$ has a mass dimension. To obtain a dimensionless quantity, it can be divided by $p^{4-d}$. The such defined quantity is expected to have an infrared finite value in two dimensions, due to the relation $2t-\kappa=0$ between the gluon exponent \pref{effexpz} and the ghost exponent \pref{effexpg} \cite{Zwanziger:2001kw,Lerche:2002ep}.

The expected leading perturbative running immediately follows from \prefr{uv2}{uv4} and \prefr{uvg2}{uvg4},
\bea
\alpha_{2d}(p)&=&\frac{g^2}{4\pi}\left(1-\frac{(c+f)g^2}{p^2}\right)\label{uva2}\\
\alpha_{3d}(p)&=&\frac{g^2}{4\pi}\left(1+\frac{19g^2}{32p}\right)\label{uva3}\\
\alpha_{4d}(p)&=&\frac{\alpha(\mu)}{1+\frac{33\alpha(\mu)}{52\pi}\ln\frac{p^2}{\mu^2}}\approx\frac{52\pi}{33\ln\frac{p^2}{\mu^2}}\label{uva4},
\eea
\no exhibiting clearly the character of a power series in the coupling $g$, and being uniquely determined once this value is set.

\begin{figure}
\includegraphics[width=\linewidth]{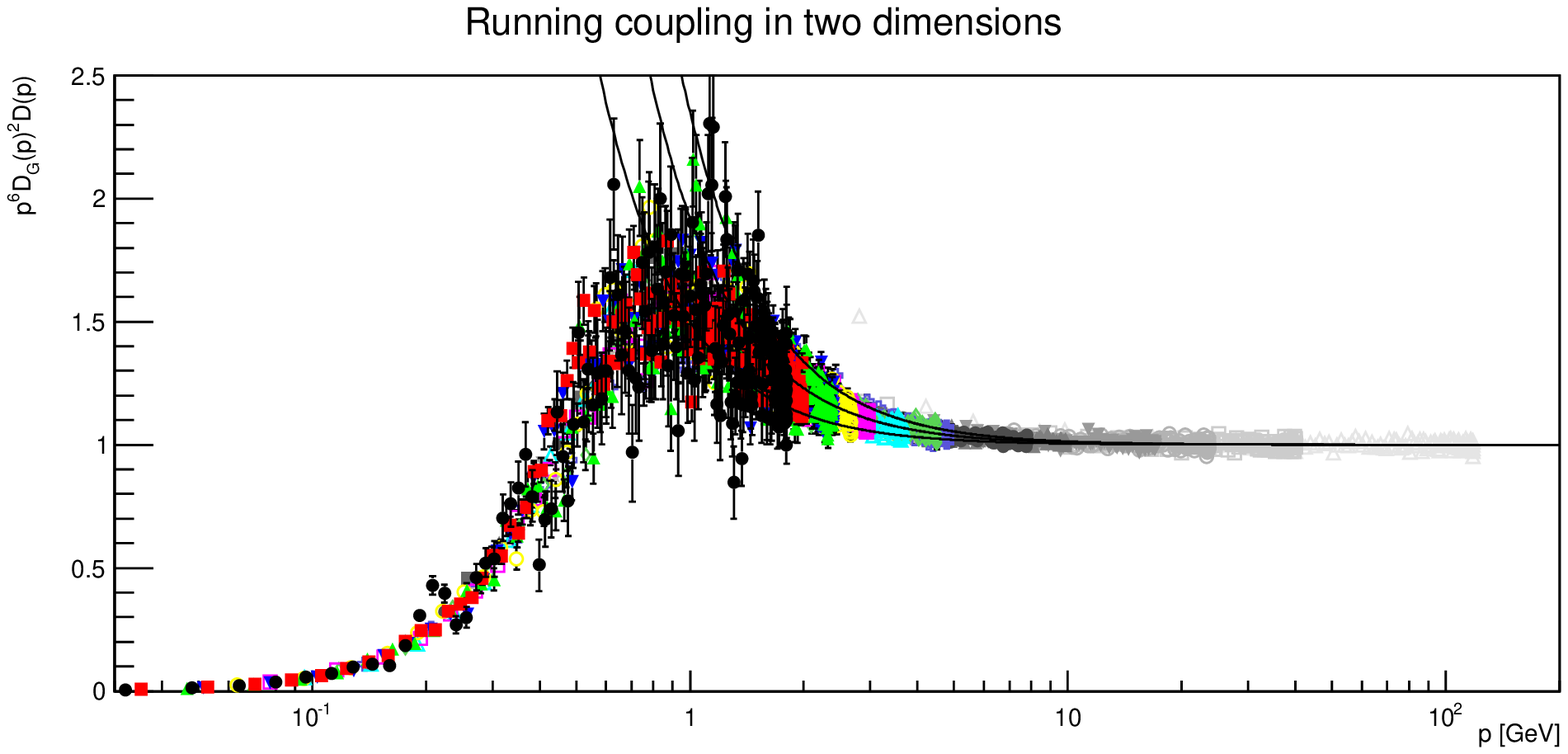}
\includegraphics[width=\linewidth]{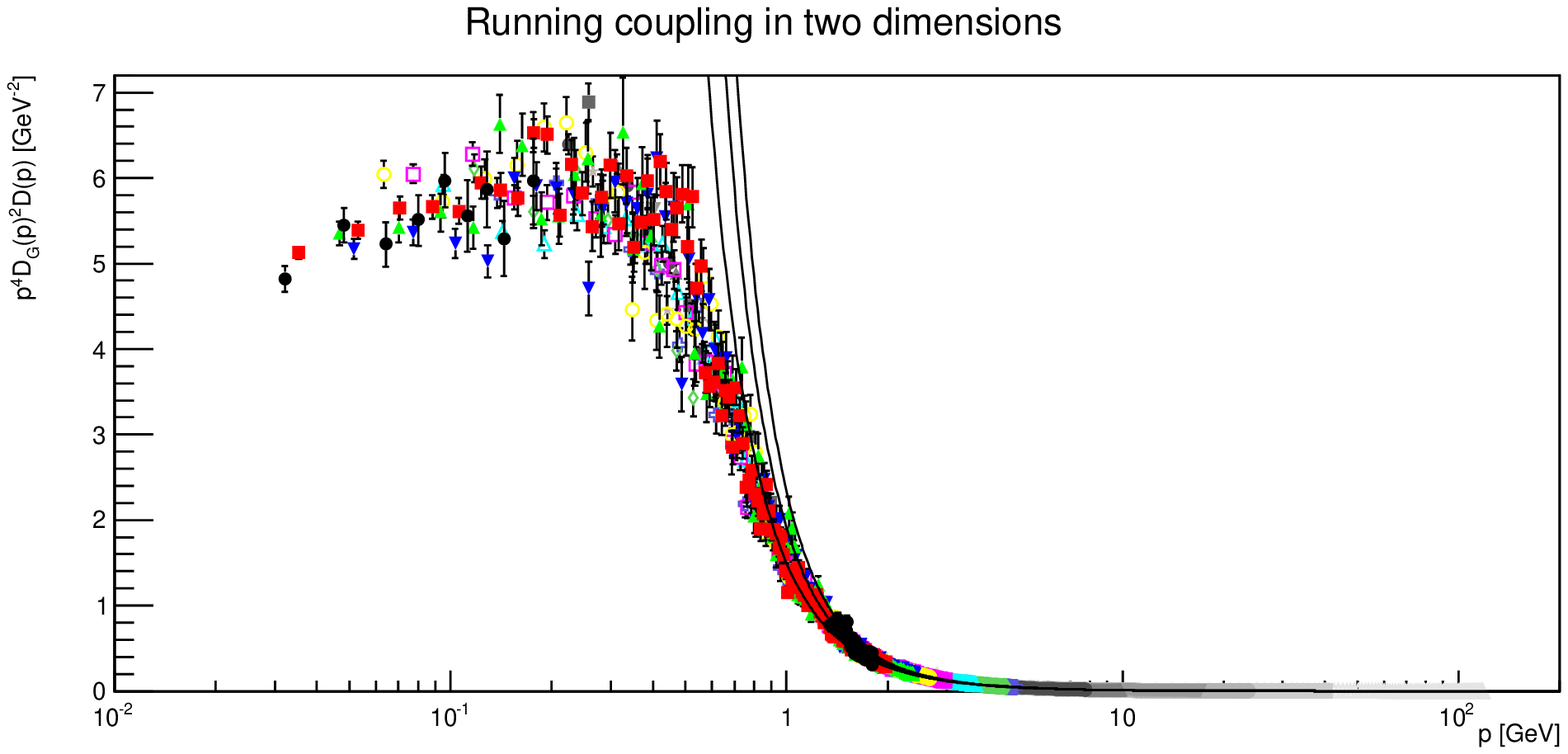}\\
\caption{\label{fig:alpha2d}The running coupling in two dimensions is shown in the top panel. The bottom panel shows the running coupling divided by $p^2$, to obtain a dimensionless quantity. All momenta are along the $x$-axis. The band is the perturbative result \pref{uva2}. Points with relative errors larger than 10\% have been suppressed. Symbols have the same meaning as in figure \ref{fig:gpuv2}.}
\end{figure}

\begin{figure}
\includegraphics[width=\linewidth]{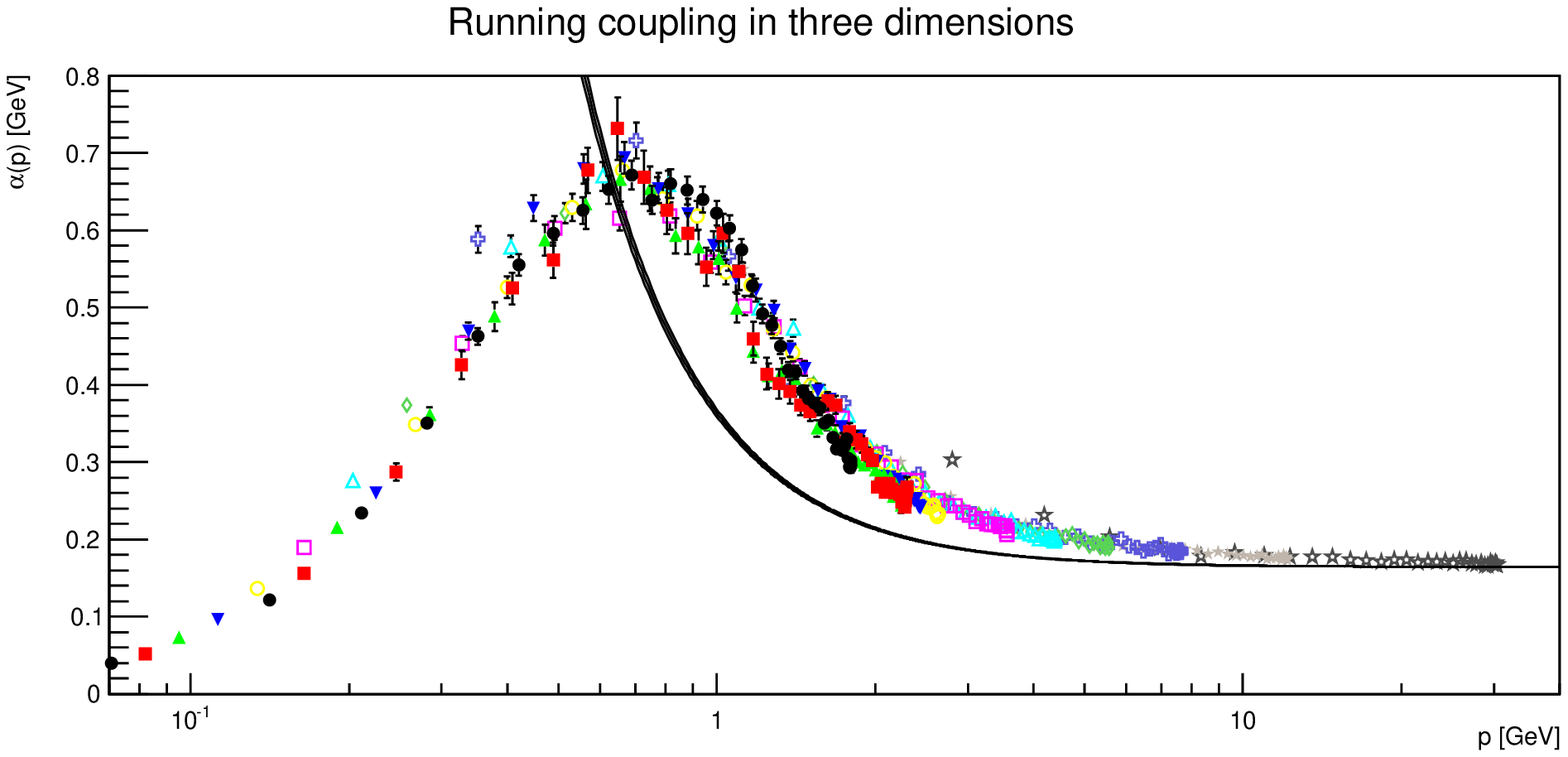}
\includegraphics[width=\linewidth]{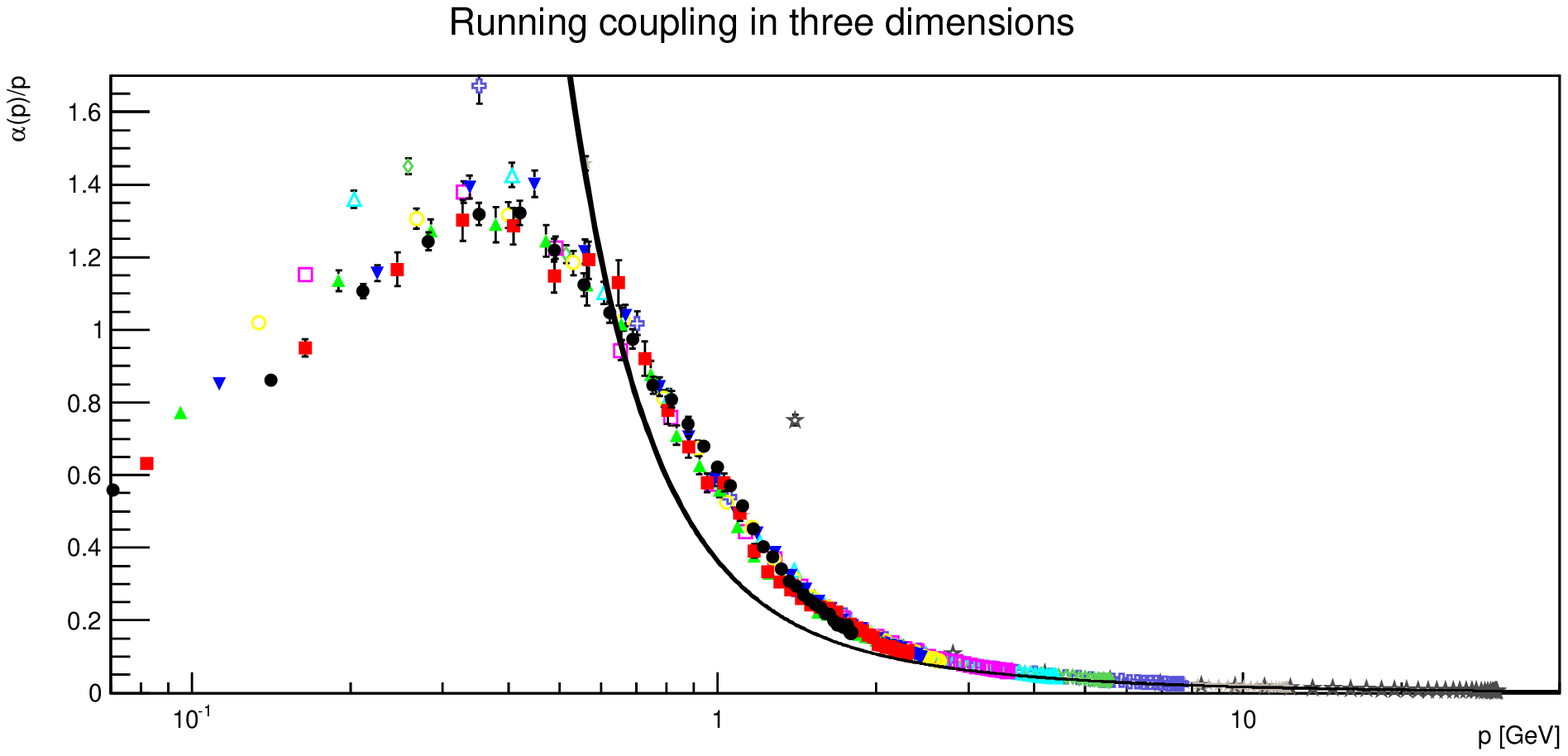}
\caption{\label{fig:alpha3d}The running coupling in three dimensions is shown in the top panel. The bottom panel shows the running coupling divided by $p$, to obtain a dimensionless quantity. All momenta are along the $x$-axis. The band is the perturbative result \pref{uva3}. Symbols have the same meaning as in figure \ref{fig:gpuv3}.}
\end{figure}

\begin{figure}
\includegraphics[width=\linewidth]{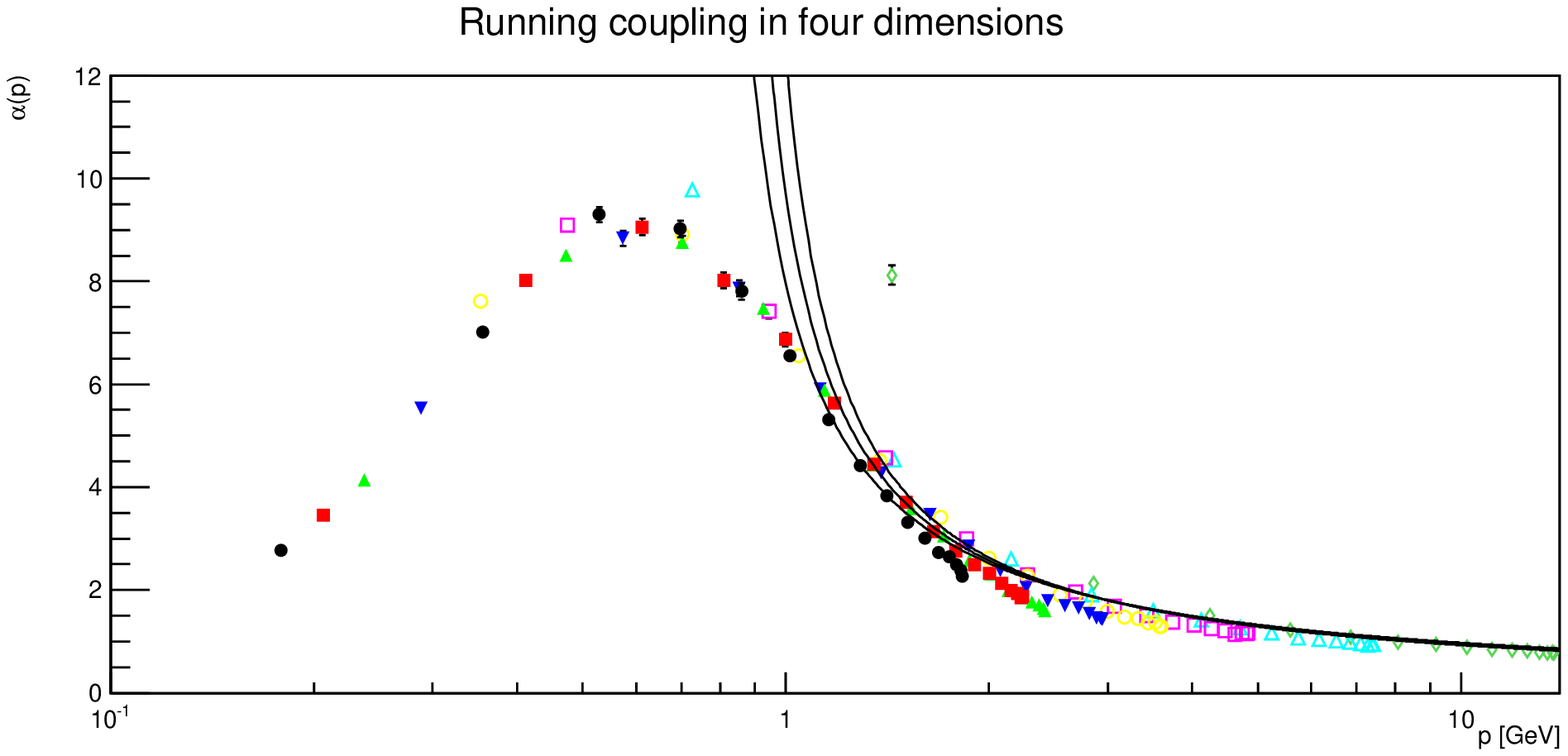}\\
\includegraphics[width=\linewidth]{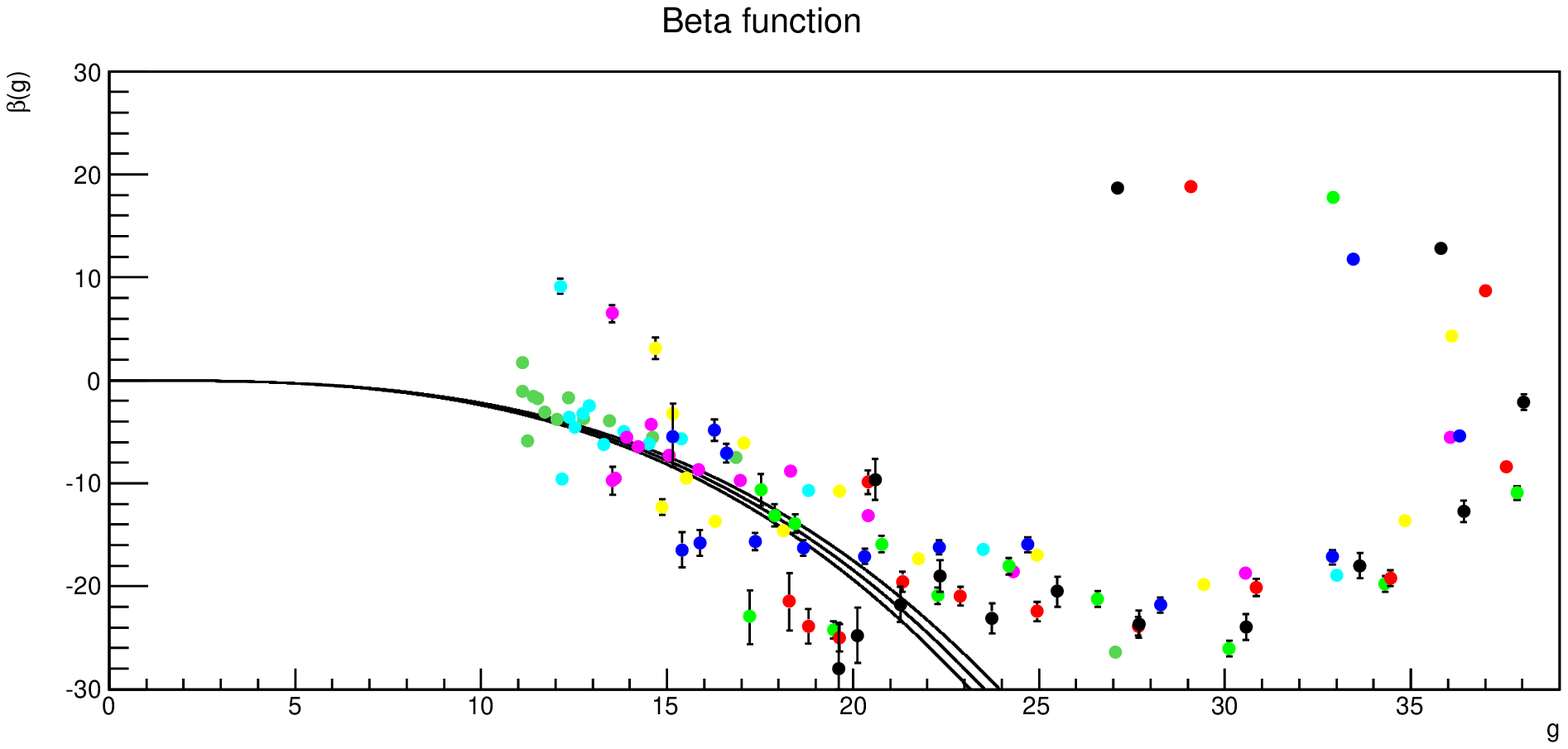}
\caption{\label{fig:alpha4d}The running coupling in four dimensions is shown in the top panel and the corresponding $\beta$-function \pref{bfunc} in the bottom panel. All momenta are along the $x$-axis. The band is the perturbative result \pref{uva4}. Symbols have the same meaning as in figure \ref{fig:gpuv4}.}
\end{figure}

The results are shown in figure \ref{fig:alpha2d}-\ref{fig:alpha4d}. In all cases the perturbative behavior is observed in a way as expected from how the perturbative behavior manifested itself in the propagators.

In four dimensions the dimensionless coupling is both infrared and ultraviolet vanishing, with a maximum in between. Of course, it is always possible to redefine the coupling by a scheme transformation \cite{Maas:2011se}, such that it adheres to the expected behavior, i.\ e.\ infrared non-vanishing and ultraviolet vanishing see \cite{Fischer:2008uz,Aguilar:2008fh}. Also shown is its corresponding $\beta$-function, implicitly defined as
\be
\beta(g(p))=p\frac{\pd g(p)}{\pd p}\label{bfunc}.
\ee
\no Due to the maximum, this $\beta$-function has, besides the Gaussian ultraviolet fixed-point, a second zero, and then tends from above towards the infrared Gaussian fixed point in this scheme. Note that in three and two dimensions the statistical accuracy was insufficient to perform the necessary numerical differentiations to obtain a meaningful result, so only the case of four dimensions is presented here.

The three-dimensional case is as expected from the propagators. The running coupling vanishes at small momenta. At the same time it becomes non-zero and constant at large momenta, which is ironically the consequence of asymptotic freedom: Because both propagators become constant and non-zero at large momenta, so must the running coupling determined from \pref{alpha}. Only after dividing out a power of momentum, the running coupling tends polynomial to zero at both large and small momenta.

The situation in two dimensions is the only one offering a slight surprise\footnote{Note that the lowest momentum point is severely affected by lattice artifacts, and therefore dropped for every volume \cite{Maas:2007uv}. This is partly a finite lattice spacing effect, but even at the present lattice spacing still severe.}. In previous investigations, with their smaller statistical reliability \cite{Maas:2007uv}, the coupling appeared to become infrared finite without any maximum. Here, a maximum is seen. This could have been anticipated from figure \ref{fig:gpex} and \ref{fig:ghpex}, as the exponents of the gluon and ghost propagator do not fulfill the necessary relation $2\kappa-t=0$ \cite{Zwanziger:2001kw,Lerche:2002ep} to obtain an infrared finite running coupling according to the definition \pref{alpha}. At the current statistics and for the presently available volumes, the results do not indicate any flattening out that would indicate that there is a maximum, but still a finite running coupling at zero momentum. On the other hand, the statistical accuracy forbids to exclude the opposite. It remains therefore an interesting open question, what its behavior is at zero momentum, especially as continuum studies favor an infrared finite coupling \cite{Dudal:2012td,Huber:2012td}.

Though this cannot be answered with the present limited set of volumes here, a speculation can be offered. It has been argued \cite{Fischer:2008uz} that in four dimensions the realization of an infrared finite running coupling in this scheme is tied to a globally well-defined, and thus non-perturbative, BRST with the same algebra as the perturbative one. Arguments have been provided that such a global BRST is only possible when averaging over all Gribov copies, especially also over all Gribov copies outside the first Gribov region \cite{Neuberger:1986xz,vonSmekal:2007ns,vonSmekal:2008es,vonSmekal:2008ws,Maas:2012ct}. In this construction, there is no restriction evident to four dimensions, and it appears therefore plausible that this is also correct in two dimensions\footnote{But note the differences to two-dimensional Coulomb gauge \cite{Reinhardt:2008ij}.}. One can then speculate further that also in two dimensions the running coupling in this scheme can only be infrared finite when performing this average. At the same time, the propagators, especially the gluon propagator, cannot show the same behavior as in higher dimensions, due to the infrared singularities \cite{Dudal:2012td}. Therefore, an integer-power-like vanishing running coupling would not be possible, and thus the present non-integer-power-law-like behavior emerges. Again, this is only a speculation, and it may well be that (again) results from larger volumes will show an infrared constant running coupling. It would then be a very interesting question why the BRST arguments from higher dimensions do not apply in two dimensions.

\section{Summary}\label{ssum}

Herein a comprehensive analysis of the 2-point functions of minimal-Landau-gauge Yang-Mills theory in two, three, and four dimensions has been presented. The results show that quantitatively discretization effects start to die out at around an inverse lattice spacing of 2-3 GeV, and are in all cases rather small. Furthermore, except for the running coupling in two dimensions and the Schwinger function, the present results otherwise confirm the qualitative behavior known from other investigations \cite{Maas:2011se}. However, the behavior of the Schwinger function shows further interesting constraints for the analytic structure, which must be taken into account in future investigations. Secondly, the running coupling in two dimensions shows an unexpected behavior. With the present resources, it was not possible to ultimately clarify whether this is a lattice artifact. In either case, its behavior will ultimately give an interesting hint for our understanding of the global properties of the Landau gauge.\\

\no{\bf Acknowledgments}

I am grateful to A.\ Sternbeck for a critical reading of the manuscript. This project was supported by the DFG under grant number MA 3935/5-1 and MA 3935/8-1 (Heisenberg program) and the FWF under grant number M1099-N16. Simulations were performed on the HPC cluster at the Universities of Jena and Graz. I am grateful to the corresponding HPC teams for the very good performance of both clusters. The ROOT framework \cite{Brun:1997pa} has been used in this project.

\appendix

\bibliographystyle{bibstyle}
\bibliography{bib}


\end{document}